\documentclass[english,twocolumn]{IEEEtran}
\usepackage[T1]{fontenc}
\usepackage{babel}
\usepackage{array}
\usepackage{booktabs}
\usepackage{multirow}
\usepackage[ruled,vlined,linesnumbered]{algorithm2e}
\usepackage{amsthm}
\usepackage{amsmath}
\usepackage{amssymb}
\usepackage{graphicx}
\usepackage[unicode=true,
 bookmarks=true,bookmarksnumbered=true,bookmarksopen=true,bookmarksopenlevel=1,
 breaklinks=false,pdfborder={0 0 0},backref=false,colorlinks=false]
 {hyperref}
\hypersetup{pdftitle={Your Title},
 pdfauthor={Your Name},
 pdfpagelayout=OneColumn, pdfnewwindow=true, pdfstartview=XYZ, plainpages=false}

\makeatletter

\providecommand{\tabularnewline}{\\}

\theoremstyle{plain}
\newtheorem{thm}{\protect\theoremname}
\theoremstyle{plain}
\newtheorem{lem}[thm]{\protect\lemmaname}

\usepackage[caption=false,font=footnotesize]{subfig}

\makeatother

\providecommand{\lemmaname}{Lemma}
\providecommand{\theoremname}{Theorem}

\begin{document}
\global\long\def\atanh{\operatorname{atanh}}

\title{Approaching the Rate-Distortion Limit with Spatial Coupling, Belief
Propagation and Decimation}

\author{Vahid~Aref, Nicolas~Macris and~Marc~Vuffray%
\thanks{Vahid~Aref is with Bell labs, Alcatel-Lucent AG and the Institute
of telecommunications, Stuttgart University, Stuttgart, Germany, e-mail:
\protect\href{mailto:vahid.aref@alcatel-lucent.com}{vahid.aref@alcatel-lucent.com}.%
}%
\thanks{Nicolas~Macris is with the School of Computer and Communication Science,
Ecole Polytechnique Fédérale de Lausanne, Lausanne, Switzerland, e-mail:
\protect\href{mailto:nicolas.macris@epfl.ch}{nicolas.macris@epfl.ch}.%
}%
\thanks{Marc~Vuffray is with the Theory Division and Center for Nonlinear
Studies, Los Alamos National Laboratory, Los Alamos NM, USA, e-mail:
\protect\href{mailto:vuffray@lanl.gov}{vuffray@lanl.gov}.%
}}
\maketitle
\begin{abstract}
We investigate an encoding scheme for lossy compression of a binary
symmetric source based on simple spatially coupled Low-Density Generator-Matrix
codes. The degree of the check nodes is regular and the one of code-bits
is Poisson distributed with an average depending on the compression
rate. The performance of a low complexity Belief Propagation Guided
Decimation algorithm is excellent. The algorithmic rate-distortion
curve approaches the optimal curve of the ensemble as the width of
the coupling window grows. Moreover, as the check degree grows both
curves approach the ultimate Shannon rate-distortion limit. The Belief
Propagation Guided Decimation encoder is based on the posterior measure
of a binary symmetric test-channel. This measure can be interpreted
as a random Gibbs measure at a ``temperature'' directly related
to the ``noise level of the test-channel''. We investigate the links
between the algorithmic performance of the Belief Propagation Guided
Decimation encoder and the phase diagram of this Gibbs measure. The
phase diagram is investigated thanks to the cavity method of spin
glass theory which predicts a number of phase transition thresholds.
In particular the dynamical and condensation ``phase transition temperatures''
(equivalently test-channel noise thresholds) are computed. We observe
that: (i) the dynamical temperature of the spatially coupled construction
saturates towards the condensation temperature; (ii) for large degrees
the condensation temperature approaches the temperature (i.e. noise
level) related to the information theoretic Shannon test-channel noise
parameter of rate-distortion theory. This provides heuristic insight
into the excellent performance of the Belief Propagation Guided Decimation
algorithm. The paper contains an introduction to the cavity method. \end{abstract}
\IEEEpeerreviewmaketitle

\section{Introduction}

\IEEEPARstart{L}{ossy} source coding is one of the oldest and most
fundamental problems in communications. The objective is to compress
a given sequence so that it can be reconstructed up to some specified
distortion. It was established long ago \cite{Goblick63} that Shannon's
rate distortion bound for binary sources (under Hamming distance)
can be achieved using linear codes. However, it is of fundamental
importance to find low complexity encoding schemes that achieve the
rate distortion limit. An early attempt used trellis codes \cite{ViterbiTrellis74},
for memoryless sources and bounded distortion measures. It is possible
to approach the Shannon limit as the trellis constraint length increases,
but the complexity of this scheme, although linear in the block length
$N$, becomes exponential in the trellis constraint length. In \cite{kostina2012fixed}
an entirely different scheme is proposed (also with linear complexity
and diverging constants) based on the concatenation of a small code
and optimal encoding of it. More recently, important progress was
achieved thanks to polar codes \cite{ArikanPolar09} which were shown
to achieve the rate-distortion bound with a successive cancellation
encoder of complexity $O(N\ln N)$ \cite{KoradaPolar10}. Further
work on the efficient construction of such codes followed \cite{Tal2013}.

Another interesting recent direction uses Low-Density Generator-Matrix (LDGM) codes
as first investigated in \cite{Martinian2003}
 for binary erasure sources
 and in  \cite{murayama04} for symmetric Bernoulli sources. 
LDGM based codes with
Poisson degrees for code-bit nodes and regular degree for check nodes,
achieve the ultimate Shannon rate-distortion limit under optimal encoding
when the check degrees grow large. This conclusion was reached
(by non-rigorous means) from the replica \cite{murayama04} and cavity \cite{ciliberti05source}
methods from statistical physics. This was later proved in~\cite{Wainwright10LDGManalysis}
 by second moment methods.
These
studies also showed that the gap to the rate-distortion bound vanishes
exponentially in the large check degree limit. 

In \cite{ciliberti05source}. 
it was recognized that using a plain message passing algorithm without
decimation is not effective in lossy compression. Indeed the estimated
marginals are either non-converging or non-biased because there exists
an exponentially large number of compressed words that lead to roughly
the same distortion. One has to supplement Belief Propagation (BP) (or Survey Propagation (SP)) with
a decimation process. This yields and encoding scheme  of low complexity%
\footnote{$O(N^{2})$ or $O(N)$ depending on the exact implementation.%
}. In this respect the lossy compression schemes
based on random graphs are an incarnation of random constraint satisfaction
problems and, from this perspective it is not too surprising that
their analysis share common features. The general idea of BP or SP
guided-decimation algorithms is to: i) Compute approximate marginals
by message passing; ii) Fix bits with the largest bias, and if there
is no biased bit take a random decision; iii) Decimate the graph and
repeat this process on the smaller graph instance. For naive choices
(say regular, or check-regular) of degree distributions the Shannon
rate-distortion limit \textit{is not} approached by such algorithms.
However it has been observed that \textit{it is} approached for degree
distributions that have been optimized for channel LDPC coding~\cite{Wainwright10LDGManalysis},
\cite{Filler07BPLDGM}, \cite{CG10LSC}. These observations are empirical:
it is not clear how to analyze the decimation process, and there is
no real principle for the choice of the degree distribution.

In this contribution we investigate a \textit{simple spatially coupled}
LDGM construction. The degree distributions that we consider are regular
on the check side and Poisson on the code-bit side. The average of
the Poisson distribution is adjusted to achieve the desired compression
rate. We explore a low complexity Belief Propagation Guided Decimation
(BPGD) encoding algorithm, that takes advantage of spatial coupling,
and approaches the Shannon rate-distortion limit for large check degrees
and any compression rate. No optimization on the degree distributions
is needed. The algorithm is based on the posterior measure of a test
binary symmetric channel (BSC). We interpret this posterior as a random
Gibbs measure with an inverse temperature parameter equal to the half-log-likelihood
parameter of the test-BSC. This interpretation allows us to use the
cavity method of spin glass theory in order to investigate the phase
diagram of the random Gibbs measure. Although the cavity method is
not rigorous, it makes definite predictions about the phase diagram
of the measure. In particular it predicts the presence of phase transitions
that allow to gain insight into the reasons for the excellent performance
of the BPGD encoder on the spatially coupled lossy compression scheme.

Spatially coupled codes were first introduced in the context of channel
coding in the form of convolutional LDPC codes~\cite{ZigFel} and
it is now well established that the performance of such ensembles
under BP decoding is consistently better than the performance of the
underlying ensembles \cite{IDLDPCC}, \cite{TerminLDPCCCthreshold},
\cite{ProtoLDPCC}. This is also true for coupled LDGM ensembles in
the context of rateless codes~\cite{Aref11UniRateless}. The key
observation is that the BP threshold of a coupled ensemble saturates
towards the maximum a posteriori MAP threshold of the underlying ensemble
as the width of the coupling window grows. A proof of this \emph{threshold
saturation phenomenon} has been accomplished in \cite{kudekar2011threshold},
\cite{kudekar2012spatially}. An important consequence is that spatially
coupled regular LDPC codes with large degrees universally achieve
capacity. Recently, more intuitive proofs based on replica symmetric
energy functionals have been given in \cite{Yelda12threshsat}, \cite{kumar2012proof}.
Spatial coupling has also been investigated beyond coding theory in
other models such as the Curie-Weiss chain, random constraint satisfaction
problems~\cite{Hassani10Couplgraphical}, \cite{HMU10}, \cite{Hassani11SAT},
and compressed sensing~\cite{Kudekar10compress}, \cite{Krzakala2012},
\cite{donoho2012information}.

Let us now describe in more details the main contents of this paper.
Summaries have appeared in \cite{amuv2012}, \cite{amv2013}. In \cite{amuv2012}
we had investigated regular spatially coupled graph constructions
with constant degrees for both check and code-bits. The performance
of the BPGD algorithm are similar to the case of Poisson degree for
code-bit nodes, on which we will concentrate here.

In section \ref{sec:framework} we set up the framework for lossy
source coding with spatially coupled LDGM ensembles for a binary symmetric
Bernoulli source and Hamming distortion. We investigate ensembles
with regular check degrees and Poisson code-bit node degrees. Important
parameters of the spatial constructions are the number of positions
$L$, the number of nodes $n$ at each position, and the window width
$w$ over which we couple the nodes. The infinite block length limit
investigated in this paper corresponds to $\lim_{L\to+\infty}\lim_{n\to+\infty}$
in the specified order. Optimal encoding consists in finding the compressed
word that minimizes the Hamming distortion between a given source
realization and the reconstructed word. Since we will use methods
from statistical mechanics, we will translate the problem in this
language. Optimal encoding can be viewed as the search for the minimum
energy configurations of a random spin Hamiltonian. Although the optimal
encoder is computationally impractical, it is important to determine
the optimal distortion of the ensemble in order to set a limit on
what cannot be achieved algorithmically for the ensemble. In this
respect, an important rigorous result that is reviewed in section
\ref{sec:framework} is that, in the infinite block length limit $\lim_{L\to+\infty}\lim_{n\to+\infty}$,
for any fixed $w$ the optimal distortion for a spatially coupled
ensemble is equal to the optimal distortion for the underlying uncoupled
ensemble (and is therefore independent of $w$). This result follows
from an equivalent one proved in \cite{Hassani11SAT} for the random
XORSAT problem for any values of the constraint density. There are
various results in the literature about the optimal encoder for the
uncoupled ensemble. So we can essentially transfer them directly to
our spatially coupled setting. 

As explained in section \ref{sec:framework} optimal encoding can
be viewed as the study of the \textit{zero temperature limit} of the
Gibbs measure associated with a Hamiltonian. This Gibbs measure forms
the basis of the BP based algorithms that we use. This Gibbs measure
is nothing else than the posterior measure of the dual test-channel
problem, and that the inverse temperature is the half-log-likelihood
parameter of a test-BSC%
\footnote{More precisely, if $p$ is the flip parameter of the BSC test-channel
then the inverse temperature is $\beta=\frac{1}{2}\ln(\frac{1-p}{p})$.%
}. The free energies of the spatially coupled and underlying ensembles
are the same \cite{Hassani11SAT} in the infinite block length limit
(fixed $w$) and therefore their \textit{static} phase transition
temperature (the condensation temperature) is also the same (see below).

The Gibbs measure (or posterior measure of the dual test-channel problem)
is the basis for setting up the BPGD algorithms. This is explained
in detail in Section \ref{sec:bpgd}. The crucial point is the use
of the spatial dimension of the graphical construction. The main idea
is that when the biases are small a random bit \textit{from the boundary
of the chain} is fixed to a random value, and as long as there exist
bits with large biases they are eliminated from the chain by fixing
them and decimating the graph. We consider two forms of BPGD. The
first one, which as it turns out performs slightly better, is based
on hard decisions. The second one uses a randomized rounding rule
for fixing the bits.

Section \ref{sec:results} reviews the simulation results and discusses
the performance for the two versions of the BPGD encoders. For both
algorithms we observe that the rate-distortion curve of the coupled
ensemble approaches the Shannon limit when $n>>L>>w>>1$ and the node
degrees get large. We cannot assess if the Shannon limit is \textit{achieved}
based on our numerical results. However we observe that in order to
avoid finite size effects the degrees have to become large only after
the other parameters grow large in the specified order. In practice
though $n=2000$, $L=64$, $w=3$ and check degrees equal to $l=3$
yield good results for a compression rate $1/2$. The performance
of the BPGD algorithms depend on the inverse temperature parameter
in the Gibbs measure, and one can optimize with respect to this parameter.
Interestingly, for the coupled ensemble, we observe that for large
degrees (when Shannon's rate-distortion limit is approached) the optimal
parameter corresponds to the information theoretic value of the flip
probability given by the Shannon distortion. This is non-trivial:
indeed it is not true for the uncoupled ensemble.

The behavior of BPGD algorithms is to some extent controlled by the
phase transitions in the phase diagram of the Gibbs measure. In section
\ref{sec:CM} we review the predictions of the cavity method, and
in particular the predictions about the dynamical and condensation
phase transition temperatures. At the condensation temperature the
free energy displays a singularity and is thus a thermodynamic or
static phase transition threshold. The dynamical temperature on the
other hand is not a singularity of the free energy. As we will see in section \ref{sec:CM}
in the framework of the cavity method it is defined via a "complexity function" which counts 
the number of "pure states". The dynamical temperature is the value at which the 
complexity jumps to a non zero value. For a number of models it is known that Markov Chain
Monte Carlo algorithms have an equilibration time which diverges at
(and below) this dynamical temperature. 
Similarly, BPGD with randomized rounding
correctly samples the Gibbs measure down to temperatures slightly
higher than the dynamical threshold. We observe a threshold saturation
phenomenon for the spatially coupled construction. First as said above,
since the condensation threshold is a singularity of the free energy
it is the same for the uncoupled and coupled ensembles for any $w$.
Second, as the window width $w$ grows the dynamical threshold saturates
towards the condensation one. In practice we observe this saturation
for values of $w$ as low as $w=3,4,5$. Thus for spatially coupled
codes the BPGD algorithm is able to correctly sample the Gibbs measure
down to a temperature approximately equal to the condensation threshold.
This explains why the algorithm performs well, indeed it is able to
operate at much lower temperatures than in the uncoupled case. A large
degree analysis of the cavity equations shows that the condensation
temperature tends to the information theoretic value corresponding
to the flip parameter of the BSC test-channel given by Shannon's distortion.
These facts, put together, provide insight as to the excellent performance
of the BPGD algorithm for the spatially coupled ensemble.

Section \ref{sec:CM_application} presents the cavity equations for
the coupled ensemble on which the results of the previous paragraph
are based. These equations are solved by population dynamics in Section
\ref{sec:Spatially_Coupling_Effect}. The cavity equations take the
form of six fixed point integral equations. However we observe by
population dynamics that two of them are satisfied by a trivial fixed
point. This is justified by a theoretical analysis in Section \ref{sec:large-degree-limit-theorems}.
When this trivial fixed point is used the remaining four equations
reduce to two fixed point integral equations which have the form of
usual density evolution equations for a BSC channel. This simplification
is interesting because although the original Gibbs measure
does not possess channel symmetry
\footnote{In the context of spin glass theory this is the Nishimori gauge symmetry.%
}, this symmetry emerges here as a solution of the cavity equations. Within
this framework the saturation of the dynamical temperature towards
the condensation one appears to be very similar than threshold saturation
in the context of channel coding with LDPC codes. A proof of threshold saturation for the present problem is 
beyond the scope of this paper, but we do give in Section \ref{sec:large-degree-limit-theorems} 
a few insights on possible ways to attack it. 

For an introduction to the cavity theory we refer the reader to the
book \cite{mezard09information}. This theory is not easy to grasp
both conceptually and technically. This paper contains a high level
introduction of the main concepts in Section \ref{sec:CM} and a summary
of the main technical ideas in Appendix \ref{sec:cavityprimer}. We
hope that this will be helpful for unfamiliar readers. The necessary
derivations and adaptations to the present setting of a spatially
coupled Gibbs measure are summarized in Appendices \ref{sec:application}
and \ref{sec:CM_population_dynamic}. The main sections \ref{sec:framework}-\ref{sec:CM}
and the conclusion can be read without explicitly going into the cavity
formalism.

\section{Coupled LDGM Ensembles for lossy compression\label{sec:framework}}

\subsection{Lossy Compression of Symmetric Bernoulli Sources\label{sub: lossy}}

Let $\underline{X}=\{X_{1},X_{2},\dots,X_{N}\}$ represent a source
of length $N$, where $X_{a}$, $a=1,\dots,N$ are i.i.d Bernoulli($1/2$)
random variables. We compress a source word $\underline{x}$ by mapping
it to one of $2^{NR}$ index words $\underline{u}\in\left\{ 0,1\right\} {}^{NR}$,
where $R\in[0,1]$ is the compression rate. This is the encoding operation.
The decoding operation maps the stored sequence $\underline{u}$ to
a reconstructed sequence $\underline{\widehat{x}}(\underline{u})\in\left\{ 0,1\right\} ^{N}$.

For a given pair $(\underline{x},\underline{\widehat{x}})$, we measure
the distortion by the relative Hamming distance 
\begin{equation}
d_{N}(\underline{x},\underline{\widehat{x}})=\frac{1}{N}\sum_{a=1}^{N}\left|x_{a}-\widehat{x}_{a}\right|.\label{eq:hamming-distortion}
\end{equation}
The quality of reconstruction is measured by the average distortion
\begin{equation}
D_{N}(R)=\mathbb{E}_{\underline{X}}[d_{N}(\underline{x},\underline{\widehat{x}})]
\end{equation}
where $\mathbb{E}_{\underline{X}}$ is the expectation with respect
to the symmetric Bernoulli source.

For the symmetric Bernoulli source considered here, it is well-known
that for any encoding-decoding scheme, the average distortion is lower
bounded by Shannon's rate-distortion curve \cite{cover12elements}
\begin{equation}
D_{{\rm sh}}(R)=h_{2}^{-1}(1-R)
\end{equation}
where $h_{2}(x)=-x\log_{2}x-(1-x)\log_{2}(1-x)$ is the binary entropy
function. The rate-distortion curve is convex decreasing with $D_{{\rm sh}}(0)=1/2$
and $D_{{\rm sh}}(1)=0$.

\subsection{Spatially Coupled Low-Density Generator Matrix Constructions}

Our lossy source coding scheme is based on a spatially coupled LDGM
code ensemble. We first describe the \emph{underlying} ensemble.

\subsubsection{Underlying Poisson LDGM$(l,R)$ Ensemble}

These are bipartite graphs with a set $C$ of $n$ check nodes of
constant degree $l$, a set $V$ of $m$ code-bit nodes of variable
degree, and a set $E$ of edges connecting $C$ and $V$. The ensemble
of graphs is generated as follows: each edge emanating from a check
node is connected uniformly at random to one of the code-bit nodes.
The degree of code-bit nodes is a random variable with Binomial distribution
${\rm Bi}(ln,1/m)$. In the asymptotic regime of large $n,m$ with
$m/n=R$ the code-bit node degrees are i.i.d Poisson distributed with
an average degree $l/R$.
Note that this construction allows the possibility to have multi-edges in the graph. 

\subsubsection{Spatially Coupled LDGM$(l,R,L,w,n)$ Ensemble}

\begin{figure}[tb]
\centering{}\includegraphics{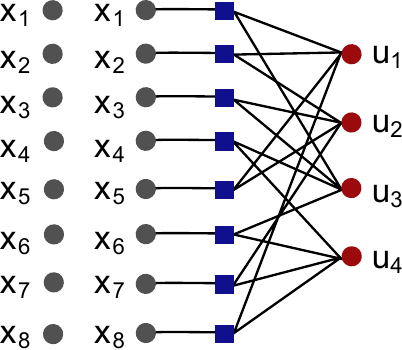}\protect\caption{\label{fig:factorgraph} A bipartite graph from the underlying LDGM$(2,0.5)$
ensemble. Here $n=8$, $m=4$ and $l=2$. Labels represent code-bits
$u_{i}$, reconstructed bits $\hat{x}_{i}$ and source bits $x_{i}$.}
\end{figure}

\begin{figure}[tb]
\centering{}\includegraphics{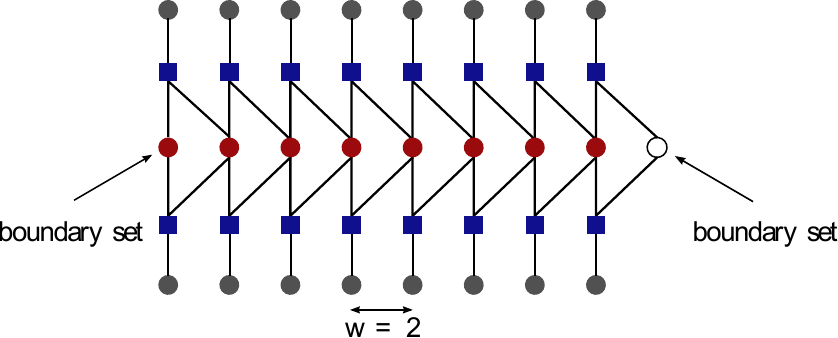}\protect\caption{\label{fig:chain} The ``protograph'' representation of the spatially
coupled LDGM$(2,0.5,L=8,w=2)$ ensemble. The code-bit nodes in boundary
sets have smaller degree than the code-bit nodes in the other sets.}
\end{figure}

We first lay out a set of positions indexed by integers $z\in\mathbb{Z}$
on a one dimensional line. This line represents a ``spatial dimension''.
We fix a ``window size'' which is an integer $w\geq1$. Consider
$L$ sets of check nodes each having $n$ nodes, and locate the sets
in positions $1$ to $L$. Similarly, locate $L+w-1$ sets of $m$
code-bit nodes each, in positions $1$ to $L+w-1$. All checks have
constant degree $l$, and each of the $l$ edges emanating from a
check at position $z\in\{1,\dots,L\}$ is connected uniformly at random
to code-bit nodes within the range $\{z,\dots,z+w-1\}$. It is easy
to see that for $z\in\{w,\dots,L-w+1\}$, in the asymptotic limit
$n\to+\infty$, the code-bit nodes have Poisson degrees with average
$l/R$. For the remaining positions close to the boundary the average
degree is reduced. More precisely for positions on the left side $z\in\{1,\dots,w-1\}$
the degree is asymptotically i.i.d Poisson with average $l/R\times z/w$.
For positions on the right side $z\in\{L+1,\dots,L+w-1\}$ the degree
is asymptotically Poisson with average $l/R\times(L+w-z)/w$. Figures
\ref{fig:factorgraph} and \ref{fig:chain} give a schematic view
of an underlying and a spatially coupled graph.

\subsubsection{Notation}

Generic graphs from the ensembles will be denoted by $\Gamma$ or
$\Gamma(C,V,E)$. We will use letters $a,b,c$ for check nodes and
letters $i,j,k$ for code-bit nodes of a given graph (from underlying
or coupled ensembles). We will often make use of the notation $\partial a$
for the set of all code-bit nodes connected to $a\in C$, i.e. $\partial a=\left\{ i\in V|\left(i,a\right)\in E\right\} $.
Similarly, for $i\in V$, $\partial i=\left\{ a\in C|\left(i,a\right)\in E\right\} $.
For spatially coupled graphs the sets of nodes at a specified position
$z$ are $C_{z}$ and $V_{z}$.

\subsection{Decoding Rule and Optimal Encoding}

We ``attach'' a code bit $u_{i}$ to each code-bit node $i\in V$.
To each check node $a\in C$ we ``attach'' two type of bits: the
reconstructed bit $\widehat{x}_{a}$ and the source bit $x_{a}$.
By definition the source sequence has length $N$. So we have $n=N$
for the underlying ensembles, and $nL=N$ for the coupled ensembles.
A compressed word $\underline{u}$ has length $m$ for the underlying
ensemble, and $m(L+w-1)$ for the coupled ensemble. Thus the compression
design rate is $R=m/n$ for the underlying ensemble, and it is $R_{{\rm cou}}=m(L+w-1)/nL=R(1+\frac{w-1}{L})$
for the coupled ensemble. The compression design rate of the coupled
ensembles is slightly higher, due to the code-bit nodes at the boundary,
but in the asymptotic regime $n,m>>L>>w$ the difference between the
design rate $R$ of the underlying ensemble vanishes.

\subsubsection{Decoding Rule}

The reconstruction mapping is given by the linear operation (modulo
$2$ sum) 
\begin{equation}
\widehat{x}_{a}(\underline{u})=\oplus_{i\in\partial a}u_{i}.\label{eq:decodingrule}
\end{equation}
In this paper we do not investigate non-linear decoding rules, although
the whole analysis developed here can be adapted to such rules. Source
coding with such ``non-linear check nodes'' have been investigated
for underlying LDGM$(l,R)$ ensembles \cite{CilMezZec06}.

\subsubsection{Optimal Encoding}

Given a source word $\underline{x}$, the optimal encoder seeks to
minimize the Hamming distortion \eqref{eq:hamming-distortion}, and
so searches among all $\underline{u}\in\{0,1\}^{NR}$ to find a configuration
$\underline{u}^{\ast}$ such that 
\begin{equation}
\underline{u}^{\ast}={\rm argmin}_{\underline{u}}d_{N}\left(\underline{x},\widehat{\underline{x}}(\underline{u})\right).\label{eq:opt-code-word}
\end{equation}
The resulting minimal distortion is 
\begin{equation}
d_{N,\min}(\underline{x})=\min_{\underline{u}}d_{N}\left(\underline{x},\widehat{\underline{x}}(\underline{u})\right).\label{eq:mindist}
\end{equation}

\subsubsection{Optimal Distortion of the Ensemble}

A performance measure is given by the \textit{optimal distortion of
the ensemble} (not to be confused with Shannon's optimal distortion)
\begin{equation}
D_{N,\text{opt}}=\mathbb{E}_{{\rm LDGM},\underline{X}}[d_{N,\text{min}}(\underline{x})]\label{eq:dopt}
\end{equation}
where $\mathbb{E}_{{\rm LDGM},\underline{X}}$ is an expectation over
the graphical ensemble at hand and the symmetric Bernoulli source
$\underline{X}$.

Finding the minimizers in \eqref{eq:opt-code-word} by exhaustive
search takes exponential time in $N$; and there is no known efficient
algorithmic procedure to solve the minimization problem. Nevertheless,
the cavity method proposes a formula for the asymptotic value of \eqref{eq:dopt}
as $N\to+\infty$. It is conjectured that this formula is exact. We
come back to this point at the end of paragraph \ref{sub:statmech}.

\subsection{Statistical Mechanics Formulation\label{sub:statmech}}

We equip the configuration space $\left\{ 0,1\right\} ^{NR}$ with
the conditional probability distribution (over $\underline{u}\in\left\{ 0,1\right\} ^{NR}$)
\begin{align}
\mu_{\beta}(\underline{u}\mid\underline{x}) & =\frac{1}{Z_{\beta}(\underline{x})}e^{-2\beta Nd_{N}(\underline{x},\widehat{\underline{x}}(\underline{u}))}\nonumber \\
 & =\frac{1}{Z_{\beta}(\underline{x})}\prod_{a\in C}e^{-2\beta\vert x_{a}-{\bigoplus_{i\in\partial a}}u_{i}\vert}\label{eq:measure-mu}
\end{align}
where $\beta>0$ is a real number and 
\begin{equation}
Z_{\beta}(\underline{x})=\sum_{\underline{u}}e^{-2\beta Nd_{N}(\underline{x},\widehat{\underline{x}}(\underline{u}))}\label{eq:partition-function}
\end{equation}
a normalizing factor. The expectation with respect to $\underline{u}$
is denoted by the bracket $\langle-\rangle$. More precisely the average
of a function $A(\underline{u})$ is 
\begin{equation}
\langle A(\underline{u})\rangle=\frac{1}{Z}\sum_{\underline{u}\in\{-1,+1\}^{N}}A(\underline u)e^{-2\beta Nd_{N}(\underline{x},\widehat{\underline{x}}(\underline{u}))}.\label{eq:thermal-average}
\end{equation}
An important function that we consider below is the
distortion of a pair $(\underline{x},\widehat{\underline{x}}(\underline{u}))$,
$A(\underline{u})=d_{N}(\underline{x},\widehat{\underline{x}}(\underline{u}))$.

Note that the minimizer $\underline{u}^{\ast}$ in \eqref{eq:opt-code-word}
maximizes this conditional distribution, 
\begin{equation}
\underline{u}^{\ast}={\rm argmax}_{\underline{u}}\mu_{\beta}\left(\underline{u}\mid\underline{x}\right).\label{eq:maxmessure}
\end{equation}
The source coding problem can thus be interpreted as an estimation
problem where $\underline{x}$ is an observation and $\underline{u}$
has to be estimated.

In this paper we prefer the statistical mechanics interpretation,
because we use related methods and concepts. Equation \eqref{eq:measure-mu}
defines the Gibbs distribution associated to a ``spin glass'' Hamiltonian
$2Nd_{N}(\underline{x},\widehat{\underline{x}}(\underline{u}))$.
This Hamiltonian is a cost-function for assignments of ``dynamical''
variables, the spins (or bits) $u_{i}\in\{0,1\}$. The Hamiltonian
is random: for each realization of the source sequence $\underline{x}$
and the graph instance we have a different realization of the cost-function.
The source and graph instance are qualified as ``quenched'' or ``frozen''
random variables, to distinguish them from dynamical variables, because
in physical systems - as well as in algorithms - they fluctuate on
vastly different time scales. The parameter $\beta$ is the ``inverse
temperature'' in appropriate units, and the normalizing factor \eqref{eq:partition-function}
is the partition function.

Finding $\underline{u}^{*}$ amounts to find the ``minimum energy
configuration''. The minimum energy per node is equal to $2d_{N,{\rm min}}$,
and it is easy to check the identity (use \ref{eq:mindist} and \ref{eq:partition-function})
\begin{equation}
2d_{N,\min}(\underline{x})=-\lim_{\beta\rightarrow\infty}\frac{1}{\beta N}\ln Z_{\beta}\left(\underline{x}\right).
\end{equation}
As this identity already shows, a fundamental role is played by the
average free energy 
\begin{equation}
f_{N}(\beta)=-\frac{1}{\beta N}\mathbb{E}_{{\rm LDGM},\underline{X}}[\ln Z_{\beta}(\underline{x})].\label{eq:free-energy}
\end{equation}
For example the average free energy allows to compute the optimal
distortion of the ensemble 
\begin{equation}
2D_{N,{\rm opt}}=\lim_{\beta\to+\infty}f_{N}(\beta).\label{eq:Dopt}
\end{equation}
There exists also another useful relationship that we will use between
average distortion and free energy. Consider the ``internal energy''
defined as 
\begin{equation}
u_{N}(\beta)=2\mathbb{E}_{{\rm LDGM},\underline{X}}[\langle d_{N}(\underline{x},\widehat{\underline{x}}(\underline{u}))\rangle]\label{eq:internal-energy}
\end{equation}
It is straightforward to check that the internal energy can be computed
from the free energy (use \eqref{eq:partition-function}, \eqref{eq:free-energy},
\eqref{eq:internal-energy}) 
\begin{equation}
u_{N}(\beta)=\frac{\partial}{\partial\beta}(\beta f_{N}(\beta))\label{eq:derivative}
\end{equation}
and that in the zero temperature limit it reduces to the average minimum
energy or optimal distortion (use \eqref{eq:mindist}, \eqref{eq:dopt},
\eqref{eq:internal-energy}) 
\begin{equation}
2D_{N,{\rm opt}}=\lim_{\beta\to+\infty}u_{N}(\beta).
\end{equation}

What is the relation between the quantities $f_{N}(\beta)$, $u_{N}(\beta)$,
and $D_{N,{\rm opt}}$ for the underlying and coupled ensembles? The
following theorem states that they are equal in the infinite block
length limit. This limit is defined as 
\[
\lim_{N\to+\infty}=\lim_{n\to+\infty}
\]
with $m/n$ fixed for the underlying ensemble; and as 
\[
\lim_{N\to+\infty}=\lim_{L\to+\infty}\lim_{n\to+\infty}
\]
with $m/n$ fixed for the coupled ensemble. We stress that for the
coupled ensemble the order of limits is important. 
\begin{thm}
\label{thm:theorem1} Consider the two ensembles LDGM$(l,R,n)$ and
LDGM$(l,R,L,w,n)$ for an even $l$ and $R$. Then the respective
limits $\lim_{N\to+\infty}f_{N}(\beta)$, $\lim_{N\to+\infty}u_{N}(\beta)$
and $\lim_{N\to+\infty}D_{N,{\rm opt}}$ exist and have identical
values for the two ensembles. 
\end{thm}
This theorem is proved in \cite{Hassani11SAT} for the max-XORSAT
problem. The proof in \cite{Hassani11SAT} does not depend on the
constraint density, so that it applies verbatim to the present setting.
We conjecture that this theorem is valid for a wider class of graph
ensembles. In particular we expect that it is valid for odd $l$ and
also for the regular LDGM ensembles (see \cite{gmu13} for similar
results concerning LDPC codes).

\begin{table}[tb]
\centering{}\protect\caption{{\small{}\label{table:distortion} Optimal distortion for LDGM$(l,R=0.5)$
ensembles computed in 
\cite{ciliberti05source}; Shannon's bound for $R=0.5$
is $D_{{sh}}\approx0.1100$.}}
\begin{tabular}{ccccc}
\toprule 
$l$ & $3$ & $4$ & $5$ & $6$\tabularnewline
\midrule 
$D_{{\rm opt}}$ & $0.1179$ & $0.1126$ & $0.1110$ & $0.1104$\tabularnewline
\bottomrule
\end{tabular}
\end{table}

It is conjectured that the one-step-replica-symmetry-breaking-formulas
(1RSB), obtained from the cavity method \cite{mezard2001cavity},
for the $N\to+\infty$ limit of the free, internal and ground state
energies are exact. Remarkably, it has been proven \cite{franz2003replica},
using an extension of the Guerra-Toninelli interpolation bounds \cite{guerra2002interpolation},
that these formulas are upper bounds. The 1RSB formulas allow to numerically
compute \cite{ciliberti05source}, using population dynamics, $D_{{\rm opt}}\equiv\lim_{N\to+\infty}D_{N,{\rm opt}}$.
As an illustration, Table \ref{table:distortion} reproduces $D_{{\rm opt}}$
for increasing check degrees. Note that $D_{{\rm opt}}$ approaches
$D_{{\rm sh}}$ as the degrees increase. One observes that with increasing
degrees the optimal distortion of the ensemble attains Shannon's rate-distortion
limit.

\section{Belief Propagation Guided Decimation\label{sec:bpgd}}

Since the optimal encoder \eqref{eq:opt-code-word} is intractable,
we investigate suboptimal low complexity encoders. In this contribution
we focus on two encoding algorithms based on the belief propagation
(BP) equations supplemented with a decimation process.

\subsubsection{Belief Propagation Equations}

Instead of estimating the block $\underline{u}$ (as in \eqref{eq:opt-code-word})
we would like to estimate bits $u_{i}$ with the help of the marginals
\begin{equation}
\mu_{i}(u_{i}\mid\underline{x})=\sum_{\underline{u}\setminus u_{i}}\mu_{\beta}(\underline{u}\mid\underline{x})\label{eq:true-marginal}
\end{equation}
where the sum is over $u_{1},\dots u_{N}$ with $u_{i}$ omitted.
However computing the exact marginals involves a sum with an exponential
number of terms and is also intractable. For sparse random graphs,
when the size of the graph is large, any finite neighborhood of a
node $i$ is a tree with high probability. As is well known, computing
the marginals on a tree-graph can be done exactly and leads to the
BP equations. It may therefore seem reasonable to compute the BP marginal
distribution in place of \eqref{eq:true-marginal}, 
\begin{equation}
\mu_{i}^{{\rm BP}}(u_{i}\mid\underline{x})=\frac{1}{2\cosh\beta\eta_{i}}e^{\beta(-1)^{u_{i}}\eta_{i}}\label{eq:bp-marginal-i}
\end{equation}
where the biases $\eta_{i}$ are computed from solutions of the BP
equations. The later are a set of fixed point equations involving
$2\left\vert E\right\vert $ real valued messages $\eta_{i\rightarrow a}$
and $\widehat{\eta}_{a\rightarrow i}$ associated to the edges $(i,a)\in E$
of the graph. We have 
\begin{align}
\begin{cases}
\widehat{\eta}_{a\rightarrow i} & =(-1)^{x_{a}}\beta^{-1}\tanh^{-1}\bigl(\tanh\beta\prod_{j\in\partial a\backslash i}\tanh\beta\eta_{j\rightarrow a}\bigr)\\
\eta_{i\rightarrow a} & =\sum_{b\in\partial i\backslash a}\widehat{\eta}_{b\rightarrow i}
\end{cases}\label{eq:bp_equation2}
\end{align}
and 
\begin{equation}
\eta_{i}=\sum_{a\in\partial i}\widehat{\eta}_{a\rightarrow i}.
\end{equation}
The derivation of these equations can be worked out by reducing the general BP equations \eqref{eq:CM_bp_equations_definition} (Appendix A) with 
the parameterization \eqref{paramappendixB} (Appendix B).

For any solution of the BP equations one may consider the estimator
\begin{align}
\widehat{u}_{i}^{{\rm BP}} & ={\rm argmax}_{u_{i}}\mu_{i}^{{\rm BP}}(u_{i}\mid\underline{x})\nonumber \\
 & =\begin{cases}
\frac{1}{2}(1-{\rm sign}\eta_{i}),\,\,\,{\rm if}\,\,\,\eta_{i}\neq0\\
{\rm Bernoulli(\frac{1}{2})},\,\,\,{\rm if}\,\,\,\eta_{i}=0
\end{cases}
\end{align}
One may then use the decoding rule \eqref{eq:decodingrule} to determine
a reconstructed word and the corresponding distortion. 

To solve the BP equations one uses an iterative method. A set of initial messages 
$\eta_{i\to a}^{(0)}$ are fixed at time $t=0$ and updated according to 
\begin{align*}
\begin{cases}
\widehat{\eta}_{a\rightarrow i}^{(t)} & =(-1)^{x_{a}}\beta^{-1}\tanh^{-1}\bigl(\tanh\beta\prod_{j\in\partial a\backslash i}\tanh\beta\eta_{j\rightarrow a}^{(t)}\bigr)\\
\eta_{i\rightarrow a}^{(t+1)} & =\sum_{b\in\partial i\backslash a}\widehat{\eta}_{b\rightarrow i}^{(t)}
\end{cases}
\end{align*}
The bias at time $t$ is simply given by $\eta_{i}^{(t)}=\sum_{a\in\partial i}\widehat{\eta}_{a\rightarrow i}^{(t)}$.

Unfortunately, even when the BP updates are converging they are not always biased. This is because 
there exist an exponentially large (in $N$) number of compressed words that lead to roughly the same distortion.
This has an undesirable consequence: it is not possible to
pick the relevant solution by a plain iterative solution of the BP equations. To get around
this problem, the BP iterations are equipped with a heuristic decimation
process explained in the next paragraph. We note that here BP always has to be equipped with a decimation process 
for all values of parameters of the algorithm, whether the BP fixed point is unique or non-unique. 
The problem here is 
akin to the class of constraint satisfaction problems.

\subsubsection{Decimation Process}

We start with a description of the \textit{first round} of the decimation
process. Let $\Gamma$, $\underline{x}$ be a graph and source instance.
Fix an initial set of messages $\eta_{i\to a}^{(0)}$ at time $t=0$.
Iterate the BP equations \eqref{eq:bp_equation2} to get a set of
messages $\eta_{i\rightarrow a}^{(t)}$ and $\widehat{\eta}_{a\rightarrow i}^{(t)}$
at time $t\geq0$. Let $\epsilon>0$ be some small positive number
and $T$ some large time. Define a \textit{decimation instant} $t_{{\rm dec}}$
as follows: 
\begin{itemize}
\item i) If the total variation of messages does not change significantly
in two successive iterations, 
\begin{equation}
\frac{1}{\left|E\right|}\sum_{(i,a)\in E}\vert\widehat{\eta}_{a\rightarrow i}^{(t)}-\widehat{\eta}_{a\rightarrow i}^{(t-1)}\vert<\epsilon\label{eq:convergence}
\end{equation}
for some $t<T$, then $t_{{\rm dec}}=t$. 
\item ii) If \eqref{eq:convergence} does not occur for all $t\leq T$ then
$t_{{\rm dec}}=T$. 
\end{itemize}
At instant $t_{{\rm dec}}$ each code-bit has a bias given by $\eta_{i}^{(t_{{\rm dec}})}$.
\textit{Select} and \textit{fix} \textit{one particular code-bit}
$i_{{\rm dec}}$ according to a decision rule 
\begin{equation}
(i_{{\rm dec}},u_{i_{{\rm dec}}})\leftarrow\mathcal{D}(\underline{\eta}^{(t_{{\rm dec}})}).\label{eq:decision}
\end{equation}
The precise decision rules that we investigate are described in the
next paragraph. At this point, update $x_{a}\leftarrow x_{a}\oplus u_{i_{{\rm dec}}}$
for all $a\in\partial i_{{\rm dec}}$, and decimate the graph $\Gamma\leftarrow\Gamma\setminus i_{{\rm dec}}$.
This defines a new graph and source instance, on which we repeat a
\textit{new round}. The initial set of messages of the new round is
the one obtained at time $t_{{\rm dec}}$ of the previous round.

\subsubsection{Belief-Propagation Guided Decimation}

The decision rule \eqref{eq:decision} involves two choices. One has
to choose $i_{{\rm dec}}$ and then set $u_{i_{{\rm dec}}}$ to some
value. Let us first describe the choice of $i_{{\rm dec}}$.

We evaluate the maximum bias 
\begin{equation}
B_{t_{{\rm dec}}}=\max_{i\in V}\vert\eta_{i}^{(t_{{\rm dec}})}\vert\label{eq:bias}
\end{equation}
at each decimation instant. If $B_{t_{{\rm dec}}}>0$, we consider
the set of nodes that maximize \eqref{eq:bias}, we choose one of
them uniformly at random, and call it $i_{{\rm dec}}$. If $B_{t_{{\rm dec}}}=0$
and we have a graph of the \textit{underlying ensemble}, we choose
a node uniformly at random from $\{1,\dots m\}$, and call it $i_{{\rm dec}}$.
If $B_{t_{{\rm dec}}}=0$ and we have a graph of the \textit{coupled
ensemble}, we choose a node uniformly at random from the $w$ left-most
positions of the current graph, and call it $i_{{\rm dec}}$. Note
that because the graph gets decimated the $w$ left-most positions
of the current graph form a moving boundary.

With the above choice of decimation node the encoding process is seeded
at the boundary each time the BP biases fail to guide the decimation
process. We have checked that if we choose $i_{{\rm dec}}$ uniformly
at random from the whole chain (for coupled graphs) the performance
is not improved by coupling. In \cite{amuv2012} we adopted periodic
boundary conditions and the seeding region was set to an arbitrary
window of length $w$ at the beginning of the process, which then
generated its own boundary at a later stage of the iterations.

We now describe two decision rules for setting the value of $u_{i_{{\rm dec}}}$
in \eqref{eq:decision}. 
\begin{enumerate}
\item \textbf{Hard Decision} 
\begin{equation}
u_{i_{{\rm dec}}}=\begin{cases}
\frac{1}{2}(1-{\rm sign}\eta_{i_{{\rm dec}}}^{(t_{{\rm dec}})}),\,\,\,{\rm if}\,\, B_{t_{{\rm dec}}}>0\\
{\rm Bernoulli}(\frac{1}{2}),\,\,\,{\rm if}\,\, B_{t_{{\rm dec}}}=0
\end{cases}\label{eq:hardrule}
\end{equation}
where $\theta(.)$ is the Heaviside step function. We call this rule
and the associated algorithm BPGD-h. 
\item \textbf{Randomized Decision} 
\begin{equation}
u_{i_{{\rm dec}}}=\begin{cases}
0,\,\,\,{\rm with~prob}\,\,\frac{1}{2}(1+\tanh\beta\eta_{i_{{\rm dec}}}^{(t_{{\rm dec}})})\\
1,\,\,\,{\rm with~prob}\,\,\frac{1}{2}(1-\tanh\beta\eta_{i_{{\rm dec}}}^{(t_{{\rm dec}})}).
\end{cases}\label{eq:randomrule}
\end{equation}
In other words, we fix a code-bit randomly with a probability given
by its BP marginal \eqref{eq:bp-marginal-i}. We call this rule and
the associated algorithm BPGD-r. 
\end{enumerate}
Algorithm \ref{alg:BPGD} summarizes the BPGD algorithms for all situations.

\begin{algorithm}[htb]
\protect\caption{\label{alg:BPGD} BP Guided Decimation Algorithm}

Generate a graph instance $\Gamma(C,V,E)$ from the underlying or
coupled ensembles. \;

Generate a Bernoulli symmetric source word $\underline{x}$.\;

Set $\eta_{i\to a}^{(0)}=0$ for all $(i,a)\in E$.\;

\While{$V\ne\emptyset$} { Set $t=0$.\;

\While{Convergence \eqref{eq:convergence} is not satisfied and
$t<T$} { Update $\hat{\eta}_{a\to i}^{(t)}$ according to \eqref{eq:bp_equation2}
for all $(a,i)\in E$.\;

Update $\eta_{i\to a}^{(t+1)}$ according to \eqref{eq:bp_equation2}
for all $(i,a)\in E$.\;

$t\leftarrow t+1$.\; } Compute bias $\eta_{i}^{(t)}=\sum_{a\in\partial i}\hat{\eta}_{a\to i}^{(t)}$
for all $i\in V$\;

Find $B=\max_{i\in V}\vert\eta_{i}^{(t)}\vert$.\;

\lIf{$B=0$\label{alg_Bcondition}} {For an instance from the
underlying ensemble randomly pick a code-bit $i$ from $V$. For a
graph from the coupled ensemble randomly pick a code-bit from the
$w$ left-most positions of $\Gamma$ and fix it randomly to $0$
or $1$.\;}

\Else{Select $i=\text{arg}\max_{i\in V}\vert\eta_{i}^{(t)}\vert$.\;

Fix a value for $u_{i}$ according to rule \eqref{eq:hardrule} or
\eqref{eq:randomrule}.\; } Update $x_{a}\leftarrow x_{a}\oplus u_{i}$
for all $a\in\partial i$.\;

Reduce the graph $\Gamma\leftarrow\Gamma\setminus\{i\}$.\; } 
\end{algorithm}

\subsubsection{Initialization and Choice of Parameters $\epsilon$, $T$}

We initialize $\eta_{i\to a}^{(0)}$ to zero just at the beginning
of the algorithm. After each decimation step, rather than resetting
messages to zero we continue with the previous messages. We have observed
that resetting the messages to zero does not lead to very good results.

The parameters $\epsilon$ and $T$ are in practice set to $\epsilon=0.01$
and $T=10$. The simulation results do not seem to change significantly
when we take $\epsilon$ smaller and $T$ larger.

\subsubsection{Choice of $\beta$}

Let us now clarify the role of $\beta$. It may seem from the discussion of the statistical mechanical formulation in section II that $\beta$ should be taken to $+\infty$. This is the case for the computation of the optimal ensemble performance. However for the BPGD algorithm this {\it is not} the best choice for $\beta$. The reason being that for large values of $\beta$  
the BP iterations do not converge and therefore one does not obtain a reliable bias. 

We indeed observe that
the performance of the BPGD algorithm does depend on the choice of
$\beta$ which enters in the BP equations \eqref{eq:bp_equation2}
and in the randomized decision rule \eqref{eq:randomrule}. It is
possible to optimize on $\beta$. This is important in order to approach
(with coupled codes) the optimal distortion of the ensemble, and furthermore
to approach the Shannon bound in the large degree limit. 

While we
do not have a first principle theory for the optimal choice of $\beta$
we provide empirical observations in section \ref{sec:results}. We
observe that knowing the dynamical and condensation (inverse) temperatures
predicted by the cavity method allows to make an educated guess for
an estimate of the optimal $\beta$. Two results (discussed at more length in the next section) are noteworthy:
(i) for coupled instances we can take larger values of $\beta$; and (ii) 
for coupled codes with large degrees the best $\beta$ approaches
the information theoretic test-channel value.

\subsubsection{Computational Complexity}

It is not difficult to see that the complexity of the plain BPGD algorithm
\ref{alg:BPGD} is $O(N^{2})$, in other words $O(n^{2})$ for underlying
and $O(n^{2}L^{2})$ for coupled ensembles. By employing window decoding~
\cite{iyengar11windowed,hassan12window}, one can reduce the complexity
of the coupled ensemble to $O(n^{2}L)$ with almost the same performance.
This can be further reduced to $O(nL)$ by noticing that the BP messages
do not change significantly between two decimation steps. As a result,
we may decimate $\delta n$ code-bits at each step for some small
$\delta$, so that the complexity becomes $O(nL/\delta)$. To summarize,
it is possible to get linear in block length complexity without significant
loss in performance.

\section{Simulations\label{sec:results}}

In this section we discuss the performance of the BPGD algorithms.
The comparison between underlying ensembles LDGM$(l,R,n)$, coupled
ensembles LDGM$(l,R,w,L,n)$ and the Shannon rate-distortion curve
is illustrated. The role played by the parameter $\beta$ is investigated.

\subsection{BPGD performance and comparison to the Shannon limit}

\begin{figure*}[tb]
\centering{}\includegraphics{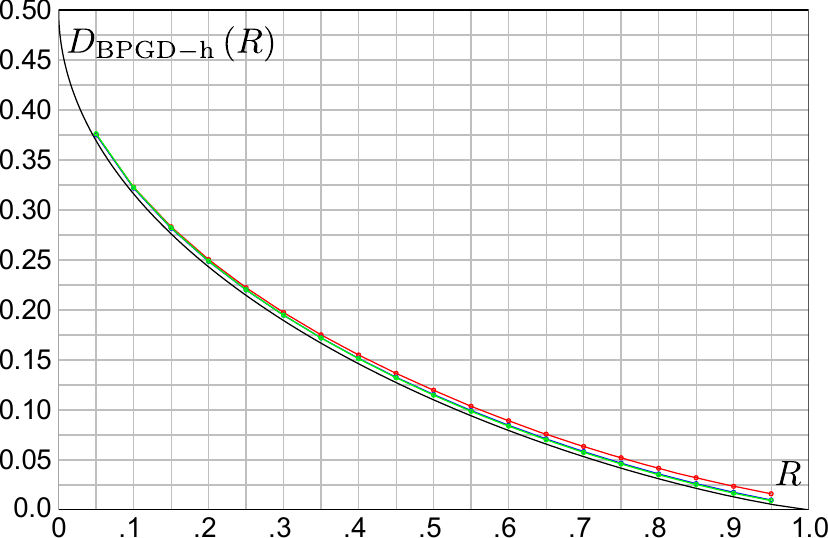}\includegraphics{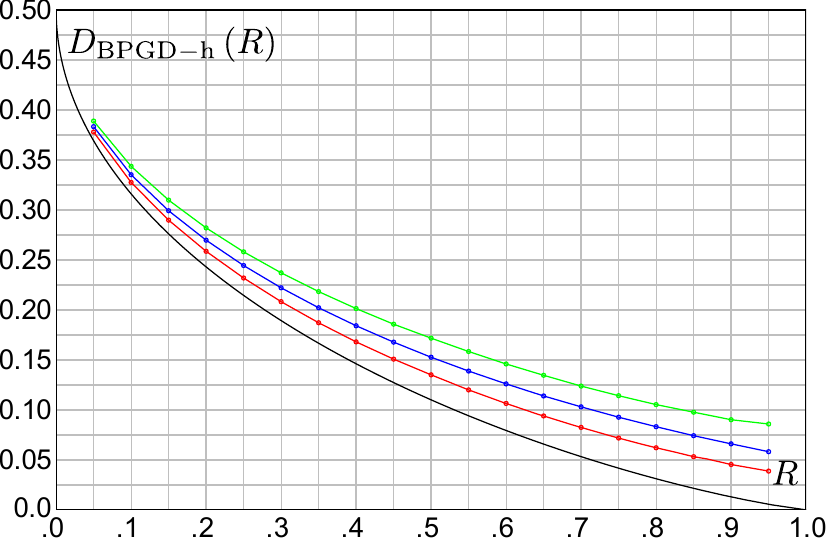}\protect\caption{ \label{fig:distortioncurve_h} The BPGD-h algorithmic distortion
versus compression rate $R$ compared to the Shannon rate-distortion
curve at the bottom. Points are obtained by optimizing over $\beta$
and averaging over $50$ instances. Left: spatially coupled LDGM$(l,R,L=64,w=3,n=2000)$
ensembles for $l=3,4,5$ (top to bottom). Right: LDGM$(l,R,N=128000)$
ensembles for $l=3,4,5$ (bottom to top).}
\end{figure*}

\begin{figure*}[tb]
\centering{}\includegraphics{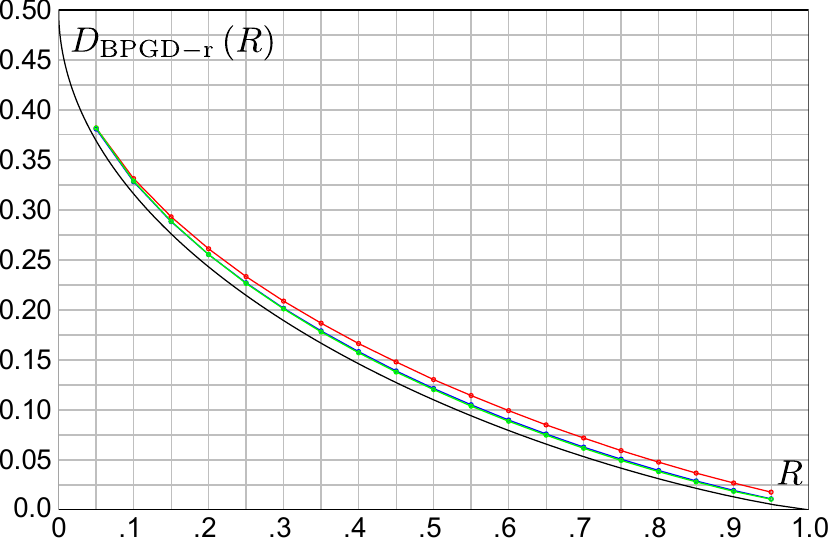}\includegraphics{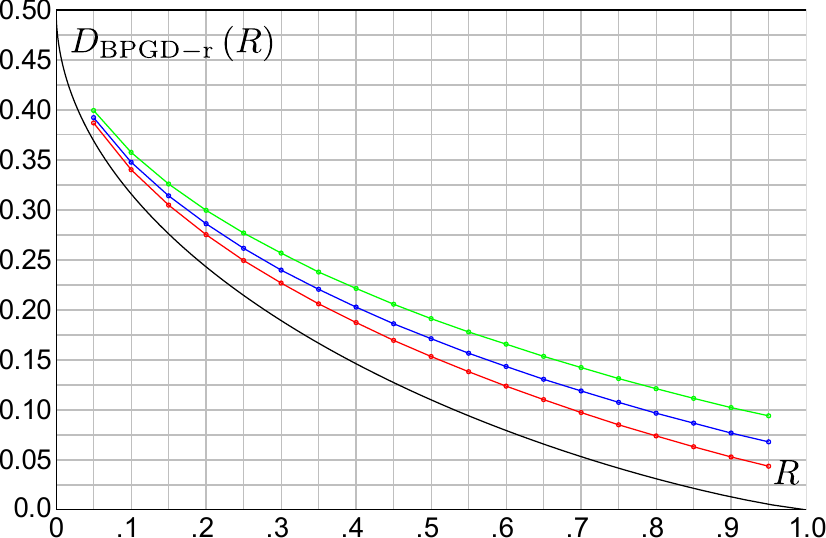}\protect\caption{ \label{fig:distortioncurve_r} The BPGD-r algorithmic distortion
versus compression rate $R$ compared to the Shannon rate-distortion
curve at the bottom. Points are obtained by choosing $\beta=\beta_{sh}=\frac{1}{2}\log(\frac{1-D_{sh}}{D_{sh}})$
and averaging over $50$ instances. Continuous lines are a guide to
the eye. Left: spatially coupled LDGM$(l,R,L=64,w=3,n=2000)$ ensembles
for $l=3,4,5$ (top to bottom). Right: LDGM$(l,R,N=128000)$ ensembles
for $l=3,4,5$ (bottom to top).}
\end{figure*}

Fig. \ref{fig:distortioncurve_h} and \ref{fig:distortioncurve_r}
display the average distortion $D_{{\rm BPGD}}(R)$ obtained by the
BPGD algorithms (with hard and randomized decision rules) as a function
of $R$, and compares it to the Shannon limit $D_{{\rm sh}}(R)$ given
by the lowest curve. The distortion is computed for fixed $R$ and
for $50$ instances, and the empirical average is taken. This average
is then optimized over $\beta$, giving one dot on the curves (continuous
curves are a guide to the eye).

We recall that the design rate of a spatially-coupled ensemble is slightly higher than the rate $R$ of its corresponding 
underlying ensemble due to the boundary nodes, i.e. $R_{\rm{cou}}=R(1+O(\frac{w-1}{L}))$. The difference between the design 
rates of both ensembles vanishes as $L\rightarrow\infty$. In order to disregard this finite size effect, we reported the algorithmic 
distortion of the coupled ensembles with respect to the rate $R$ of their corresponding underlying ensembles.

The plots on the right are for the underlying ensembles with $l=3,4,5$
and $n=128000$. We observe that as the check degree increases the
BPGD performance gets worse. But recall from Table \ref{table:distortion}
that with increasing degrees the optimal distortion of the ensemble
(not shown explicitly on the plots) gets better and approaches the
Shannon limit. Thus the situation is similar to the case of LDPC codes
where the BP threshold gets worse with increasing degrees, while the
MAP threshold approaches Shannon capacity.

The plots on the left show the algorithmic performance for the coupled
ensembles with $l=3,4,5$, $n=2000$, $w=3$, and $L=64$ (so again
a total length of $N=128000$). We see that the BPGD performance approaches
the Shannon limit as the degrees increase. One obtains a good performance,
for a range of rates, without any optimization on the degree sequence
of the ensemble, and with simple BPGD schemes.

The simulations, suggest the following. Look at the regime $n>>L>>w>>1$.
When these parameters go to infinity in the specified order \textit{for
the coupled ensemble} $D_{{\rm BPGD}}(R)$ approaches $D_{{\rm opt}}(R)$.
In words, the algorithmic distortion approaches the optimal distortion
of the ensemble. When furthermore $l\to+\infty$ after the other parameters
$D_{{\rm BPGD}}(R)$ approaches $D_{{\rm sh}}(R)$. At this point
it is not possible to assess from the simulations whether these limits
are exactly attained.

\subsection{The choice of the parameter $\beta$}

We discuss the empirical observations for the dependence of the curves
$D_{{\rm BPGD}}(\beta,R)$ on $\beta$ at fixed rate. We illustrate
our results for $R=1/2$ and with the underlying LDGM$(l=5,R=0.5,N=128000)$
and coupled LDGM$(l=5,R=0.5,w=3,L=64,n=2000)$ ensembles.

\begin{figure}[tb]
\centering{}\includegraphics{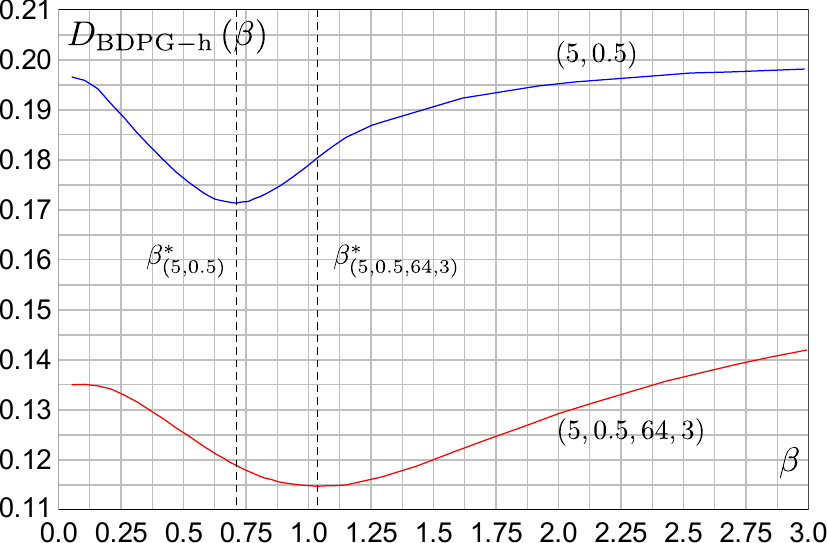}\protect\caption{ \label{fig:distortionpoisson5} The BPGD-h algorithmic distortion
versus $\beta$. Results are obtained for coupled LDGM$(5,0.5,L=64,w=3,n=2000)$
and LDGM$(5,0.5,128000)$ ensemble. Results are averaged over 50 instances.
The minimum distortion occurs at $\beta_{(5,0.5,64,3)}^{*}\approx1.03\pm0.01$
and $\beta_{(5,0.5)}^{*}\approx0.71\pm0.01$. }
\end{figure}

On Fig. \ref{fig:distortionpoisson5} we plot the distortion $D_{{\rm BPGD-h}}(\beta,R=1/2)$
of the hard decision rule. For \textit{all} values of $0<\beta<3$,
the algorithmic distortion $D_{{\rm BPGD-h}}(\beta,R)$ of the coupled
ensemble is below the corresponding curve of the underlying ensemble.
The most important feature is a clear minimum at a value $\beta^{*}$
which is rate dependent. The rate distortion curve for the hard decision
rule on Figure \ref{fig:distortioncurve_h} is computed at this $\beta^{*}$
and is the result of the optimization 
\begin{equation}
D_{{\rm BPGD-h}}(R)=\min_{\beta>0}D_{{\rm BPGD-h}}(\beta,R).
\end{equation}
We observe that the optimal value $\beta_{\text{cou}}^{*}$ for the
coupled ensemble is always larger than $\beta_{\text{un}}^{*}$ for
the underlying ensemble. 

An additional observation is the following. As the degree
$l$ increases $\beta_{\text{un}}^{*}$ tends to zero, and $\beta_{\text{cou}}^{*}$
approaches $\beta_{\text{sh}}(R)$ where 
\begin{equation}
\beta_{\text{sh}}(R)\equiv\frac{1}{2}\ln\biggl(\frac{1-D_{\text{sh}}(R)}{D_{\text{sh}}(R)}\biggr).\label{eq:betashannon}
\end{equation}
This is the information theoretic value corresponding to the half-loglikelihood
parameter of a test-BSC with the noise tuned at capacity. This observation is interesting because it shows
that for large $l$, with the coupled ensemble, one does not really need to optimize over $\beta$, but it suffices to fix 
$\beta = \beta_{\rm sh}(R)$. Theoretical motivation for this choice is discussed in Section \ref{sec:CM}.

\begin{figure}[tb]
\centering{}\includegraphics{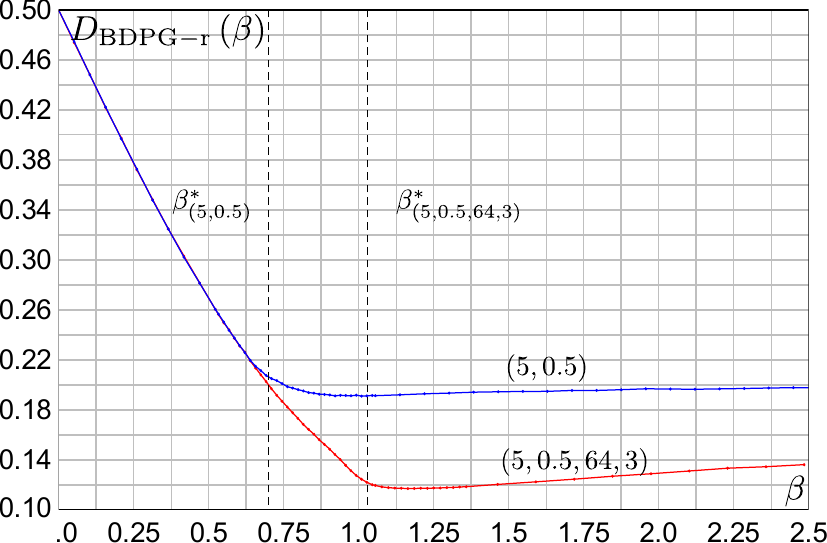}\protect\caption{\label{fig:distortionpoisson5_random} The BPGD-r algorithmic distortion
versus $\beta$. Results are obtained for coupled LDGM$(5,0.5,L=64,w=3,n=2000)$
and LDGM$(5,0.5,128000)$ ensemble. Results are averaged over 50 instances.
The values $\beta^{*}$ of Figure \ref{fig:distortionpoisson5} are
reported for comparison.}
\end{figure}

On Figure \ref{fig:distortionpoisson5_random} we plot the curve $D_{{\rm BPGD-r}}(\beta,R=1/2)$
for the randomized algorithm. The behavior of the underlying and coupled
ensemble have the same flavor. The curves are first decreasing with
respect to $\beta$ and then flatten. The minimum is reached in the
flattened region and as long as $\beta$ is chosen in the flat region,
the optimized distortion is not very sensitive to this choice. We
take advantage of this feature, and compute the rate distortion curve
of the randomized decision rule at a predetermined value of $\beta$.
This has the advantage of avoiding optimizing over $\beta$. Again, for the coupled case 
a good choice is to take
$\beta_{\text{sh}}(R)$ given by Equ. \ref{eq:betashannon}. With
these considerations the distortion curve on Figure \ref{fig:distortioncurve_r}
is 
\begin{equation}
D_{{\rm BPGD-r}}(R)=D_{{\rm BPGD-r}}(\beta_{\text{sh}},R).
\end{equation}

\subsection{Convergence}

\begin{figure}[tb]
\centering{} \includegraphics{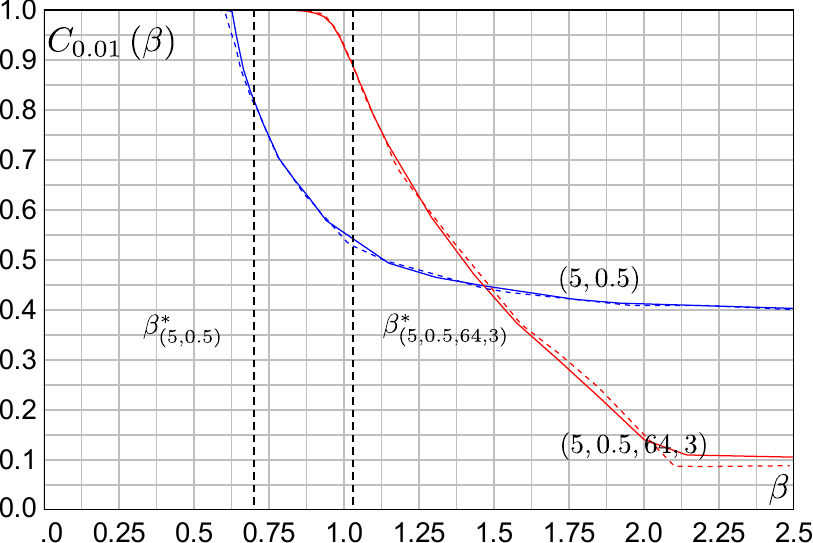}\protect\caption{ \label{fig:convergence510} $C_{0.01}(\beta)$ versus $\beta$. Empirical
convergence probability for underlying LDGM$(5,0.5,128000)$ and coupled
LDGM$(5,0.5,L=64,w=3,n=2000)$ ensembles. Solid (resp. dashed) lines
are for the hard (resp. random) decision rule. Results are averaged
over $50$ instances.}
\end{figure}

We have tested the convergence of the BPGD algorithms for both decision
rules. We compute an \textit{empirical probability of convergence}
$C_{\epsilon,T}(\beta)$ defined as the fraction of decimation rounds
that results from the convergence condition \eqref{eq:convergence}.
In other words $C_{\epsilon,T}(\beta)=1$ means that at every round
of the decimation process the BP update rules converge in less than
$T$ iterations to a fixed point of the BP equations \eqref{eq:bp_equation2}
up to a precision $\epsilon$. Figure \ref{fig:convergence510} shows
$C_{\epsilon,T}(\beta)$ at $(\epsilon,T)=(0.01,10)$ for the underlying
and coupled ensembles. The hard decision rule is represented by solid
lines and the random decision rule by dashed lines. The first observation
is that both decision rules have identical behaviors. This is not
a priori obvious since the decimation rules are different, and as
a result the graph evolves differently for each rule during the decimation
process. This suggest that the convergence of the algorithms essentially
depends on the convergence of the plain BP algorithm. The second observation
is that the values of $\beta$ where $C_{\epsilon,T}(\beta)$ drops
below one are roughly comparable to the values where $D_{{\rm BPGD-r}}$
flattens and where $D_{{\rm BPGD-h}}$ attains its minimum.

\section{The Phase Diagram: Predictions of the Cavity Method\label{sec:CM}}

It is natural to expect that the behavior of belief propagation based
algorithms should be in a way or another related to the phase diagram
of the Gibbs distribution \eqref{eq:measure-mu}. The phase diagram
can be derived by using the cavity method. As this is pretty involved,
in the present section we provide a high level picture. The cavity
equations are presented in section \ref{sec:CM_application}. We give
a primer on the cavity method in appendix \ref{sec:cavityprimer}
and the technical derivations for the present problem are given in
appendices \ref{sec:application}, \ref{sec:CM_population_dynamic}.

As we vary $\beta$ the nature of the Gibbs measure and the geometry
of the space of its typical configurations changes at special \textit{dynamical}
and \textit{condensation} thresholds $\beta_{d}$ and $\beta_{c}$.
In paragraph \ref{sub:dyncond} we explain what these thresholds are
and what is their significance. We discuss how they are affected by
spatial coupling in paragraph \ref{sub:compcoup}. Finally in paragraph
\ref{sub:insight} we discuss some heuristic insights that allow to
understand why Shannon's limit is approached with the BPGD algorithm
for coupled ensembles with large check degrees.

In this section $f$ and $u$ denote the limits $\lim_{N\to+\infty}f_{N}$
and $\lim_{N\to+\infty}u_{N}$.

\subsection{Dynamical and Condensation Thresholds\label{sub:dyncond}}

The cavity method assumes that the random Gibbs distribution \eqref{eq:measure-mu}
can, in the limit of $N\to+\infty$, be decomposed into a convex superposition
of ``extremal measures'' 
\begin{equation}
\mu_{\beta}(\underline{u}\mid\underline{x})=\sum_{p=1}^{\mathcal{N}}w_{p}\,\mu_{\beta,p}(\underline{u}\mid\underline{x})\label{eq:pure-states}
\end{equation}
each of which occurs with a weight $w_{p}=e^{-\beta N(f_{p}-f)}$,
where $f_{p}$ is a free energy associated to the extremal measure
$\mu_{\beta,p}$.
Since the weights $w_{p}$ have to sum to $1$, we have 
\begin{align}
e^{-\beta Nf}\approx\sum_{p=1}^{\mathcal{N}}e^{-\beta Nf_{p}}\approx e^{-\beta N\min_{\varphi}(\varphi-\beta^{-1}\Sigma(\varphi;\beta))}\label{eq:exponent}
\end{align}
where $e^{N\Sigma(\varphi;\beta)}$ counts the number of extremal
states $\mu_{\beta,p}$ with free energy $f_{p}\approx\varphi$. 

Such convex decompositions of the Gibbs distribution into bona fide
extremal measures are under mathematical control for ``simple''
models such as the (deterministic) Ising model on a square grid \cite{georgii2011gibbs}.
But for spin glass models is it not known how to construct or even
precisely define the extremal measures. One important conceptual difference
with respect to the Ising model, which has a small number of extremal
states, is that for spin glasses one envisions the possibility of
having an exponentially large in $N$ number of terms in the decomposition
\eqref{eq:pure-states}.

In the context of sparse graph models it is further assumed that there
are ``extremal'' Bethe measures which are a good proxy for the ``extremal
measures''. The Bethe measures are those measures that have marginals
given by BP marginals. When the BP equations have many fixed point
solutions there are many possible Bethe measures and one must have a criterion to choose among them. 
This is provided by the Bethe free energy. The Bethe free energy is the functional whose stationary point equations 
(gradient equal zero) yield the BP equations. 
Heuristically, the
extremal  Bethe measures correspond to the solutions of the BP equations that are minima of the Bethe free energy
\footnote{Remarkably, it is not very important to be able to precisely select
these minima because at low temperatures one expects that
they outnumber the other ones.
}. Similarly, it is assumed that the Bethe free energies corresponding
to solutions of the BP equations are good proxy's for the free energies
$f_{p}$. Moreover one expects that the later concentrate.

Once
one chooses to replace $f_{p}$ by the Bethe free energies, the counting
function $\Sigma(\varphi;\beta)$ and the free energy $f$ can be
computed through a fairly technical procedure, and a number of remarkable
predictions about the decomposition \eqref{eq:pure-states} emerge.

\begin{figure}[tb]
\centering{}{\small{}\includegraphics{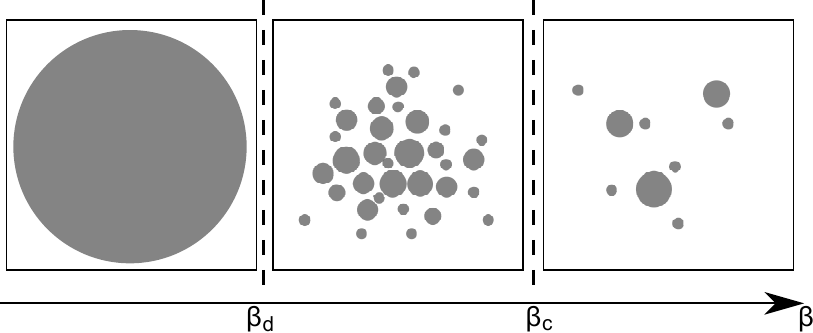}}\protect\caption{{\small{}\label{fig:phasediagram} Pictorial representation of the
decomposition of the Gibbs distribution into a convex superposition
of extremal states. Balls represent extremal states (their size represents
their internal entropy). For $\beta<\beta_{d}$ there is one extremal
state. For $\beta_{d}<\beta<\beta_{c}$ there are exponentially many
extremal states (with the same internal free energy $\varphi_{{\rm int}}$)
that dominate to the convex superposition. For $\beta>\beta_{c}$
there is a finite number of extremal states that dominate the convex
superposition. }}
\end{figure}

The cavity method predicts the existence of two sharply defined thresholds
$\beta_{d}$ and $\beta_{c}$ at which the nature of the convex decomposition
\eqref{eq:pure-states} changes drastically. Figure \ref{fig:phasediagram}
gives a pictorial view of the transitions associated with the decomposition
\eqref{eq:pure-states}. For $\beta<\beta_{d}$ the measure $\mu_{\beta}(\underline{u}\mid\underline{x})$
is extremal, in the sense that $\mathcal{N}=1$ in \eqref{eq:pure-states}.
For $\beta_{d}<\beta<\beta_{c}$ the measure is a convex superposition
of an exponentially large number of extremal states. The exponent
$\varphi-\beta^{-1}\Sigma(\varphi;\beta)$ in \eqref{eq:exponent}
is minimized at a value $\varphi_{{\rm int}}(\beta)$ such that $\Sigma(\varphi_{{\rm int}}(\beta);\beta)>0$.
Then 
\begin{equation}
\Sigma(\beta)\equiv\Sigma(\varphi_{{\rm int}}(\beta);\beta)=\beta(\varphi_{{\rm int}}(\beta)-f(\beta))\label{eq:complexity}
\end{equation}
is strictly positive and gives the growth rate (as $N\to+\infty$)
of the number of extremal states that dominate the convex superposition
of pure states \eqref{eq:pure-states}. This quantity is called the
complexity. It turns out that the complexity is a decreasing function
of $\beta$ which becomes negative at $\beta_{c}$ where it looses
its meaning. To summarize, above $\beta_{d}$ and below $\beta_{c}$
an exponentially large number of extremal states with the same free
energy $\varphi_{{\rm int}}$ contribute significantly to the Gibbs
distribution. For $\beta>\beta_{c}$ the number of extremal states
that dominate the measure is finite. One says that the measure is
\textit{condensed} over a small number of extremal states. In fact,
there may still be an exponential number of extremal states but they
do not contribute significantly to the measure because their weight
is exponentially smaller than the dominant ones.

There exist a mathematically more precise definition of $\beta_{d}$
and $\beta_{c}$ in terms of correlation functions. When these correlation
functions are computed within the framework of the cavity method the
results for $\beta_{d}$ and $\beta_{c}$ agree with those given by
the complexity curve $\Sigma(\beta)$. Although these definitions
nicely complete the perspective, we refrain from giving them here
since we will not use them explicitly.

What is the significance of the transitions at $\beta_{d}$ and $\beta_{c}$?
The condensation threshold is a thermodynamic phase transition point:
the free energy $f(\beta)$ and internal energy $u(\beta)$ are not
analytic at $\beta_{c}$. At $\beta_{d}$ the free and internal energies
have no singularities: in particular their \textit{analytical expressions}
do not change in the whole range $0<\beta<\beta_{c}$. At $\beta_{d}$
the (phase) transition is dynamical: Markov chain Monte Carlo algorithms
have an equilibration time that diverges when $\beta\uparrow\beta_{d}$,
and are unable to sample the Gibbs distribution for $\beta>\beta_{d}$.
For more details we refer to \cite{mezard09information}.

\subsection{Complexity and Thresholds of the Underlying and Coupled ensembles\label{sub:compcoup}}

We have computed the complexity and the thresholds from the cavity
theory. These have been computed both from the full cavity equations
of Section \ref{sub:cavmeth} and from the simplified ones of Section
\ref{sub:futhersimpl}. Tables \ref{table:betad=000026c_simpl} and
\ref{table:betad=000026c_simpl_w} illustrate the results.

\begin{table}[tbh]
\centering{}\protect\caption{\label{table:betad=000026c_simpl} The numerical values of $\beta_{d}$
and $\beta_{c}$ for coupled Poisson LDGM$(l,R=0.5,L,w=3)$ ensembles
with $l=3,4,$ and $5$ and different values of $L$. The results
are obtained by population dynamics (see Sect. \ref{sec:Spatially_Coupling_Effect}.}
\begin{tabular}{cccccc}
\toprule 
\multirow{2}{*}{$l$} & \multirow{2}{*}{$\beta$} & uncoupled & \multicolumn{3}{c}{$L$}\tabularnewline
\cmidrule{4-6} 
 &  & coupled & $32$ & $64$ & $128$\tabularnewline
\midrule
\multirow{2}{*}{$3$} & $\beta_{d}$ & $0.883$ & $0.942$ & $0.941$ & $0.941$\tabularnewline
 & $\beta_{c}$ & $0.940$ & $0.958$ & $0.948$ & $0.946$\tabularnewline
\midrule 
\multirow{2}{*}{$4$} & $\beta_{d}$ & $0.875$ & $1.010$ & $1.010$ & $1.009$\tabularnewline
 & $\beta_{c}$ & $1.010$ & $1.038$ & $1.023$ & $1.017$\tabularnewline
\midrule
\multirow{2}{*}{$5$} & $\beta_{d}$ & $0.832$ & $1.032$ & $1.030$ & $1.029$\tabularnewline
 & $\beta_{c}$ & $1.032$ & $1.067$ & $1.048$ & $1.039$\tabularnewline
\bottomrule
\end{tabular}
\end{table}

Since the free energies of the coupled and underlying ensembles are
the same in the limit of infinite length (known from theorem \ref{thm:theorem1})
and the condensation threshold is a singularity of the free energy
(known from the cavity method), we can conclude on theoretical grounds
that 
\begin{equation}
\lim_{L\to+\infty}\beta_{c}(L,w)=\beta_{c}(w=1).
\end{equation}
Table \ref{table:betad=000026c_simpl} shows that the condensation
threshold $\beta_{c}(L,w)$ of the coupled ensemble is higher than
$\beta_{c}(w=1)$ and decreases as $L$ increases. The finite size
effects are still clearly visible at lengths $L=128$ and are more
marked for larger $w$. This is not surprising since we expect the
finite size corrections to be of order $O(w/L)$.

Let us now discuss the behavior of the dynamical threshold. Table
\ref{table:betad=000026c_simpl_w} displays the results for the ensembles
LDGM$(l=5,R=0.5)$ and LDGM$(l=5,R=0.5,L,w)$. 
\begin{table}[tbh]
\centering{}\protect\caption{\label{table:betad=000026c_simpl_w} The numerical values of $\beta_{d}$
and $\beta_{c}$ for coupled Poisson LDGM$(5,R=0.5,L,w)$ ensembles
with different values of $L$ and $w$. The results are obtained by
population dynamics (see Sect. \ref{sec:Spatially_Coupling_Effect}).}
\begin{tabular}{ccccc}
\toprule 
\multirow{2}{*}{$L$} & \multirow{2}{*}{$\beta$} & \multicolumn{3}{c}{$w$}\tabularnewline
 &  & $2$ & $3$ & $4$\tabularnewline
\midrule
\multirow{2}{*}{$128$} & $\beta_{d}$ & $1.028$ & $1.029$ & $1.030$\tabularnewline
 & $\beta_{c}$ & $1.038$ & $1.039$ & $1.043$\tabularnewline
\midrule 
\multirow{2}{*}{$256$} & $\beta_{d}$ & $1.023$ & $1.027$ & $1.029$\tabularnewline
 & $\beta_{c}$ & $1.035$ & $1.037$ & $1.038$\tabularnewline
\bottomrule
\end{tabular}
\end{table}

The column $w=1$ gives the dynamical and condensation thresholds
of the underlying ensemble, $\beta_{d}(w=1)$ and $\beta_{c}(w=1)$.
We see that for each fixed $L$ the dynamical threshold increases
as a function of $w$. Closer inspection suggests that 
\begin{equation}
\lim_{w\to+\infty}\lim_{L\to+\infty}\beta_{d}(L,w)=\beta_{c}(w=1).\label{eq:saturationequation}
\end{equation}
Equ. \ref{eq:saturationequation} indicates a threshold saturation
phenomenon: for the coupled ensemble the phase of non-zero complexity
shrinks to zero and the condensation point remains unchanged. This
is analogous to the saturation of the BP threshold of LDPC codes towards
the MAP threshold \cite{kudekar2012spatially}. It is also analogous
to the saturation of spinodal points in the Curie-Weiss chain \cite{HMU10}.
Similar observations have been discussed for constraint satisfaction
problems in \cite{Hassani11SAT}.

\subsection{Comparison of $\beta^{*}$ with $\beta_{d}$}

We systematically observe that the optimal algorithmic value $\beta^{*}$
of the BPGD-h algorithm is always lower, but somewhat close to $\beta_{d}$.
For example for the uncoupled case $l=5$ we have $(\beta^{*},\beta_{d})\approx(0.71,0.832)$.
For the coupled ensembles with $(L=64,w=3)$ we have $(\beta^{*},\beta_{d})\approx(1.03,1.038)$.
In fact, in the coupled case we observe $\beta^{*}\approx\beta_{d}\approx\beta_{c}$.
Thus for the coupled ensemble BPGD-h operates well even close to the
condensation threshold.

This is also the case for BPGD-r as we explain in the next paragraph.
We use this fact in the next section to explain the good performance
of the algorithm for coupled instances.

\subsection{Sampling of the Gibbs distribution with BPGD-r}

Threshold saturation, equation \eqref{eq:saturationequation}, indicates
that for $L$ large, the phase of non-zero complexity, occupies a
very small portion of the phase diagram close to $\beta_{c}$. This
then suggests that for coupled ensembles Markov chain Monte Carlo
dynamics, and BPGD-r algorithms are able to correctly sample the Gibbs
measure for values of $\beta$ up to $\approx\beta_{c}$. Let us discus
in more detail this aspect of the BPGD-r algorithm.

By the Bayes rule: 
\begin{align}
\mu_{\beta}(\underline{u}\mid\underline{x})=\prod_{i=1}^{m}\mu_{\beta}(u_{i}\vert\underline{x},u_{1},\dots,u_{i-1}).
\end{align}
Thus we can sample $\underline{u}$ by first sampling $u_{1}$ from
$\mu_{\beta}(u_{1}\vert\underline{x})$, then $u_{2}$ from $\mu_{\beta}(u_{2}\vert\underline{x},u_{1})$
and so on. Then, computing $x_{a}=\oplus_{i\in\partial a}u_{i}$ and
the resulting average distortion, yields half the internal energy
$u(\beta)/2$. With the BPGD-r algorithm the average distortion is
computed in the same way except that the sampling is done with the
BP marginals. So as long as the BP marginals are a good approximation
of the true marginals, the average distortion $D_{{\rm BPGD-r}}(\beta)$
should be close to $u(\beta)/2$. This can be conveniently tested
because the cavity method predicts the simple formula%
\footnote{For $\beta>\beta_{c}$ the formula is different. Indeed, $\beta_{c}$
is a static phase transition point.%
} $u(\beta)/2=(1-\tanh\beta)/2$ for $\beta<\beta_{c}$.

\begin{figure}[tb]
\centering{} \includegraphics{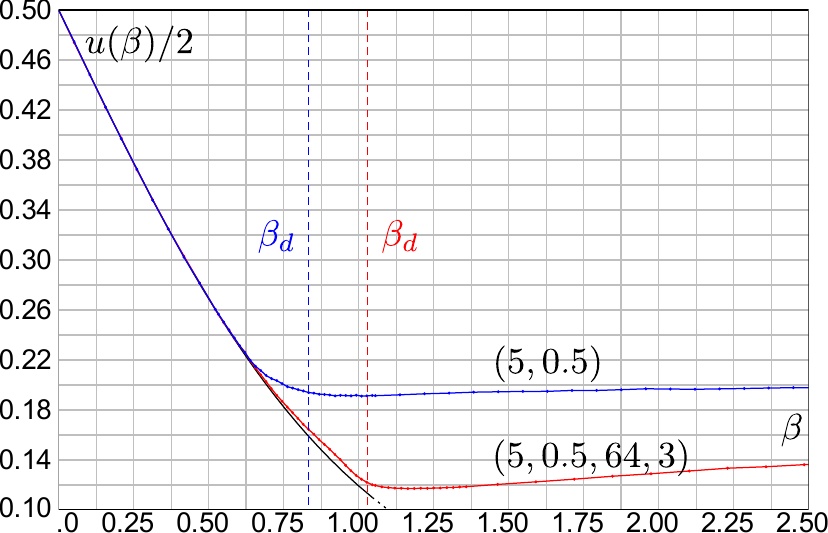}\protect\caption{ \label{fig:randomized510} The performance of the BPGD-r algorithm.
The plot shows that the algorithm can approximate average distortion
quite precisely for $\beta<\beta^{\prime}\approx\beta_{d}$. The black
curve shows the average distortion $u(\beta)/2=(1-\tanh\beta)/2$
for $\beta<\beta_{c}$. The results are obtained for the the underlying
LDGM$(5,0.5,128000)$ and coupled LDGM$(5,0.5,64,3,2000)$ ensembles.
The results are averaged over $50$ instances. Numerical values of
various thresholds are $\beta_{{\rm d,un}}=0.832$, $\beta_{{\rm d,cou}}=1.030$,
$\beta_{{\rm c}}=1.032$. }
\end{figure}

On Fig. \ref{fig:randomized510} we observe $D_{{\rm BPGD-r}}(\beta)\approx(1-\tanh\beta)/2$
for $\beta<\beta^{\prime}$, with a value of $\beta^{\prime}$ lower
but comparable to $\beta_{d}$. In particular for a coupled ensemble
we observe $\beta^{\prime}\approx\beta_{d}\approx\beta_{c}$. So Fig.
\ref{fig:randomized510} strongly suggests that BPGD-r correctly samples
the Gibbs distribution of coupled instances all the way up to $\approx\beta_{c}$,
and that BP correctly computes marginals for the same range.

\subsection{Large Degree Limit\label{sub:insight}}

According to the information theoretic approach to rate-distortion
theory, we can view the encoding problem, as a decoding problem for
a random linear code on a test-BSC(p) test-channel with noise $p=D_{{\rm sh}}(R)$.
Now, the Gibbs distribution \eqref{eq:measure-mu} with $\beta=\frac{1}{2}\ln(1-p)/p$
is a MAP-decoder measure for a channel problem with the noise tuned
to the Shannon limit. Moreover, for large degrees the LDGM ensemble
is expected to be equivalent to the random linear code ensemble. These
two remarks suggest that, since in the case of coupled ensembles with
large degrees the BPGD-h encoder with optimal $\beta^{*}$, approaches
the rate-distortion limit, we should have 
\begin{equation}
\beta^{*}\approx\frac{1}{2}\ln\frac{1-p}{p}\equiv\frac{1}{2}\ln\frac{1-D_{{\rm sh}}(R)}{D_{{\rm sh}}(R)}.
\end{equation}
In fact this is true. Indeed on the one hand, as explained above,
for coupled codes we find $\beta^{*}\approx\beta_{d}\approx\beta_{c}$
(even for finite degrees). On the other hand an analytical \textit{large
degree} analysis of the cavity equations in section \ref{sub:LDL}
allows to compute the complexity and to show the remarkable relation
\begin{equation}
\beta_{c}\approx\frac{1}{2}\ln\frac{1-D_{{\rm sh}}(R)}{D_{{\rm sh}}(R)},\,\,\,{\rm for}\,\,\, l>>1.
\end{equation}
These remarks also show that the rate-distortion curve can be interpreted
as a line of condensation thresholds for each $R$.

\section{Cavity Equations for LDGM Coupled Ensembles\label{sec:CM_application}}

We display the set of fixed point equations needed to compute the
complexity \eqref{eq:complexity} of the coupled ensemble. To get
the equations for the underlying ensembles one sets $w=1$ and drops
the positional $z$ dependence in all quantities.

In order to derive the fixed point equations one first writes down
the cavity equations for a single instance of the graph and source
word. These involve a set of messages on the edges of the graph. These
messages are \textit{random probability distributions}. If one assumes
independence of messages flowing into a node, it is possible to write
down a set of integral fixed point equations - the cavity equations
- for the \textit{probability distributions of the messages}. It turns
out that the cavity equations are much harder to solve numerically
than usual density evolution equations because of ``reweighting factors''.
Fortunately for $\beta<\beta_{c}$ it is possible to eliminate the
reweighting factor, thus obtaining a much simpler set of six integral
fixed point equations. This whole derivation is quite complicated
and for the benefit of the reader, we choose to present it three stages
in appendices \ref{sec:cavityprimer}, \ref{sec:application} and
\ref{sec:CM_population_dynamic}. The calculations are adapted from
the methods of \cite{semerjian} for the $K$-SAT problem in the SAT
phase.

Paragraphs \ref{sub:cavmeth} and \ref{sub:complexityintermsofdensities}
give the set of six integral fixed point equations and the complexity
(derived in appendices \ref{sec:cavityprimer}, \ref{sec:application}
and \ref{sec:CM_population_dynamic}).

We will see that in the present problem for $\beta<\beta_{c}$, not
only one can eliminate the reweighting factors, but there is a further
simplification of the cavity equations. With this extra simplification
the cavity equations reduce to standard density evolution equations
associated to a coupled LDGM code over a test-BSC-channel. This is
explained in paragraph \ref{sub:futhersimpl}.

\subsection{Fixed Point Equations of the Cavity Method for $\beta\leq\beta_{c}$\label{sub:cavmeth}}

Our fixed point equations involve six distributions $q_{z}(h)$, $\widehat{q}_{z}(\widehat{h})$,
$q_{z}^{\sigma}(\eta|h)$ and $\widehat{q}_{z}^{\sigma}(\widehat{\eta}|\widehat{h})$
with $\sigma=\pm1$. The subscript $z$ indicates that the distributions
are position dependent, $z=1,\dots,L+w-1$. A hat (resp. no hat) indicates
that this is the distribution associated to messages that emanate
from a check node (resp. code-bit node). All messages emanating from
a node have the \textit{same} distribution. Thus the distributions
depend only on the position of the node and not on the direction of
the edge.

It is convenient to define two functions $g$ and $\widehat{g}$ (see
the BP equations \eqref{eq:bp_equation2}) 
\[
\begin{cases}
g(\widehat{h}_{1},...\widehat{h}_{r-1})=\sum_{i=1}^{r-1}\widehat{h}_{i}\\
\widehat{g}\left(h_{1},...h_{l-1}\mid J\right)=J\beta^{-1}\tanh^{-1}\bigl(\tanh\beta\prod_{i=1}^{l-1}\tanh\beta h_{i}\bigr)
\end{cases}
\]
where $J\equiv(-1)^{x}$ is the random variable representing the source
bits. Furthermore we set $P(r)=\frac{(l/R)^{r}}{r!}e^{-l/R}$ for
the Poisson degree distribution of code-bit nodes.

Distributions $q_{z}\left(h\right)$, $\widehat{q}_{z}\left(\widehat{h}\right)$
satisfy a set of closed equations%
\footnote{We use the convention that if $z$ is out of range the corresponding
distribution is a unit mass at zero.%
} 
\begin{align}
q_{z}\left(h\right)= & \sum_{r=0}^{\infty}\frac{P(r)}{w^{r}}\sum_{y_{1},\dots y_{k}=0}^{w-1}\int\prod_{a=1}^{r}d\widehat{h}_{a}\widehat{q}_{z-y_{a}}(\widehat{h}_{a})\nonumber \\
 & \times\delta(h-g(\widehat{h}_{1},...,\widehat{h}_{r}))\label{eq:DE_h_updates}
\end{align}
and 
\begin{align}
\widehat{q}_{z}(\widehat{h})= & \frac{1}{w^{l-1}}\sum_{y_{1},\dots,y_{l-1}=0}^{w-1}\int\prod_{i=1}^{l-1}dh_{i}q_{z+y_{i}}\left(h_{i}\right)\nonumber \\
 & \times\frac{1}{2}\sum_{J=\pm1}\delta(\widehat{h}-\widehat{g}(h_{1},...,h_{l-1}\mid J)).\label{eq:DE_hhat_updates}
\end{align}
Let $\sigma_{i}=\pm1$ denote auxiliary ``spin'' variables. We introduce
the conditional measure over $\sigma_{1},\dots,\sigma_{l-1}$, 
\begin{align}
\nu_{1} & (\sigma_{1},...,\sigma_{l-1}|J\sigma ,h_{1},...,h_{l-1})\nonumber \\
 & =\frac{1+J\sigma \tanh\beta\prod_{i=1}^{l-1}\sigma_{i}}{1+J\sigma \tanh\beta\prod_{i=1}^{l-1}\tanh\beta h_{i}}\prod_{i=1}^{l-1}\frac{1+\sigma_{i}\tanh\beta h_{i}}{2}.\label{eq:DE_nu_1}
\end{align}
The equations for distributions $q_{z}^{\sigma}(\eta|h)$ and $\widehat{q}_{z}^{\sigma}(\widehat{\eta}|\widehat{h})$
are 
\begin{align}
q_{z}^{\sigma}(\eta|h) & q_{z}(h)=\sum_{r=0}^{\infty}\frac{P(r)}{w^{r}}\sum_{y_{1},\dots,y_{r}=0}^{w-1}
\times\int\prod_{a=1}^{r}d\widehat{h}_{a}\widehat{q}_{z-y_{a}}(\widehat{h}_a)
\nonumber \\
& \times \delta(h-g(\widehat{h}_{1},...,\widehat{h}_{r}))
\nonumber \\
 & \times\int\prod_{a=1}^{r}d\widehat{\eta}_{a}\widehat{q}_{z-y_{a}}^{\sigma}(\widehat{\eta}_{a}|\widehat{h}_{a})
 \times\delta(\eta-g(\widehat{\eta}_{1},...,\widehat{\eta}_{r}))\label{eq:DE_h_eta_updates}
\end{align}
and 
\begin{align}
\widehat{q}_{z}^{\sigma}(\widehat{\eta}| \widehat{h})&\widehat{q}_{z}(\widehat{h})=\frac{1}{w^{l-1}}\sum_{y_{1},\dots,y_{l-1}=0}^{w-1}\int\prod_{i=1}^{l-1}dh_{i}q_{z+y_{i}}(h_{i})\nonumber \\
 & \times\frac{1}{2}\sum_{J=\pm1}\sum_{\sigma_{1},...,\sigma_{l-1}=\pm1}\nu_{1}(\sigma_{1},...,\sigma_{l-1}|J\sigma,h_{1},...,h_{l-1})\nonumber \\
 & \times\delta(\widehat{h}-\widehat{g}(h_{1},...,h_{l-1}\mid J))\nonumber \\
 & \times\int\prod_{i=1}^{l-1}d\eta_{i}q_{z+y_{i}}^{\sigma_{i}}(\eta_{i}|h_{i})\delta(\widehat{\eta}-\widehat{g}(\eta_{1},...,\eta_{l-1}\mid J)).\label{eq:DE_hhat_etahat_updates}
\end{align}
Equations \eqref{eq:DE_h_updates}, \eqref{eq:DE_hhat_updates}, \eqref{eq:DE_h_eta_updates},
\eqref{eq:DE_hhat_etahat_updates} constitutes a closed set of fixed
point equations for six probability distributions.

Let us pause for a moment to give some information on these distributions and an interpretation of the equations that relate them. 

When there is a proliferation of BP fixed points,
usual density evolution does not track correctly the average behavior of the BP messages. In the formalism of the cavity method (see Appendix A) one 
introduces new messages called {\it cavity messages} which are 
random valued distributions over the space of BP fixed points (for a fixed instance). 
They satisfy
``cavity message passing equations'' (see equ. \eqref{eq:CM_Cavity_equations}). The fixed point equations presented here 
\eqref{eq:DE_h_updates}, \eqref{eq:DE_hhat_updates}, \eqref{eq:DE_h_eta_updates},
\eqref{eq:DE_hhat_etahat_updates}, describe 
the behavior of the ``distributions'' of these cavity messages. More precisely
 the {\it averages of the cavity messages} - themselves 
random quantities -
satisfy message passing BP equations (see equ. \eqref{eq:bphfields}). 
The quantities $q_z(h)$ and $\widehat{q}_z(\widehat{h})$ are the distributions of the {\it averages of the cavity messages} 
(see equ. \eqref{avhhat}) and therefore satisfy the 
``usual'' density evolution equations. The quantities 
$q_z^\sigma(\eta\vert h)$ and $\widehat{q}_z^\sigma(\hat \eta \vert \widehat{h})$ are 
 conditionnal averages of the random cavity messages (see equ.\label{qsigmas}, \eqref{eq:qz_definition}). The conditioning corresponds to fix the average of 
 the cavity message.
 
The equations \eqref{eq:DE_h_eta_updates} and \eqref{eq:DE_hhat_etahat_updates} have an interesting interpretation 
as a  reconstruction problem on a tree (see \cite{Mezard2006} where the case of coloring is treated in detail and a brief discussion of 
more general models is presented). Consider a rooted tree of 
depth $t > 0$ created at random from a stochastic branching process where 
variable nodes have $r-1$ descendants with probability $P(r)$ (except for the root node which has $r$ descendants) and check nodes 
have $l-1$ descendants. Each check node ``broadcasts'' the variable $\sigma$ that is immediately above it, to
its $l-1$ descendants which receive the vector $(\sigma_1,\cdots,\sigma_{l-1})$ with probability 
$\nu_{1}(\sigma_{1},...,\sigma_{l-1}|J\sigma ,h_{1},...,h_{l-1})$. This broadcasting process induces 
a probability distribution on the configurations of the variables at the leaf nodes of the tree. 
The aim of the reconstruction problem is to infer the value of the root node given the configuration at the leafs at depth $t$.
The analysis of the reconstruction problem on a tree suggests that the equations \eqref{eq:DE_h_eta_updates} and \eqref{eq:DE_hhat_etahat_updates} 
possess non-trivial fixed points if and only if the iterations of these equations with the initial condition\footnote{Here we adopt the notation $\delta_{+\infty}$ for a unit mass distribution
at infinity.
} 
\begin{align}
q_{z}^{\sigma_i}(\eta|h)=\delta_{+\infty}(\sigma_i\eta_i),\label{eq:tree_reconstruction_initial_condition}
\end{align}
converges to a non-trivial fixed point.
 This has the advantage of removing the ambiguity of the initial conditions in order to solve iteratively the fix point equations for $q_{z}^{\sigma}(\eta|h)$ and $\widehat{q}_{z}^{\sigma}(\widehat{\eta}|\widehat{h})$.

\subsection{Complexity in Terms of Fixed Point Densities\label{sub:complexityintermsofdensities} }

Let 
\begin{align*}
\begin{cases}
Z_{1}(h_{1},...,h_{l}\mid J)=1+J(\tanh\beta)\prod_{i=1}^{l}\tanh\beta h_{i}\\
Z_{2}(\widehat{h}_{1},...,\widehat{h}_{r})=\frac{1}{2}\sum_{\sigma=\pm1}\prod_{i=1}^{r}(1+\sigma\tanh\beta\widehat{h}_{i}).
\end{cases}
\end{align*}
We are now ready to give the expression for the complexity in terms
of the densities $q_{z}(h)$, $\widehat{q}_{z}(\widehat{h})$, $q_{z}^{\sigma}(\eta|h)$
and $\widehat{q}_{z}^{\sigma}(\widehat{\eta}|\widehat{h})$. Recall
formula \eqref{eq:complexity} which expresses the complexity as $\Sigma(\beta)=\beta(\varphi_{{\rm int}}(\beta)-f(\beta))$.
In the formulas below it is understood that $n\to+\infty$.

The expression of $f$ is the simplest 
\begin{align}
-\beta & f=\ln(1+e^{-2\beta})+(R-1)\ln2\nonumber \\
 & -\frac{l-1}{L}\sum_{z=1}^{L}\frac{1}{w^{l}}\sum_{y_{1},\dots,y_{l}=0}^{w-1}\int\prod_{i=1}^{l}dh_{i}q_{z+y_{i}}(h_{i})\nonumber \\
 & \times\frac{1}{2}\sum_{J=\pm1}\ln Z_{1}(h_{1},...,h_{l}\mid J)\nonumber \\
 & +\frac{R}{L+w-1}\sum_{z=1}^{L+w-1}\sum_{r=0}^{\infty}\frac{P(r)}{w^{r}}\nonumber \\
 & \times\sum_{y_{1},\dots,y_{r}=0}^{w-1}\int\prod_{a=1}^{r}d\widehat{h}_{a}\widehat{q}_{z-y_{a}}(\widehat{h}_{a})\ln Z_{2}(\widehat{h}_{1},...,\widehat{h}_{r}).\label{eq:f_RSB}
\end{align}
To express $\varphi_{{\rm int}}$ we first need to define the conditional
measure over $\sigma=\pm1$ 
\begin{align*}
\nu_{2} & (\sigma|\widehat{h}_{1},...,\widehat{h}_{k})\\
 & =\frac{\prod_{a=1}^{k}(1+\sigma\tanh\beta\widehat{h}_{a})}{\prod_{a=1}^{k}(1+\tanh\beta\widehat{h}_{a})+\prod_{a=1}^{k}(1-\tanh\beta\widehat{h}_{a})}.
\end{align*}
We have 
\begin{align}
-\beta & \varphi_{{\rm int}}=\ln(1+e^{-2\beta})+(R-1)\ln2\nonumber \\
 & -\frac{l-1}{L}\sum_{z=1}^{L}\frac{1}{w^{l}}\sum_{y_{1},\dots,y_{l}=0}^{w-1}\int\prod_{i=1}^{l}dh_{i}q_{z+y_{i}}\left(h_{i}\right)\nonumber \\
 & \times\frac{1}{2}\sum_{J=\pm1}\sum_{\sigma_{1},...,\sigma_{l}=\pm1}\nu_{1}(\sigma_{1},...,\sigma_{l}|J,h_{1},...,h_{l})\nonumber \\
 & \times\int\prod_{i=1}^{l}d\eta_{i}q_{z+y_{i}}^{\sigma_{i}}(\eta_{i}|h_{i})\ln Z_{1}(\eta_{1},...,\eta_{l}\mid J)\nonumber \\
 & +\frac{R}{L+w-1}\sum_{z=1}^{L+w-1}\sum_{r=0}^{\infty}\frac{P(r)}{w^{r}}\nonumber \\
 & \times\sum_{y_{1},...,y_{r}=0}^{w-1}\int\prod_{a=1}^{r}d\widehat{h}_{a}\widehat{q}_{z-y_{a}}(\widehat{h}_{a})\sum_{\sigma}\nu_{2}(\sigma|\widehat{h}_{1},...,\widehat{h}_{r})\nonumber \\
 & \times\int\prod_{a=1}^{r}d\widehat{\eta}_{a}\widehat{q}_{z-y_{a}}^{\sigma}(\widehat{\eta}_{a}|\widehat{h}_{a})\ln Z_{2}(\widehat{\eta}_{1},...,\widehat{\eta}_{r}).\label{eq:DE_phi_rsb}
\end{align}

Thanks to \eqref{eq:f_RSB}, \eqref{eq:DE_phi_rsb} the complexity
$\Sigma(\beta;L,w)$ of the coupled ensemble is computed, one reads
off the dynamical and condensation thresholds $\beta_{d}(L,w)$ and
$\beta_{c}(L,w)$. The corresponding quantities for the underlying
ensemble are obtained by setting $L=w=1$.

\subsection{Further Simplications of Fixed Point Equations and Complexity\label{sub:futhersimpl}}

It is immediate to check that $q_{z}(h)=\delta(h)$ and $\widehat{q}_{z}(\widehat{h})=\delta(\widehat{h})$
is a trivial fixed point of \eqref{eq:DE_h_updates}, \eqref{eq:DE_hhat_updates}.
When we solve these equations by population dynamics with a uniform
initial condition over $[-1,+1]$ for $\widehat{h}$, we find that
for fixed degrees and $\beta$ fixed in a finite range depending on
the degrees, the updates converge towards the trivial fixed point.
Up to numerical precision, the values of $h,\widehat{h}$ are concentrated
on $0$. It turns out that the range of $\beta$ for which this is
valid is wider than the interval $[0,\beta_{c}]$. At first sight
this may seem paradoxical, and one would have expected that this range
of $\beta$ is equal to $[0,\beta_{c}]$. In fact, one must recall
that beyond $\beta_{c}$ the equations of paragraph \ref{sub:cavmeth}
are not valid (see Appendix \ref{sec:cavityprimer}), so there is
\textit{no} paradox. Theorem \ref{thm:h_converge} in section \ref{sec:large-degree-limit-theorems}
shows that, for a wide class of initial conditions and given $\beta$,
for large enough degree $l$ the iterative solution of \eqref{eq:DE_h_updates},
\eqref{eq:DE_hhat_updates} tends to the trivial point. This theorem,
together with the numerical evidence, provides a strong justification for the following simplification.

We assume that for $\beta<\beta_{c}$, equations \eqref{eq:DE_h_updates},
\eqref{eq:DE_hhat_updates} have a unique solution $q_{z}(h)=\delta(h)$
and $\widehat{q}_{z}(\widehat{h})=\delta(\widehat{h})$.
Note that the initial condition \eqref{eq:tree_reconstruction_initial_condition} satisfies a symmetry 
$q^\sigma(\eta\vert 0) = q^{-\sigma}(-\eta\vert 0) = \delta_{+\infty}(\eta\sigma)$ (even for $h\neq 0$). 
Now for $h=\hat{h}=0$ the iterations of \eqref{eq:DE_h_eta_updates} and \eqref{eq:DE_hhat_etahat_updates}
preserve this symmetry. In other words the solutions of these equations (for $h=\hat h=0$) found from a symmetric initial condition
satisfy $q_{z}^{\sigma=1}(\eta|0)=q_{z}^{\sigma=-1}(-\eta|0)$,
$\widehat{q}_{z}^{\sigma=1}\left(\widehat{\eta}|0\right)=\widehat{q}_{z}^{\sigma=-1}(-\widehat{\eta}|0)$.

Therefore we look only for symmetrical solutions, and set
\[
q_{z}^{+}(\eta)=q_{z}^{\sigma=+1}(\eta|0),\qquad{\rm and}\qquad\widehat{q}_{z}^{+}(\widehat{\eta})=\widehat{q}_{z}^{\sigma=+1}(\widehat{\eta}|0)
\]
Then the equations \eqref{eq:DE_h_eta_updates}, \eqref{eq:DE_hhat_etahat_updates}
simplify drastically, 
\begin{align}
q_{z}^{+}(\eta) & =\sum_{r=0}^{\infty}\frac{P(r)}{w^{r}}\sum_{y_{1},\dots,y_{r}=0}^{w-1}\nonumber \\
 & \times\int\prod_{a=1}^{r}d\widehat{\eta}_{a}\widehat{q}_{z-y_{a}}^{+}(\widehat{\eta}_{a})\delta(\eta-g(\widehat{\eta}_{1},...,\widehat{\eta}_{r}))\label{eq:DE_zero_eta_updates}\\
\widehat{q}_{z}^{+}\left(\widehat{\eta}\right) & =\frac{1}{w^{l-1}}\sum_{y_{1},...,y_{l-1}=0}^{w-1}\int\prod_{i=1}^{l-1}d\eta_{i}q_{z+y_{i}}^{+}(\eta_{i})\nonumber \\
 & \times\sum_{J=\pm1}\frac{1+J\tanh\beta}{2}\delta(\widehat{\eta}-\widehat{g}(\eta_{1},...\eta_{l-1}\mid J)).\label{eq:DE_zero_eta_updates1}
\end{align}
Remarkably, these are the standard density evolution equations for
an LDGM code over a test-BSC-channel with half-log-likelihood parameter
equal to $\beta$.

The free energy \eqref{eq:f_RSB} now takes a very simple form 
\begin{equation}
-\beta f=\ln(1+e^{-2\beta})+(R-1)\ln2.\label{eq:free-simple}
\end{equation}
At this point let us note that this simple formula has been proven
by the \textit{interpolation method} \cite{guerra2004high}, for small
enough $\beta$. Since it is expected that there is no (static) thermodynamic
phase transition for $\beta<\beta_{c}$, the free energy is expected
to be analytic for $\beta<\beta_{c}$. Thus by analytic continuation,
formula \eqref{eq:free-simple} should hold for all $\beta<\beta_{c}$.
This also provides justification for the triviality assumption
made above for the fixed point. Indeed, a non-trivial fixed point
leading to the same free energy would entail miraculous cancellations.

When we compute the complexity, expression \eqref{eq:free-simple}
cancels with the first line in $\varphi_{{\rm int}}$ (see equ. \eqref{eq:DE_phi_rsb}).
We find 
\begin{align*}
\Sigma(\beta;L,w) & =\frac{l-1}{L}\sum_{z=1}^{L}\frac{1}{w}\sum_{y=0}^{w-1}\Sigma_{e}[q_{z+y}^{+},\widehat{q}_{z}^{+}]\\
 & -\frac{l}{L}\sum_{z=1}^{L}\Sigma_{v}\left[\widehat{q}_{z}^{+}\right]+\frac{R}{L+w-1}\sum_{z=1}^{L+w-1}\Sigma_{v}\left[q_{z}^{+}\right],
\end{align*}
where 
\begin{align*}
\Sigma_{v}[q^{+}] & =\int d\eta\, q^{+}(\eta)\ln(1+\tanh\beta\eta)\\
\Sigma_{e}[q^{+},\widehat{q}^{+}] & =\int d\eta d\widehat{\eta}\, q^{+}(\eta)\widehat{q}^{+}(\widehat{\eta})\ln(1+\tanh\beta\eta\tanh\beta\widehat{\eta}).
\end{align*}
For the underlying ensemble ($L=w=1$) the complexity reduces to 
\begin{equation}
\Sigma(\beta)=(l-1)\Sigma_{e}[q^{+},\widehat{q}^{+}]-l\Sigma_{v}[\widehat{q}^{+}]+R\Sigma_{v}[q^{+}].\label{eq:compl_underlying}
\end{equation}

The average distortion or internal energy (see \eqref{eq:internal-energy},
\eqref{eq:derivative}) at temperature $\beta$ is obtained by differentiating
\eqref{eq:free-simple}, which yields the simple formula $(1-\tanh\beta)/2$.
This is nothing else than the (bottom) curve plotted in Figure \ref{fig:randomized510}.
It has to be noted that this expression is only valid for $\beta<\beta_{c}$.
To obtain the optimal distortion of the ensemble $D_{{\rm opt}}$
(see table \ref{table:distortion}) one needs to recourse to the full
cavity formulas in order to take the limit $\beta\to+\infty$.

\subsection{Large degree limit\label{sub:LDL}}

Inspection of the fixed point equations \eqref{eq:DE_zero_eta_updates}
and \eqref{eq:DE_zero_eta_updates1} shows that the distributions

\begin{equation}
q^{+}(\eta)=\delta_{+\infty}(\eta),\,{\rm and}\,\,\widehat{q}^{+}(\widehat{\eta})=\sum_{J=\pm1}\frac{1+J\tanh\beta}{2}\delta(\widehat{\eta}-J)\label{eq:large-degree-fixed-point}
\end{equation}
are a fixed point solution for the underlying model ($w =1$) in the limit $l\to+\infty$, $R$ fixed.
This is (partially) justified by theorem \ref{thm:Poisson} in section
\ref{sec:large-degree-limit-theorems}. The fixed point \eqref{eq:large-degree-fixed-point}
leads to a complexity for the underlying model for $l\to+\infty$,
\begin{align*}
\lim_{l\to+\infty}\Sigma(\beta)= & \left(R-1\right)\ln2\\
 & -\sum_{J=\pm1}\frac{1+J\tanh\beta}{2}\ln\bigl(\frac{1+J\tanh\beta}{2}\bigr).
\end{align*}

On this expression one can read the large degree limit of the dynamical and condensation thresholds for the underlying ensemble. In this limit
the complexity is non-zero all the way up to $\beta=0$ (infinite temperature) so one finds that $\lim_{l\to +\infty}\beta_d = 0$.
The condensation threshold on the other hand, $\lim_{l\to+\infty}\beta_{c}$, is obtained
by setting the complexity to zero 
\begin{equation}
1-R=\lim_{l\to+\infty}h_{2}\bigl(\frac{1+\tanh\beta_{c}}{2}\bigr),
\end{equation}
which is equivalent to 
\begin{equation}
\lim_{l\to+\infty}\beta_{c}=\beta_{{\rm sh}}\equiv\frac{1}{2}\ln\bigl(\frac{1-D_{\mathrm{sh}}(R)}{D_{\mathrm{sh}}(R)}\bigr).\label{eq:large-degree-beta-c}
\end{equation}
In the large degree limit the condensation threshold is equal to the
half-log-likelihood of a BSC test-channel with probability of error
$D_{\mathrm{sh}}(R)$, i.e. tuned to capacity.

Notice that since the condensation thresholds for both the underlying and the spatially-coupled ensembles are equal, Equation \eqref{eq:large-degree-beta-c} is also true for coupled ensembles.  Moreover the average
distortion or internal energy is given for both ensembles by 
\begin{equation}
\frac{1}{2}u(\beta)=\left\{ \begin{array}{cc}
\frac{1}{2}(1-\tanh\beta) & \beta<\beta_{{\rm sh}}(R)\\
D_{\mathrm{sh}}(R) & \beta\geq\beta_{{\rm sh}}(R)
\end{array}\right.
\end{equation}
The above equation is a consequence of the monotonicity of $u\left(\beta\right)$
and the saturation of the condensation threshold toward the Shannon
threshold. We conclude this section with a proof of this fact.

Using \eqref{eq:partition-function}, \eqref{eq:thermal-average}
and \eqref{eq:free-energy}, it is not hard to show that the derivative
with respect to $\beta$ of the internal energy for finite size $N$
has a sign opposite to  the variance of the distortion 
\begin{equation}
\frac{d}{d\beta}u_{N}\left(\beta\right)=-4N\mathbb{E}_{{\rm LDGM},\underline{X}}[\langle d_{N}(\underline{x},\widehat{\underline{x}})\rangle^{2}-\langle d_{N}(\underline{x},\widehat{\underline{x}})^{2}\rangle].
\end{equation}
This proves that for every $N$ the internal energy $u_{N}\left(\beta\right)$
is a non-increasing function with respect to $\beta\in\left[0,\infty\right[$.
It also proves, thanks to Equation \eqref{eq:derivative}, that the
free energy $-\beta f_{N}$ is a convex function with respect to
$\beta\in\left[0,\infty\right[$. The cavity method predicts that
in the thermodynamic limit $N\rightarrow+\infty$ the quantity $\beta f_{N}$
converges to \eqref{eq:free-simple} for $\beta\leq\beta_{c}$. This prediction combined with the fact that $-\beta f_{N}$ is convex implies that
the internal energy $u\left(\beta\right)=\lim_{N\rightarrow+\infty}u_{N}\left(\beta\right)$
converges\footnote{See for instance \cite[p.~203]{Urruty2001}  to understand why convexity enables us to exchange the thermodynamical limit and the derivative.} to 
\begin{equation}
u\left(\beta\right)=\frac{d}{d\beta}(\beta f)=(1-\tanh\beta),
\end{equation}
for $\beta\leq\beta_{c}$. Thus in the limit of large degree the internal
energy becomes equal to twice the Shannon distortion at the condensation
threshold
\begin{equation}
\lim_{l\rightarrow+\infty}u\left(\beta_{c}\right)=u\left(\beta_{{\rm sh}}\right)=2D_{\mathrm{sh}}(R).
\end{equation}
But since $2D_{\mathrm{sh}}(R)$ is a lower bound for $\liminf_{\beta\to+\infty}u\left(\beta\right)$
(thanks to the rate-distortion theorem) and $u\left(\beta\right)$
is a non-increasing function, we conclude that $u\left(\beta\right)\equiv2D_{\mathrm{sh}}(R)$
for $\beta\geq\beta_{{\rm sh}}.$

\section{Population Dynamics Computation of the Complexity\label{sec:Spatially_Coupling_Effect}}

In this section, we describe the population dynamics solutions of
the various fixed point equations.

Let us first discuss the solution of \eqref{eq:DE_h_updates}, \eqref{eq:DE_hhat_updates},
\eqref{eq:DE_h_eta_updates} and \eqref{eq:DE_hhat_etahat_updates}.
To represent the densities $q_{z}(h)$, $q_{z}^{\pm}(\eta|h)$, $\widehat{q}_{z}(\widehat{h})$,
and $\widehat{q}_{z}^{\pm}(\widehat{\eta}|\widehat{h})$ we use two
populations: a code-bit population and a check population. The code-bit
population is constituted of $L+w-1$ sets labeled by $z\in[1,L+w-1]$.
Each set, say $z$, has a population of size $n$, constituted of
triples: $(h_{(z,i)},\eta_{(z,i)}^{+},\eta_{(z,i)}^{-})$, $1\leq i\leq n$.
The total size of the code-bit population is $(L+w-1)n$. Similarly,
we have a population of triples with size $Ln$ for check nodes, i.e.
$(\widehat{h}_{(z,a)},\widehat{\eta}_{(z,a)}^{+},\widehat{\eta}_{(z,a)}^{-}),z=1,\dots,L,a=1,\dots,n$.
As inputs, they require the population size $n$, the maximum number
of iterations $t_{\text{max}}$, and the specifications of the coupled
LDGM ensemble $l,r,L,w$. First we solve the two equations \eqref{eq:DE_h_updates}
and \eqref{eq:DE_hhat_updates} with Algorithm \ref{alg:PopDyn_in}.

\begin{algorithm}[tbh]
\protect\caption{\label{alg:PopDyn_in}Population Dynamics for \eqref{eq:DE_h_updates}
and \eqref{eq:DE_hhat_updates}}

\For{ $z=1$ to $L+w-1$} {\For{$i=1$ to $n$}{Draw $\widehat{h}_{(z,i)}$
uniformly from $[-1,+1]$;}} \For {$t\in\{1,\dots,t_{max}\}$}
{ \For{$z=1$ to $L+w-1$} { \For{$i=1$ to $n$} { \textit{Generate
a new $h_{(z,i)}$};

Choose $l-1$ pair indices $a_{1},\dots,a_{l-1}$ uniformly from $nw$
pairs $(y,j)$, $y\in[z-w+1,z]$ and $j\in\{1,...,n\}$;

\If{ for some index $k$, $a_{k}=(y,j)$ and $y<1$} {Set $\widehat{h}_{a_{k}}=0$;}

Set $h_{(z,i)}=\sum_{k=1}^{l-1}\widehat{h}_{a_{k}}$; } } \For{$z=1$
to $L$} { \For{$a=1$ to $n$} { \textit{Generate $J$ randomly
and generate a new $\widehat{h}_{(z,a)}$};

Choose $r-1$ indices $i_{1},\dots,i_{r-1}$ uniformly from $nw$
pairs $(y,j)$, $y\in[z,z+w-1]$ and $j\in\{1,...,n\}$;

Compute $\widehat{h}_{(z,a)}$ according to \eqref{eq:DE_hhat_updates};
} }

}
\end{algorithm}

Then we solve \eqref{eq:DE_h_eta_updates} and \eqref{eq:DE_hhat_etahat_updates}
with the Algorithm%
\footnote{ In the next to last line marked ({*}) the chosen index is not in
a valid range. In an instance of a coupled ensemble, this happens
at the boundary, in which the corresponding node has smaller degree.
In the message passing equation we discard these indices or equivalently
assume that their triples are $(0,0,0)$.%
} \ref{alg:PopDyn}. 
\begin{algorithm}[tbh]
\protect\caption{\label{alg:PopDyn}Population Dynamics for \eqref{eq:DE_h_eta_updates}
and \eqref{eq:DE_hhat_etahat_updates}}

\For{ $z=1$ to $L$} {\For{$i=1$ to $n$}{Set $\eta_{(z,i)}^{\pm}=\pm\infty$
and draw $h_{(z,i)}$ from $q_{z}(h)$;}} \For {$t\in\{1,\dots,t_{max}\}$}
{ \For{$z=1$ to $L$} { \For{$a=1$ to $n$} { \textit{Generate
$J$ randomly and generate a new triple $(\widehat{h}_{(z,a)},\widehat{\eta}_{(z,a)}^{+},\widehat{\eta}_{(z,a)}^{-})$}:

Choose $r-1$ indices $i_{1},\dots,i_{r-1}$ uniformly from $nw$
pairs $(y,j)$, $y\in[z,z+w-1]$ and $j\in\{1,...,n\}$;

Compute $\widehat{h}_{(z,a)}$ according to \eqref{eq:DE_hhat_updates};

Generate a configuration $\sigma_{1},\dots,\sigma_{r-1}$ from $\nu_{1}(\dots\vert+J,h_{i_{1}},\dots,h_{i_{r-1}})$
in \eqref{eq:DE_nu_1};

Compute $\widehat{\eta}_{(z,a)}^{+}$ by plugging $\eta_{i_{1}}^{\sigma_{1}},\dots,\eta_{i_{r-1}}^{\sigma_{r-1}}$
in \eqref{eq:DE_hhat_etahat_updates};

Generate a configuration $\sigma_{1},\dots,\sigma_{r-1}$ from $\nu_{1}(\dots\vert-J,h_{i_{1}},\dots,h_{i_{r-1}})$
in \eqref{eq:DE_nu_1};

Compute $\widehat{\eta}_{(z,a)}^{-}$ by plugging $\eta_{i_{1}}^{\sigma_{1}},\dots,\eta_{i_{r-1}}^{\sigma_{r-1}}$
in \eqref{eq:DE_hhat_etahat_updates}; } } \For{$z=1$ to $L+w-1$}
{ \For{$i=1$ to $n$} { \textit{Generate a new triple $(h_{(z,i)},\eta_{(z,i)}^{+},\eta_{(z,i)}^{-})$}:

Choose $l-1$ pair indices $a_{1},\dots,a_{l-1}$ uniformly from $nw$
pairs $(y,j)$, $y\in[z-w+1,z]$ and $j\in\{1,...,n\}$;

\If{ for some index $k$, $a_{k}=(y,j)$ and $y<1$} {Set $(\widehat{h}_{a_{k}},\widehat{\eta}_{a_{k}}^{+},\widehat{\eta}_{a_{k}}^{-})=(0,0,0)$;({*})}

Set $h_{(z,i)}=\sum_{k=1}^{l-1}\widehat{h}_{a_{k}}$ and $\eta_{(z,i)}^{\pm}=\sum_{k=1}^{l-1}\widehat{\eta}_{a_{k}}^{\pm}$;
} } } 
\end{algorithm}

From the final populations obtained after $t_{{\rm max}}$ iterations
it is easy to compute the complexity and the thresholds $\beta_{d}$,
$\beta_{c}$.

It is much simpler to solve the simplified fixed point equations \eqref{eq:DE_zero_eta_updates},
\eqref{eq:DE_zero_eta_updates1}. The population dynamics algorithm
is almost the same than in Table \ref{alg:PopDyn_in}. The only difference
is that $J$ is generated according to the p.d.f $(1+J\tanh\beta)/2$
instead of Ber$(1/2)$. The big advantage is that there is no need
to generate the $2^{r-1}$ configurations $\sigma_{1},...,\sigma_{r-1}$
which reduces the complexity of each iteration.

As expected the complexity obtained in either way is the same up to
numerical precision. Numerical values of the dynamical and condensation
thresholds are presented in tables \ref{table:betad=000026c_simpl}
and \ref{table:betad=000026c_simpl_w}. Results are obtained with
population sizes $n=30000$ (uncoupled), $n=500-1000$ (coupled),
and iteration number $t_{\text{max}}=3000$.

\section{Two Theorems and Discussion of Threshold Saturation\label{sec:large-degree-limit-theorems}}

Theorem \ref{thm:h_converge} provides theoretical support for the
simplifications of the cavity equations discussed in section \ref{sub:futhersimpl}. 
\begin{thm}
\label{thm:h_converge} Consider the fixed point equations \eqref{eq:DE_h_updates}
and \eqref{eq:DE_hhat_updates} for the individual Poisson LDGM$(l,R)$
ensemble with a fixed $\beta$. Take any initial continuous density
$\hat{q}^{(0)}(\hat{h})$ and consider iterations $\hat{q}^{(t)}(\hat{h})$.
There exists $l_{0}\in\mathbb{N}$ such that for $l>l_{0}$, $\lim_{t\to\infty}\widehat{h}^{(t)}=0$
almost surely. 
\end{thm}
The proof%
\footnote{It can be extended to other irregular degree distributions.%
} is presented in Appendix \ref{sec:proof_h_converge}. Note that $l_{0}$
depends on $\beta$ and $R$. However we expect that as long as $\beta<\beta_{c}$
the result holds for all $l\geq3$ and $R$. This is corroborated
by the numerical observations. When we solve equations \eqref{eq:DE_h_updates}
and \eqref{eq:DE_hhat_updates} by population dynamics with $\hat{q}^{(0)}(\hat{h})$
the uniform distribution, we observe that for a finite range of $\beta$
depending on $(l,R)$, the densities $q^{(t)}(h),\widehat{q}^{(t)}(\widehat{h})$
tend to a Dirac distribution at the origin. The range of $\beta$
for which this occurs always contains the interval $[0,\beta_{c}]$
irrespective of $(l,R)$. These observations also hold for many other
initial distributions. We note that these observations break down
for $\beta$ large enough.

Theorem \ref{thm:Poisson} partially justifies \eqref{eq:large-degree-fixed-point}
which is the basis for the computation of the complexity in the large
degree limit in section \ref{sub:LDL}. 
\begin{thm}
\label{thm:Poisson} Consider the fixed point equations \eqref{eq:DE_zero_eta_updates}
and \eqref{eq:DE_zero_eta_updates1} associated to the individual
Poisson LDGM$(l,R)$ ensemble for some $l$, $R$ and $\beta$ ($w=1$
in the equations). Let $\widehat{\eta}^{(t)}$ be a random variable
distributed according to $\widehat{q}^{+(t)}(\widehat{\eta})$ at
iteration $t$. Assume that the initial density is 
\[
\widehat{q}^{+(0)}(\hat{\eta})=\sum_{J=\pm1}\frac{1+J\tanh(\beta)}{2}\delta(\hat{\eta}-J).
\]
Then, 
\begin{itemize}
\item i) For all $t$, 
\begin{equation}
\widehat{q}^{+(t)}\left(-\widehat{\eta}\right)=e^{-2\beta\widehat{\eta}}\widehat{q}^{+(t)}\left(\widehat{\eta}\right),\label{eq:Nishimori}
\end{equation}
\begin{equation}
q^{+(t)}\left(-\eta\right)=e^{-2\beta\eta}q^{+(t)}\left(\eta\right).
\end{equation}

\item ii) For any $\delta>0$, $\epsilon>0$ and $B>0$ , there exits $l_{1}$
such that for $l>l_{1}$ and all $t$. 
\begin{equation}
\mathbb{P}\left\{ 1-\epsilon\leq\widehat{\eta}^{(t)}\leq1\right\} >\frac{e^{2\beta}}{1+e^{2\beta}}(1-\delta),\label{eq:etahat_concen}
\end{equation}
\begin{equation}
\mathbb{P}\left\{ -1\leq\widehat{\eta}^{(t)}\leq-1+\epsilon\right\} >\frac{1}{1+e^{2\beta}}(1-\delta).\label{eq:etahat_concen2}
\end{equation}

\end{itemize}
\end{thm}
The proof is presented in Appendix \ref{proof:Poisson}.

We now wish to briefly discuss the issue of threshold saturation. One of the main observations 
of this work is the saturation of the dynamical inverse temperature threshold towards
the condensation threshold: $\lim_{w\to +\infty}\lim_{L\to +\infty} \beta_d(L, w) = \beta_c(w=1)$.
This is analogous to threshold saturation in coding theory where the Belief Propagation threshold of the coupled code ensemble saturates towards 
the MAP threshold. In this latter case we have proofs of this phenomenon for the rather general case of irregular LDPC codes (with bounded degrees) and 
binary-input memoryless-output symmetric channels \cite{kudekar2012spatially}, \cite{kumar2012proof}, \cite{KYMP2014}. 
The proof in \cite{KYMP2014} is based on the analysis of a potential function given by the replica-symmetric formula
(an average form of the Bethe free energy)
for the (infinite length) conditional input-output entropy of the code ensemble. 
We expect that, for the present problem, a proof of threshold saturation could be based on a potential function given by the complexity functional 
introduced in Section \ref{sec:CM_application}. Theorem 2 hints that the only solutions (for $\beta$ in the range of interest) of equations 
\eqref{eq:DE_h_updates}
and \eqref{eq:DE_hhat_updates} is trivial. Then the complexity functional reduces to a simplified form as explained in Section \ref{sub:futhersimpl}.
It is possible to check by explicit functional differentiation that the stationary point equations for this functional are precisely
the fixed point equations \eqref{eq:DE_zero_eta_updates}, \eqref{eq:DE_zero_eta_updates1}, and as already pointed out 
these are the density evolution relations for an LDGM code over a test-BSC-channel with half-log-likelihood parameter $\beta$. 
A proof of threshold saturation could eventually be achieved along these lines, using the techniques of the recent paper \cite{KYMP2014}, which also addresses 
LDGM codes.

\section{conclusion\label{sec:conclusion}}

Let us briefly summarize the main points of this paper. We have investigated
a simple spatially coupled LDGM code ensemble for lossy source coding.
No optimization on the degree distribution is required: the check
degree is regular and the code-bit degree is Poisson. We have shown
that the algorithmic rate-distortion curve of a low complexity encoder
based on BPGD allows to approach the ultimate Shannon rate-distortion
curve, for all compression rates, when the check degree grows large.
The inverse temperature parameter (or equivalently test-channel parameter)
of the encoder may be optimized. However we have observed numerically,
and have argued based on large degree calculations, that a good universal
choice is $\beta_{{\rm sh}}(R)$, given by tuning the test channel
to capacity. We recall that for the underlying (uncoupled) ensemble
the same encoder does not perform well, indeed as the degree grows
large, the difference between the algorithmic rate-distortion and
Shannon rate-distortion curves grows. Insight into the excellent performance
of the BPGD algorithm for spatially coupled ensemble is gained by
studying the phase diagram of the Gibbs measure on which the BPGD
encoder is based. We have found, by applying the cavity method to
the spatially coupled ensemble, that the dynamical (inverse temperature)
threshold $\beta_{d}$ saturates towards the condensation (inverse
temperature) threshold $\beta_{c}$. For this reason the BPGD encoder
can operate close to the condensation threshold $\beta_{c}$, which
itself tends in the large degree limit to $\beta_{{\rm sh}}(R)$,
the test channel parameter tuned at capacity. For the underlying (uncoupled)
ensemble the dynamical threshold moves in the opposite direction in
the large degree limit so that the BPGD algorithm cannot operate close
to the Shannon limit.

We mention some open questions that are left out by the present study
and which would deserve more investigations.

For fixed degrees the best value of the inverse temperature $\beta_{*}$
of the BPGD algorithm is close to, but systematically lower, than
the dynamical temperature $\beta_{d}$. While the value of $\beta_{d}$
can be calculated by the cavity theory, here we determine $\beta_{*}$
by purely empirical means and it is not clear what are the theoretical
principles that allow to determine its value. As the graph is decimated
the degree distribution changes and the effective dynamical temperature
of the decimated graphs should evolve to slightly different values.
It is tempting to conjecture that $\beta_{*}$ is the limit of such
a sequence of dynamical temperatures. A related phenomenon has been
observed for the dynamical threshold with respect to clause density
for random constraint satisfaction problems in their SAT phase \cite{Tersenghi}.

The decimation process used in this paper is hard to analyze rigorously
because it is not clear how to keep track of the statistics of the
decimated graph. As a consequence it is also not clear how to compute the optimal value of the 
inverse temperature along the decimation process (we fix this value once for all). Progress on this problem could maybe be achieved by 
redesigning the decimation process, however how to achieve this is at the moment not clear. We would like to point out that a related
process has been investigated in recent works \cite{coja2011belief}
for the $K$-SAT problem in the large $K$ limit up to the dynamical
threshold(in the SAT phase). These methods could be of use also in
the present case.

In this contribution we have investigated a linear decoding rule. Source coding with non-linear rules are of interest and have been 
studied in  \cite{CilMezZec06}. It is an open question to look at the algorithmic performance of such codes in the framework of spatial coupling.

Finally, while a rigorous control of the full cavity method is, in
general, beyond present mathematical technology, there are sub-problems
for which progress can presumably be made. For example in the present
case we have observed that the cavity equations reduce (in the dynamical
phase $\beta_{d}<\beta<\beta_{c}$) to density evolution equations
for an LDGM code on a BSC. The saturation of the dynamical temperature
$\beta_{d}$ to the condensation temperature $\beta_{c}$ appears
to be very similar to the threshold saturation phenomenon of channel
coding theory. We have by now a host of mathematical methods pertaining
to this effect for LDPC on general binary memoryless channels \cite{kudekar2012spatially},
\cite{kumar2012proof}. We think that these methods could be adapted
to prove the saturation of $\beta_{d}$ towards $\beta_{c}$. One
extra difficulty faced in the present problem is that the ``trivial''
fixed point of density evolution equations of LDPC codes is not always
present in the LDGM case.

\appendices{}

\section{A Primer on the cavity Method\label{sec:cavityprimer}}

We give a brief introduction to the cavity method for general spin
systems on sparse graphs. As explained in Sect. \ref{sub:dyncond},
turning this formalism into a rigorous mathematical theory is a long
standing open problem. However, it allows to compute many quantities
of interest. In appendices \ref{sec:application} and \ref{sec:CM_population_dynamic}
we specialize to the probability distribution \eqref{eq:measure-mu}.

The treatment given here applies to \textit{single instances}. Let
$\Gamma=\left(V,C,E\right)$ a factor graph which is assumed to be
locally tree like. We attach spins $\sigma_{j}$, $j\in V$ to variable
nodes, and constraint functions $\psi_{a}\left(\{\sigma_{i},i\in\partial a\}\right)$,
$a\in C$ to check nodes. We sometimes use the notation $\underline{\sigma}_{\partial a}=\{\sigma_{i},i\in\partial a\}$
as a shorthand. The formalism developed in this appendix is valid
for general spin variables belonging to a finite alphabet $\sigma_{j}\in\mathcal{X}$.
The constraint functions depend only on the set of spins connected
to $a$. We are interested in the thermodynamic limit where $\vert V\vert=N$
and $\vert C\vert=M$ tend to infinity and the ratio $M/N$ is kept
fixed. We consider the general class of Gibbs distributions of the
form 
\begin{equation}
\mu\left(\underline{\sigma}\right)=\frac{1}{Z}\prod_{a\in C}\psi_{a}\left(\{\sigma_{i},i\in\partial a\}\right),\label{eq:CM_markov_random_field}
\end{equation}
where $Z$ is the partition function. The free energy of an instance
is defined as usual 
\begin{equation}
\phi\left(\beta\right)=-\frac{1}{N\beta}\ln Z\left(\beta\right)\,.
\end{equation}
One of the goals of the cavity method is to compute this free energy
in the limit $N\to+\infty$.

Let us first outline the general strategy. For locally tree like graphs,
one can compute the marginals for a given node by restricting the
measure to a tree. In the absence of long range correlations%
\footnote{More precisely point-to-set correlations \cite{mezard09information}.%
} the marginal does not depend on the boundary conditions at the leaf
nodes, and the BP equations have one relevant solution. The BP marginals
then constitute a good description of the measure \eqref{eq:CM_markov_random_field}.
In particular, the true free energy is well approximated by replacing
this solution in the Bethe free energy functional. As the control
parameters vary long range correlations may appear. In such a situation
the marginals computed on a tree will depend on the boundary conditions
at the leaf nodes, and the BP equations will have many relevant solutions
yielding nearly the same Bethe free energy. The cavity method assumes
that the measure \eqref{eq:CM_markov_random_field} is then described
by a convex superposition of ``extremal measures''. There may be
a large number of extremal measures. A good proxy for the extremal
measures is given by the BP marginals. The convex superposition of
extremal measures yields a new statistical model on the same factor
graph, the so-called \textit{level-one model}. Assuming that the level
one model does not display long range correlations, one can solve
it using BP equations and the Bethe free energy. Otherwise, the cavity
method iterates the previous considerations and constructs a level-two
model. However, this usually becomes bewildering and one stops at
the first level. In the following paragraphs we give a concrete implementation
of these ideas.

The BP equations are a set of fixed point equations satisfied by messages
$\{\nu_{i\to a},\widehat{\nu}_{a\to i}\}=(\underline{\nu},\widehat{\underline{\nu}})$,
\begin{equation}
\widehat{\nu}_{a\rightarrow i}=\widehat{g}_{\mathrm{BP}}\left(\left\{ \nu_{j\rightarrow a}\right\} _{j\in\partial a\backslash i}\right),~\nu_{i\rightarrow a}=g_{\mathrm{BP}}\left(\left\{ \widehat{\nu}_{b\rightarrow i}\right\} _{b\in\partial i\backslash a}\right),\label{eq:CM_bp_equations_definition}
\end{equation}
where 
\begin{align*}
\widehat{g}_{\mathrm{BP}}\left(\left\{ \nu_{j\rightarrow a}\right\} _{j\in\partial a\backslash i}\right) & =\frac{\sum_{\underline{\sigma}_{\partial a\setminus i}}\psi_{a}\left(\underline{\sigma}_{\partial a}\right)\prod_{j\in\partial a\setminus i}\nu_{j\rightarrow a}\left(\sigma_{j}\right)}{\sum_{\underline{\sigma}_{\partial a}}\psi_{a}\left(\underline{\sigma}_{\partial a}\right)\prod_{j\in\partial a\setminus i}\nu_{j\rightarrow a}\left(\sigma_{j}\right)}\\
g_{\mathrm{BP}}\left(\left\{ \widehat{\nu}_{b\rightarrow i}\right\} _{b\in\partial i\backslash a}\right) & =\frac{\prod_{b\in\partial i\setminus a}\widehat{\nu}_{b\rightarrow i}\left(\sigma_{i}\right)}{\sum_{\sigma_{i}}\prod_{b\in\partial i\setminus a}\widehat{\nu}_{b\rightarrow i}\left(\sigma_{i}\right)}.
\end{align*}
When there is only one relevant solution, the BP marginal for $\sigma_{j}$
is $\nu(\sigma_{j})=\prod_{a\in\partial j}\nu_{a\to j}(\sigma_{j})$.
The set of messages is a proxy for the measure \eqref{eq:CM_markov_random_field}
in the sense that in principle one can ``reconstruct'' the measure
from this set. The Bethe free energy functional which approximates
$\phi(\beta)$ is given by 
\begin{align}
\phi^{\mathrm{Bethe}}\left(\underline{\nu},\underline{\widehat{\nu}}\right)=\frac{1}{N}\biggl\{\sum_{i\in V}\phi_{i}+\sum_{a\in C}\phi_{a}-\sum_{\left(i,a\right)\in E}\phi_{ai}\biggr\}\label{eq:CM_definition_bethe_free_energy}
\end{align}
where 
\begin{align*}
 & \phi_{i}\left(\left\{ \widehat{\nu}_{b\rightarrow i}\right\} _{b\in\partial i}\right)=-\frac{1}{\beta}\ln\sum_{\sigma_{i}}\prod_{b\in\partial i}\widehat{\nu}_{b\rightarrow i}\left(\sigma_{i}\right)\\
 & \phi_{a}\left(\left\{ \nu_{j\rightarrow a}\right\} _{j\in\partial a}\right)=-\frac{1}{\beta}\ln\sum_{\underline{\sigma}_{\partial a}}\psi_{a}\left(\underline{\sigma}_{\partial a}\right)\prod_{j\in\partial a}\nu_{j\rightarrow a}\left(\sigma_{j}\right)\\
 & \phi_{ai}\left(\nu_{i\rightarrow a},\widehat{\nu}_{a\rightarrow i}\right)=-\frac{1}{\beta}\ln\sum_{\sigma_{i}}\nu_{i\rightarrow a}\left(\sigma_{i}\right)\widehat{\nu}_{a\rightarrow i}\left(\sigma_{i}\right).
\end{align*}

As explained before, in the presence of long range correlations this
formalism is too simplistic. The cavity method assumes that: (i) the
Gibbs distribution \eqref{eq:CM_markov_random_field} is a convex
sum of extremal measures; (ii) to leading exponential order, the number
of solutions of the BP equations is equal to the number of extremal
measures; (iii) the free energy of an extremal measure is well approximated
by the Bethe free energy of the BP fixed point. These assumptions
suggest that the Gibbs distribution \eqref{eq:CM_markov_random_field}
is well approximated by the following convex superposition 
\begin{equation}
\mu\left(\underline{\sigma}\right)\approx\frac{1}{Z}\sum_{\left(\underline{\nu},\underline{\widehat{\nu}}\right)\in{\rm BP}}e^{-\beta N\phi^{\mathrm{Bethe}}\left(\underline{\nu},\underline{\widehat{\nu}}\right)}\mu_{\left(\underline{\nu},\underline{\widehat{\nu}}\right)}\left(\underline{\sigma}\right)\label{eq:CM_measure_loopy_graph_convex_sum}
\end{equation}
The measures $\mu_{(\underline{\nu},\underline{\widehat{\nu}})}$
are the ones whose marginals are given by the BP marginals computed
from $(\underline{\nu},\underline{\widehat{\nu}})$. They play the
role of the ``extremal measures''. The sum is over solutions of
the BP equations. In principle one should sum only over stable solutions,
i.e. local minima of the Bethe free energy. However at low temperatures
these are expected to be exponentially more numerous than the other
critical points and it is assumed to be a good approximation to sum
over all BP solutions. The normalization factor yields the partition
function 
\begin{equation}
Z\approx\sum_{\left(\underline{\nu},\underline{\widehat{\nu}}\right)\in{\rm BP}}e^{-\beta N\phi^{\mathrm{Bethe}}\left(\underline{\nu},\underline{\widehat{\nu}}\right)}.\label{eq:partition-fct}
\end{equation}

In order to compute this partition function and uncover the properties
of the convex decomposition \eqref{eq:CM_measure_loopy_graph_convex_sum}
we introduce the level-one statistical mechanical model. The dynamical
variables of this model are the BP messages $\left(\underline{\nu},\underline{\widehat{\nu}}\right)$.
According to \eqref{eq:CM_measure_loopy_graph_convex_sum}, \eqref{eq:partition-fct}
the probability distribution over $(\underline{\nu},\underline{\widehat{\nu}})$
is 
\begin{equation}
\mu_{\mathrm{level-1}}\left(\underline{\nu},\underline{\widehat{\nu}}\right)=\frac{e^{-\beta N\phi^{\mathrm{Bethe}}\left(\underline{\nu},\underline{\widehat{\nu}}\right)}}{Z_{\mathrm{level-1}}}\mathbb{I}\left(\left(\underline{\nu},\underline{\widehat{\nu}}\right)\in{\rm BP}\right),\label{eq:CM_Parisi_model_measure}
\end{equation}
and 
\begin{equation}
Z_{\mathrm{level-1}}=\sum_{\left(\underline{\nu},\underline{\widehat{\nu}}\right)\in{\rm BP}}e^{-\beta N\phi^{\mathrm{Bethe}}\left(\underline{\nu},\underline{\widehat{\nu}}\right)},\label{eq:CM_Parisi_part_function}
\end{equation}
The level-one free energy is defined as usual, 
\begin{equation}
\phi_{{\rm level-1}}(\beta)=-\frac{1}{\beta N}\ln Z_{\mathrm{level-1}}.\label{eq:free-level}
\end{equation}
From \eqref{eq:partition-fct} it should be clear that $\phi(\beta)\approx\phi_{{\rm level-1}}(\beta)$.
The average Bethe free energy, or level-one \textit{internal energy},
is given by 
\begin{equation}
\varphi_{\mathrm{int}}(\beta)=\frac{1}{N}\langle\phi^{{\rm Bethe}}[\underline{\nu},\widehat{\underline{\nu}}]\rangle_{{\rm level-1}}\label{eq:internal}
\end{equation}
Here the bracket denotes the average with respect to \eqref{eq:CM_Parisi_model_measure}.

One also needs to compute the Shannon-Gibbs entropy $\Sigma(\beta)$
of $\mu_{{\rm level-1}}$. An important ``trick'' is to replace
the \textit{explicit} $\beta$ dependence in \eqref{eq:CM_Parisi_model_measure},
\eqref{eq:CM_Parisi_part_function}, \eqref{eq:free-level} by $\beta x$
where $x$ is for the moment an arbitrary parameter%
\footnote{Note that there is also an \textit{implicit} $\beta$ dependence in
$\phi^{{\rm Bethe}}[\underline{\nu},\widehat{\underline{\nu}}]$.%
}. This parameter turns out to play a crucial role and is called the
Parisi parameter. This gives us an $x$-dependent level-one auxiliary
model 
\begin{equation}
\mu_{\mathrm{level-1}}\left(\underline{\nu},\underline{\widehat{\nu}};x\right)=\frac{e^{-\beta xN\phi^{\mathrm{Bethe}}\left(\underline{\nu},\underline{\widehat{\nu}}\right)}}{Z_{\mathrm{level-1}(x)}}\mathbb{I}\left(\left(\underline{\nu},\underline{\widehat{\nu}}\right)\in{\rm BP}\right),\label{eq:CM_Parisi_model_measure_x}
\end{equation}
and 
\begin{equation}
Z_{\mathrm{level-1}}(x)=\sum_{\left(\underline{\nu},\underline{\widehat{\nu}}\right)\in{\rm BP}}e^{-\beta xN\phi^{\mathrm{Bethe}}\left(\underline{\nu},\underline{\widehat{\nu}}\right)},\label{eq:CM_Parisi_part_function_x}
\end{equation}
and also 
\begin{equation}
\phi_{{\rm level-1}}(\beta;x)=-\frac{1}{\beta xN}\ln Z_{\mathrm{level-1}}(x).\label{eq:free-level_x}
\end{equation}
It is then a matter of simple algebra to check that the Shannon-Gibbs
entropy $\Sigma(\beta)$ is given by 
\begin{equation}
\Sigma(\beta)=\Sigma(\beta;x)\equiv\beta x^{2}\frac{\partial}{\partial x}\phi_{\mathrm{level-1}}\left(\beta;x\right)\vert_{x=1},\label{eq:complexityyy}
\end{equation}
and that 
\begin{equation}
\Sigma(\beta)=\beta(\varphi_{\mathrm{int}}(\beta)-\phi_{\mathrm{level-1}}(\beta)).\label{eq:SSS}
\end{equation}
Considering formulas \eqref{eq:CM_Parisi_part_function}, \eqref{eq:internal}
and \eqref{eq:SSS}, it is not hard to argue that $e^{N\Sigma(\beta)}$
is (to leading exponential order) the number of BP solutions with
free energy $\varphi_{\mathrm{int}}(\beta)$ contributing to the sum
\eqref{eq:CM_Parisi_part_function}. The quantity $\Sigma(\beta)$
(a kind of entropy) is called the complexity. It is the growth rate
of the number of extremal measures dominating the convex decomposition
\eqref{eq:CM_measure_loopy_graph_convex_sum}.

We explain later on how to concretely compute $\phi_{{\rm level-1}}(\beta)$,
$\varphi_{\mathrm{int}}(\beta)$ and $\Sigma(\beta)$. Let us immediately
describe how $\Sigma(\beta)$ informs us about the convex decomposition
of the Gibbs distribution. For a large class of problems one finds
that $\Sigma(\beta)=0$ for $\beta<\beta_{d}$, which signals that
only one extremal measure contributes to the Gibbs distribution. At
$\beta_{d}$ the complexity jumps to a non-zero value and then decreases
as a function of $\beta$ till $\beta_{c}$ after which it takes negative
values. In the range $\beta_{d}<\beta<\beta_{c}$ where $\Sigma(\beta)>0$
an exponentially large (with respect to $N$) number of extremal measures
with the same internal free energy $\varphi_{\mathrm{int}}(\beta)$
contribute to the Gibbs distribution. Beyond $\beta_{c}$ one finds
a negative complexity: this is inconsistent with the fact that it
is an entropy. In order to enforce this constraint correctly one is
forced to take the Parisi parameter $0<x<1$ in \eqref{eq:complexityyy}.
More precisely, one sets $x$ to the largest possible value (less
than $1$) such that $\Sigma(\beta)=0$. With this prescription%
\footnote{One can argue that the Parisi parameter is a kind of ``Lagrange multiplier''
that enforces the non-negativity of the complexity in the level-one
model.%
} for the correct value of $x$ when $\beta>\beta_{c}$, one computes
the internal free energy and the free energy and the complexity from
the $x$-dependent level-one model. The complexity is zero by construction
which means that there exist at most a sublinear (believed to be finite)
number of extremal measures contributing to the Gibbs distribution.
This phenomenon is called condensation.

The nature of the thresholds $\beta_{d}$ and $\beta_{c}$ has been
discussed in Sect. \eqref{sub:dyncond} and we do not come back to
this issue here.

\begin{figure}[ptb]
\centering{} \includegraphics[width=3.525in,height=0.8268in]{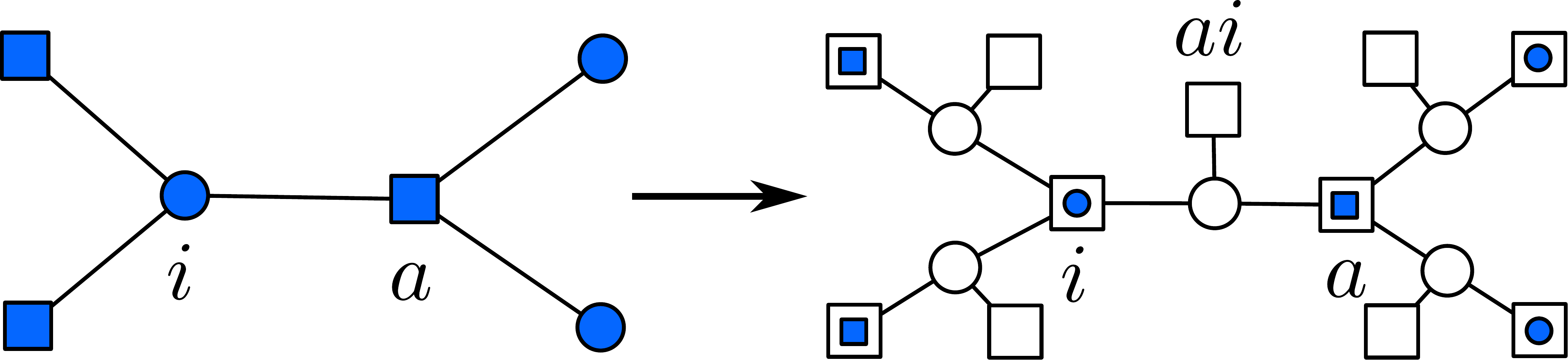}
\protect\caption{\label{fig:CM_factor_graph_parisi_graph_changes} On the left, an
example of an original graph $\Gamma$. On the right its corresponding
graph $\Gamma_{1}$ for the level-one model.}
\end{figure}

We now show how the ($x$-dependent) level-one model is solved in
practice. The main idea is to apply again the BP and Bethe equations
for this model. The first step is to recognize that, if $\Gamma=\left(V,C,E\right)$
is the original factor graph, then the level-one model has the factor
graph $\Gamma_{1}=\left(V_{1},C_{1},E_{1}\right)$ described on Fig.
\ref{fig:CM_factor_graph_parisi_graph_changes}. 

\begin{figure}[ptb]
\centering{}\includegraphics[width=3.2232in,height=0.915in]{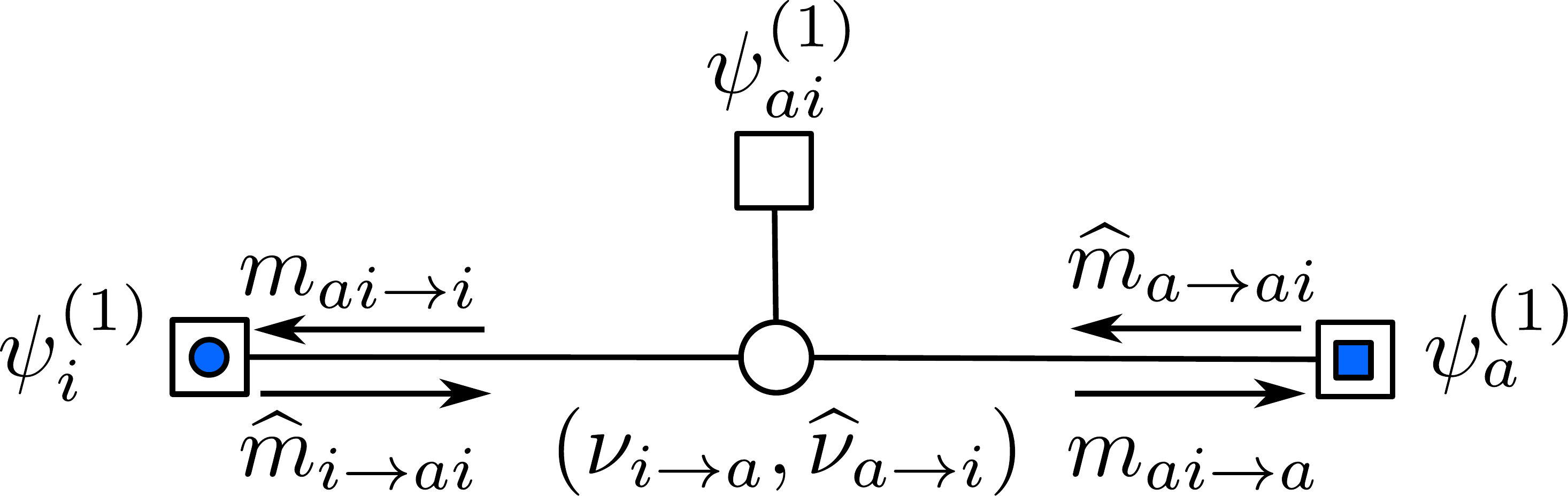}
\protect\caption{\label{fig:CM_Parisi_model_BP_messages} Messages are labeled by $m$
if they are outgoing from a variable node in $V_{1}$ and by $\widehat{m}$
if they are outgoing from a function node in $C_{1}$.}
\end{figure}

A variable node $i\in V$, becomes a function node $i\in C_{1}$,
with the function 
\begin{equation}
\psi_{i}^{(1)}=\prod_{a\in\partial i}\mathbb{I}\left(\nu_{i\rightarrow a}=g_{\mathrm{BP}}\right)e^{-x\beta\phi_{i}}.\label{eq:CM_parisi_factor_i}
\end{equation}
A function node $a\in C$ remains a function node $a\in C_{1}$ with
factor 
\begin{equation}
\psi_{a}^{(1)}=\prod_{i\in\partial a}\mathbb{I}\left(\widehat{\nu}_{a\rightarrow i}=\widehat{g}_{\mathrm{BP}}\right)e^{-x\beta\phi_{a}}.\label{eq:CM_parisi_factor_a}
\end{equation}
An edge $\left(a,i\right)\in E$, becomes a variable node $\left(a,i\right)\in V_{1}$.
The dynamical variables are now couples of distributions $\left(\nu_{a\rightarrow i},\widehat{\nu}_{a\rightarrow i}\right)$.
There is also an extra function node attached to each variable node
of the new graph, or equivalently attached to each edge of the old
graph. The corresponding function is 
\begin{equation}
\psi_{ai}^{(1)}=e^{x\beta\phi_{ai}}.\label{eq:CM_parisi_factor_ai}
\end{equation}
With these definitions, Equ. \eqref{eq:CM_Parisi_model_measure_x}
can be written as 
\begin{equation}
\mu_{\mathrm{level-1}}(\underline{\nu},\underline{\widehat{\nu}};x)=\frac{1}{Z_{{\rm level-1}}(x)}\prod_{i\in V}\psi_{i}^{(1)}\prod_{a\in C}\psi_{i}^{(1)}\prod_{ai\in E}\psi_{ai}^{(1)}.\label{eq:level-1-1}
\end{equation}
For the distributions $\left(\underline{\nu},\underline{\widehat{\nu}}\right)$
that satisfy the BP equations (\ref{eq:CM_bp_equations_definition}),
some algebra leads to the useful formulas 
\begin{align*}
\begin{cases}
e^{-x\beta\left(\phi_{a}-\phi_{ai}\right)} & =\widehat{z}_{a\rightarrow i}^{x}\\
e^{-x\beta\left(\phi_{i}-\phi_{ai}\right)} & =z_{i\rightarrow a}^{x}
\end{cases}
\end{align*}
where 
\begin{align*}
\begin{cases}
z_{i\rightarrow a} & =\sum_{\sigma_{i}}\prod_{b\in\partial i\setminus a}\widehat{\nu}_{b\rightarrow i}\left(\sigma_{i}\right)\\
\widehat{z}_{a\rightarrow i} & =\sum_{\sigma_{\partial a}}\psi_{a}\left(\sigma_{\partial a}\right)\prod_{\partial j\in a\setminus i}\widehat{\nu}_{j\rightarrow a}\left(\sigma_{i}\right)
\end{cases}
\end{align*}
The BP equations for \eqref{eq:level-1-1} involve four kind of messages
as shown on figure \ref{fig:CM_Parisi_model_BP_messages}. 

Messages from a (new) function node to a (new) variable node satisfy
\begin{align*}
 & \widehat{m}_{a\rightarrow ai}\simeq\sum_{\left(\underline{\nu},\underline{\widehat{\nu}}\right)\setminus\left(\nu_{i\rightarrow a},\widehat{\nu}_{a\rightarrow i}\right)}\psi_{a}^{(1)}\prod_{aj\in\partial a\setminus ai}m_{aj\rightarrow a}\\
 & =e^{-x\beta\phi_{ai}}\sum_{\underline{\nu}\setminus\nu_{i\rightarrow a}}\mathbb{I}\left(\widehat{\nu}_{a\rightarrow i}=\widehat{g}_{\mathrm{BP}}\right)\left(\widehat{z}_{a\rightarrow i}\right)^{x}\prod_{aj\in\partial a\setminus ai}m_{aj\rightarrow a}
\end{align*}
and 
\begin{align*}
 & \widehat{m}_{i\rightarrow ai}\simeq\sum_{\left(\underline{\nu},\underline{\widehat{\nu}}\right)\setminus\left(\nu_{i\rightarrow a},\widehat{\nu}_{a\rightarrow i}\right)}\psi_{i}^{(1)}\prod_{bi\in\partial i\setminus ai}\widehat{m}_{bi\rightarrow i}\\
 & =e^{-x\beta\phi_{ai}}\sum_{\underline{\widehat{\nu}}\setminus\widehat{\nu}_{a\rightarrow i}}\mathbb{I}\left(\nu_{i\rightarrow a}=g_{\mathrm{BP}}\right)\left(z_{i\rightarrow a}\right)^{x}\prod_{bi\in\partial i\setminus ai}\widehat{m}_{bi\rightarrow i}.
\end{align*}
where the symbol $\simeq$ means equal up to a normalization factor.
Messages from a (new) function node to a (new) variable node satisfy
\begin{equation}
\begin{cases}
m_{ai\rightarrow i}\simeq e^{x\beta\phi_{ai}}\widehat{m}_{a\rightarrow ai}\\
m_{ai\rightarrow a}\simeq e^{x\beta\phi_{ai}}\widehat{m}_{i\rightarrow ai}.
\end{cases}
\end{equation}
Notice that $m_{ai\rightarrow i}$ (resp. $m_{ai\rightarrow a}$)
becomes independent of $\widehat{\nu}_{a\rightarrow i}$ (resp. $\nu_{i\rightarrow a}$).
This allows us to make a simplification by defining the following
distributions 
\begin{align*}
\begin{cases}
Q_{i\rightarrow a}\left(\nu_{i\rightarrow a}\right) & =m_{ai\rightarrow a}\left(\nu_{i\rightarrow a},\widehat{\nu}_{a\rightarrow i}\right)\\
\widehat{Q}_{a\rightarrow i}\left(\widehat{\nu}_{a\rightarrow i}\right) & =m_{ai\rightarrow i}\left(\nu_{i\rightarrow a},\widehat{\nu}_{a\rightarrow i}\right).
\end{cases}
\end{align*}
Distributions $Q$ and $\widehat{Q}$ are called cavity messages,
and live on the edges of the original factor graph $\Gamma=(V,C,E)$.
From now on we can forget about the factor graph $\Gamma_{1}=(V_{1},C_{1},E_{1})$.
The cavity messages satisfy 
\begin{align}
\widehat{Q}_{a\rightarrow i}\left(\widehat{\nu}_{a\rightarrow i}\right) & \simeq\sum_{\underline{\nu}}\mathbb{I}\left(\widehat{\nu}_{a\rightarrow i}=\widehat{g}_{\mathrm{BP}}\right)\widehat{z}_{a\rightarrow i}^{x}\prod_{j\in\partial a\setminus i}Q_{j\rightarrow a}\left(\nu_{j\rightarrow a}\right)\nonumber \\
Q_{i\rightarrow a}\left(\nu_{i\rightarrow a}\right) & \simeq\sum_{\underline{\widehat{\nu}}}\mathbb{I}\left(\nu_{i\rightarrow a}=g_{\mathrm{BP}}\right)z_{i\rightarrow a}^{x}\prod_{b\in\partial i\setminus a}\widehat{Q}_{b\rightarrow i}\left(\widehat{\nu}_{b\rightarrow i}\right).\label{eq:CM_Cavity_equations}
\end{align}
The Bethe free energy functional of the level-one model can be expressed
as a functional of the cavity messages (one way to determine this
functional is to write down the functional whose critical points are
given by Equ. \eqref{eq:CM_Cavity_equations}). This is an approximation
for the true free energy \eqref{eq:free-level} of the level-one model.
\begin{align}
\phi_{{\rm level-1}}^{{\rm Bethe}}(\underline{Q},\underline{\widehat{Q}};x) & :=\frac{1}{N}\biggl\{\sum_{i\in V}\mathcal{F}_{i}\nonumber \\
 & +\sum_{a\in C}\mathcal{F}_{a}^{{\rm Bethe}}-\sum_{(i,a)\in E}\mathcal{F}_{ai}\biggr\}\label{eq:CM_Parisi_bethe_free_energy}
\end{align}
where 
\begin{align}
 & \mathcal{F}_{i}\left(\left\{ \widehat{Q}_{b\rightarrow i}\right\} _{b\in\partial i}\right)=-\frac{1}{x\beta}\ln\sum_{\underline{\widehat{\nu}}}e^{-x\beta\phi_{i}}\prod_{b\in\partial i}\widehat{Q}_{b\rightarrow i}\nonumber \\
 & \mathcal{F}_{a}\left(\left\{ Q_{j\rightarrow a}\right\} _{j\in\partial a}\right)=-\frac{1}{x\beta}\ln\sum_{\underline{\nu}}e^{-x\beta\phi_{a}}\prod_{j\in\partial a}Q_{j\rightarrow a}\nonumber \\
 & \mathcal{F}_{ai}\left(Q_{i\rightarrow a},\widehat{Q}_{a\rightarrow i}\right)=-\frac{1}{x\beta}\ln\sum_{\nu,\widehat{\nu}}e^{-x\beta\phi_{ai}}Q_{i\rightarrow a}\widehat{Q}_{a\rightarrow i}.
\end{align}

In principle one has to solve the cavity equations \eqref{eq:CM_Cavity_equations}
for $0<x\leq1$, and compute the $x$-dependent free energy $\phi_{\mathrm{level-1}}^{\mathrm{Bethe}}$.
From this free energy we obtain the complexity by computing the derivative
in equation \eqref{eq:complexityyy}. This allows to determine the
thresholds $\beta_{d}$ and $\beta_{c}$. For $\beta<\beta_{c}$ the
free energy is given by $\phi_{\mathrm{level-1}}^{\mathrm{Bethe}}\vert_{x=1}$.
This function has no singularities, which means that there are no
static (thermodynamic) phase transitions for $\beta<\beta_{c}$. In
this phase one has $\Sigma(\beta;x=1)\geq0$. For $\beta>\beta_{c}$
one enforces a zero complexity by setting the Parisi parameter to
a value $0<x_{*}<1$ s.t. $\Sigma(\beta;x_{*})=0$. The free energy
is not analytic at $\beta_{c}$, due to the change of $x$ parameter.
This a static phase transition threshold.

In practice, as long as we are interested only in the range $\beta<\beta_{c}$
we can set $x=1$. It is then possible to simplify the cavity equations
\eqref{eq:CM_Cavity_equations} and the level-1 free energy \eqref{eq:CM_Parisi_bethe_free_energy}.
In the next appendix we perform these simplifications for the case
at hand.

\section{Application of the cavity equations to the lossy source coding\label{sec:application}}

We apply the formalism of appendix \ref{sec:cavityprimer} to the
measure $\mu_{\beta}(\underline{u}|\underline{x})$ (see Equ.\eqref{eq:measure-mu}).
Instead of working with the alphabet $\{0,1\}$, we find it convenient
to use the mapping $\sigma_{i}=(-1)^{u_{i}}$ and $J_{a}=(-1)^{x_{a}}$
to the alphabet $\{-1,+1\}$. The measure \eqref{eq:measure-mu} is
of the form \eqref{eq:CM_markov_random_field} with 
\begin{equation}
\psi_{a}(\{\sigma_{i},i\in\partial a\})=e^{-\beta(1-J_{a}\prod_{i\in\partial a}\sigma_{i})}.
\end{equation}

The probability distributions $\nu_{i\to a}(\sigma_{i})$ and $\widehat{\nu}_{a\to i}(\sigma_{i})$
are entirely characterized by their means, $\tanh\beta\eta_{i\to a}$
and $\tanh\beta\widehat{\eta}_{a\to i}$, as follows (we drop the
subscripts) 
\begin{equation}\label{paramappendixB}
\nu(\sigma)=\frac{1+\sigma\tanh\beta\eta}{2}.
\end{equation}
With this parameterization, the BP equations \eqref{eq:CM_bp_equations_definition}
for the model \eqref{eq:measure-mu} become 
\begin{align}
\begin{cases}
\widehat{\eta}_{a\rightarrow i} & =\widehat{g}_{\mathrm{BP}}(\{\eta_{j\rightarrow a}\}_{j\in\partial a\backslash i}\mid J_{a})\\
\eta_{i\rightarrow a} & =g_{\mathrm{BP}}(\{\widehat{\eta}_{b\rightarrow i}\}_{b\in\partial i\backslash a}),
\end{cases}\label{eq:CM_application_BP_equations_eta}
\end{align}
where
\begin{eqnarray}
\widehat{g}_{\mathrm{BP}}\left(\left\{ \eta_{j\rightarrow a}\right\} _{j\in\partial a\setminus i}\mid J_{a}\right) & = & \frac{J_{a}}{\beta}\atanh(\tanh\beta\nonumber \\
 &  & \times\prod_{j\in\partial a\setminus i}\tanh\beta\eta_{j\rightarrow a})
\end{eqnarray}
 and 
\begin{align}
g_{\mathrm{BP}}(\{\widehat{\eta}_{b\rightarrow i}\}_{b\in\partial i\backslash a})=\sum_{b\in\partial i\backslash a}\widehat{\eta}_{bi}.
\end{align}

The Bethe free energy per variables \eqref{eq:CM_definition_bethe_free_energy}
reads 
\begin{align}
\phi^{\mathrm{Bethe}}\left(\underline{\eta},\widehat{\underline{\eta}}\right) & =-\beta^{-1}(\ln\left(1+e^{-2\beta}\right)+(R-1)\ln2)\nonumber \\
 & -\frac{1}{\beta N}\sum_{a\in C}\left(1-\left\vert \partial a\right\vert \right)\ln Z_{1}(\{\eta_{j\rightarrow a}\}_{j\in\partial a}\mid J_{a})\nonumber \\
 & -\frac{R}{\beta M}\sum_{i\in V}\ln Z_{2}(\{\widehat{\eta}_{b\rightarrow i}\}_{b\in\partial i}),\label{eq:CM_application_bethe_free_energy}
\end{align}
where 
\begin{align*}
\begin{cases}
Z_{1}(\{\eta_{j\rightarrow a}\}_{j\in\partial a}\mid J_{a})=1+J_{a}(\tanh\beta)\prod_{i\in\partial a}\tanh\beta\eta_{i\rightarrow a}\\
Z_{2}(\{\widehat{\eta}_{a\rightarrow i}\}_{a\in\partial i})=\frac{1}{2}\sum_{s\in\{-1,1\}}\prod_{a\in\partial i}(1+s\tanh\beta\widehat{\eta}_{a\rightarrow i}).
\end{cases}
\end{align*}

Since we have parameterized the BP messages by real numbers, the cavity
messages $Q_{i\to a}$, $\widehat{Q}_{a\to i}$ become distributions
on $\eta_{i\to a}$, $\widehat{\eta}_{a\to i}$. The cavity equations
\eqref{eq:CM_Cavity_equations} reduce to 
\begin{align}
 & Q_{i\rightarrow a}(\eta_{i\rightarrow a})\simeq\int\prod_{b\in\partial i\backslash a}d\widehat{\eta}_{b\rightarrow i}\widehat{Q}_{b\rightarrow i}(\widehat{\eta}_{b\rightarrow i})\nonumber \\
 & \times Z_{2}^{x}(\{\widehat{\eta}_{b\rightarrow i}\}_{b\in\partial i\setminus a})\delta\biggl(\eta_{i\rightarrow a}-g_{\mathrm{BP}}(\{\widehat{\eta}_{b\rightarrow i}\}_{b\in\partial i\backslash a})\biggr)\label{eq:CM_application_cavity_equations1}
\end{align}
and 
\begin{align}
 & \widehat{Q}_{a\rightarrow i}(\widehat{\eta}_{a\rightarrow i})\simeq\int\prod_{j\in\partial a\backslash i}d\eta_{j\rightarrow a}Q_{j\rightarrow a}(\eta_{j\rightarrow a})\nonumber \\
 & \times Z_{1}^{x}(\{\eta_{j\rightarrow a}\}_{b\in\partial i\setminus a})\delta\biggl(\widehat{\eta}_{a\rightarrow i}-\widehat{g}_{\mathrm{BP}}(\{\eta_{j\rightarrow a}\}_{j\in\partial a\backslash i}\mid J_{a})\biggr).\label{eq:CM_application_cavity_equations2}
\end{align}
For the Bethe free energy of the level-one model one finds 
\begin{align}
 & \phi_{\mathrm{level-1}}^{\mathrm{Bethe}}\left(\underline{\eta},\widehat{\underline{\eta}};x\right)=-\beta^{-1}(\ln(1+e^{-2\beta})+(R-1)\ln2)\nonumber \\
 & -\frac{1}{\beta xN}\sum_{a\in C}(1-\vert\partial a\vert)\ln\biggl\{\int\prod_{i\in\partial a}d\eta_{i\rightarrow a}Q_{i\rightarrow a}(\eta_{i\rightarrow a})\nonumber \\
 & \,\,\,\,\,\,\,\,\,\times Z_{1}^{x}(\{\eta_{i\rightarrow a}\}_{i\in\partial a}\mid J_{a})\biggr\}\nonumber \\
 & -\frac{R}{\beta xM}\sum_{i\in V}\ln\biggl\{\int\prod_{a\in\partial i}d\widehat{\eta}_{a\rightarrow i}\widehat{Q}_{a\rightarrow i}(\widehat{\eta}_{a\rightarrow i})\nonumber \\
 & \,\,\,\,\,\,\,\,\,\times Z_{2}^{x}(\{\widehat{\eta}_{a\rightarrow i}\}_{a\in\partial i})\biggr\}.\label{eq:CM_application_Bethe_Parisi_free_energy_general_x}
\end{align}

We are interested in the range $\beta<\beta_{c}$ for which the Parisi
parameter is set to $x=1$. In this case the above equations greatly
simplify. We first define average cavity messages 
\begin{align}
\begin{cases}
h_{i\rightarrow a} & ={\rm Av}[Q_{i\rightarrow a}]\\
\widehat{h}_{a\rightarrow i} & ={\rm Av}[\widehat{Q}_{a\rightarrow i}],
\end{cases}
\end{align}
where the functional ${\rm Av}[P]$ is 
\begin{equation}
{\rm Av}[P]=\frac{1}{\beta}\atanh\biggl\{\int d\eta P(\eta)\tanh\beta\eta\biggr\}.\label{eq:CM_H_transform}
\end{equation}
Thus $\tanh\beta h_{i\rightarrow a}$ and $\tanh\beta\widehat{h}_{a\to i}$
are real valued messages and are averages of $\tanh\beta\eta_{i\to a}$
and $\tanh\beta\widehat{\eta}_{a\to i}$ with respect to the cavity
distributions $Q_{i\to a}(\eta_{i\to a})$ and $\widehat{Q}_{a\to i}(\widehat{\eta}_{a\to i})$
respectively. The free energy of the level-one model for $x=1$ can
be expressed in terms of these real valued messages, and one finds
\begin{align}
 & \phi_{\mathrm{level-1}}^{\mathrm{Bethe}}\left(\underline{h},\widehat{\underline{h}}\right)=-\beta^{-1}(\ln(1+e^{-2\beta})+(R-1)\ln2)\nonumber \\
 & -\frac{1}{\beta N}\sum_{a\in C}(1-\vert\partial a\vert)\ln Z_{1}(\{h_{j\rightarrow a}\}_{j\in\partial a}\mid J_{a})\nonumber \\
 & -\frac{R}{\beta M}\sum_{i\in V}\ln Z_{2}(\{\widehat{h}_{b\rightarrow i}\}_{b\in\partial i}).\label{eq:CM_application_Bethe_Parisi_free_energy}
\end{align}
Remarkably, is the same than the original Bethe free energy functional
$\phi^{\mathrm{Bethe}}\left(\underline{\eta},\widehat{\underline{\eta}}\right)$
defined in \eqref{eq:CM_application_bethe_free_energy}, but now evaluated
for the average fields $h_{i\to a}$ and $\widehat{h}_{a\to i}$.
From the cavity equations \eqref{eq:CM_application_cavity_equations1}-\eqref{eq:CM_application_cavity_equations2}
for $x=1$, one can deduce that the average fields $h_{i\to a}$ and
$\widehat{h}_{a\to i}$ satisfy 
\begin{align}
\begin{cases}
\widehat{h}_{a\rightarrow i} & =\widehat{g}_{\mathrm{BP}}(\{h_{j\rightarrow a}\}_{j\in\partial a\backslash i}\mid J_{a})\\
h_{i\rightarrow a} & =g_{\mathrm{BP}}(\{\widehat{h}_{b\rightarrow i}\}_{b\in\partial i\backslash a}).
\end{cases}\label{eq:bphfields}
\end{align}
Thus the average fields satisfy the BP\ equations \eqref{eq:CM_application_BP_equations_eta}.

To summarize, when $x=1$, $\phi_{\mathrm{level-1}}^{\mathrm{Bethe}}$
equals $\phi^{\mathrm{Bethe}}$ computed at a certain appropriate
BP fixed point. This fixed point corresponds to messages $\tanh\beta h_{i\to a}$,
$\tanh\beta\widehat{h}_{a\to i}$ which are an average of the BP solutions
$\tanh\beta\eta_{i\to a}$, $\tanh\beta\widehat{\eta}_{a\to i}$ over
the cavity distributions $Q_{i\to a}(\eta_{i\to a})$ and $\widehat{Q}_{a\to i}(\widehat{\eta}_{a\to i})$.
The messages $\tanh\beta\eta_{i\to a}$, $\tanh\beta\widehat{\eta}_{a\to i}$
describe the ``extremal states'' whereas the messages $\tanh\beta h_{i\to a}$,
$\tanh\beta\widehat{h}_{a\to i}$ describe their convex superposition.

\section{Density evolution for the cavity equations of lossy source coding\label{sec:CM_population_dynamic}}

The discussion in appendices \ref{sec:cavityprimer} and \ref{sec:application}
is valid for a single instance. It is expected that the free energy,
internal free energy and complexity concentrate on their ensemble
average, and in practice one computes their ensemble average. The
ensemble average is performed over the graph ensemble and the Bernoulli
source. In the present context this leads to the complicated set of
fixed point equations \eqref{eq:DE_h_updates}-\eqref{eq:DE_hhat_etahat_updates}
that links six densities.

To perform the ensemble average we assume that the cavity messages
$Q_{i\rightarrow a}(\eta_{i\rightarrow a})$ and $\widehat{Q}_{a\rightarrow i}(\widehat{\eta}_{a\rightarrow i})$
can be considered as i.i.d. realizations of random variables $Q_{z}(\eta)$
and $\widehat{Q}_{z}(\widehat{\eta})$. The random variables depend
only on the \textit{position} $z$ along the spatial dimension and
not on the \textit{direction} of the edges $i\to a$ and $a\to i$.
The distributions of these random variables are denoted $\mathcal{Q}_{z}$
and $\widehat{\mathcal{Q}}_{z}$. Note that the cavity messages are
already distributions over real numbers, so that $\mathcal{Q}_{z}$
and $\widehat{\mathcal{Q}}_{z}$ are distributions of distributions.
From the cavity equations \eqref{eq:CM_application_cavity_equations1},
\eqref{eq:CM_application_cavity_equations2} it is easy to formally
write down the set of integral equations that these distributions
of distributions satisfy.

We can write down probability distributions for the average fields
$h_{i\to a}$ and $h_{a\to i}$, 
\begin{align}\label{avhhat}
\begin{cases}
q_{z}(h) & =\int\mathcal{D}\mathcal{Q}_{z}[Q]\delta(h-Av[Q])\\
\widehat{q}_{z}(\widehat{h}) & =\int\mathcal{D}\widehat{\mathcal{Q}}_{z}[\widehat{Q}]\delta(\widehat{h}-Av[\widehat{Q}]).
\end{cases}
\end{align}
With the independence assumption on the cavity messages, relations
\eqref{eq:bphfields} imply that these distributions satisfy \eqref{eq:DE_h_updates}
and \eqref{eq:DE_hhat_updates}. Furthermore from \eqref{eq:CM_application_Bethe_Parisi_free_energy}
we deduce formula \eqref{eq:f_RSB} for the average level-one free
energy.

We define the conditional distributions $q_{z}(\eta|h)$ and $\widehat{q}_{z}(\widehat{\eta}|\widehat{h})$
\begin{align}\label{qsigmas}
\begin{cases}
q_{z}(\eta|h)q_{z}(h) & =\int\mathcal{D}\mathcal{Q}_{z}[Q]Q(\eta)\delta(h-Av[Q])\\
\widehat{q}_{z}(\widehat{\eta}|\widehat{h})\widehat{q}_{z}(\widehat{h}) & 
=\int\mathcal{D}\widehat{\mathcal{Q}}_{z}[\widehat{Q}]\widehat{Q}(\widehat{\eta})\delta(\widehat{h}-Av[\widehat{Q}]),
\end{cases}
\end{align}
and for $\sigma=\pm1$, 
\begin{align}
\begin{cases}
q_{z}^{\sigma}(\eta|h) & =\frac{1+\sigma\tanh\beta\eta}{1+\sigma\tanh\beta h}q_{z}(\eta|h)\\
\widehat{q}_{z}^{\sigma}(\widehat{\eta}|\widehat{h}) & =\frac{1+\sigma\tanh\beta\widehat{\eta}}{1+\sigma\tanh\beta\widehat{h}}q_{z}(\widehat{\eta}|\widehat{h}).
\end{cases}\label{eq:qz_definition}
\end{align}
These distributions satisfy \eqref{eq:DE_h_eta_updates}-\eqref{eq:DE_hhat_etahat_updates}.

With the six distributions $q_{z}(h)$, $\widehat{q}_{z}(\widehat{h})$,
$q_{z}^{\sigma=\pm1}(\eta|h)$ and $\widehat{q}_{z}^{\sigma=\pm1}(\widehat{\eta}|\widehat{h})$
we can compute the complexity. We use (see \eqref{eq:SSS}) 
\begin{equation}
\Sigma(\beta)=\beta(\varphi_{{\rm int}}^{{\rm Bethe}}(\beta)-\phi_{{\rm level-1}}^{{\rm Bethe}}(\beta)).
\end{equation}
Since we already know that $\phi_{{\rm level-1}}^{{\rm Bethe}}(\beta)$
is given by \eqref{eq:f_RSB}, it remains to compute the internal
free energy in the Bethe approximation. For this purpose we use 
\begin{equation}
\varphi_{{\rm int}}^{{\rm Bethe}}(\beta)=\frac{\partial}{\partial x}(x\phi_{{\rm level-1}}^{{\rm Bethe}}(\beta;x))\vert_{x=1}.
\end{equation}
We compute the $x$-derivative on \eqref{eq:CM_application_Bethe_Parisi_free_energy_general_x},
and average over the cavity distributions, the graph ensemble and
the Bernoulli source. After some algebra one finds that $\varphi_{{\rm int}}^{{\rm Bethe}}(\beta)$
is given by \eqref{eq:DE_phi_rsb}.

\section{Proof of Theorem \ref{thm:h_converge}\label{sec:proof_h_converge}}

We first state two useful lemmas 
\begin{lem}
\label{lem:poisson} Let the random variable $X$ is distributed according
to a Poisson distribution with mean $\lambda$. 
\begin{align*}
 & \mathbb{P}(X<\frac{\lambda t}{2})<\exp(-\frac{\lambda t}{10}),\,\,\,\,\, t\leq1,\\
 & \mathbb{P}(X>\frac{3\lambda t}{2})<\exp(-\frac{\lambda t}{10}),\,\,\,\,\, t\geq1.
\end{align*}
\end{lem}
\begin{IEEEproof}
Use the Chernoff bound. \end{IEEEproof}
\begin{lem}
\label{lem:lem2} Let 
\begin{align*}
\epsilon_{1} & =\beta\frac{3l}{2R}(\tanh\epsilon_{0})^{(lR^{3})^{1/4}}\\
\delta_{1} & =\exp(-\frac{l}{10R})+\frac{l}{R}\exp\bigl(-\frac{\epsilon_{0}\sqrt{Rl}}{\beta\sqrt{3\pi}}\bigr).
\end{align*}
with $\epsilon_{0}=\min(1/2,\beta/2)$. Consider the recursions for
$t\geq1$ 
\begin{align*}
\epsilon_{t+1} & =(t+1)\beta\frac{3l}{2R}(\tanh\epsilon_{t})^{(lR^{3})^{1/4}},\\
\delta_{t+1} & =\exp\bigl(-\frac{l}{10R}(t+1)\bigr)+\frac{l}{R}(2\sqrt{\delta_{t}})^{l-1}.
\end{align*}
There exist an integer $l_{0}$ (depending only on $R$ and $\beta$)
such that for $l\geq l_{0}$, 
\begin{itemize}
\item i) $\epsilon_{t}\leq\frac{1}{2^{t+1}}$ for $t\geq0$. 
\item ii) $\delta_{t}<2\exp(-\frac{l}{5R}t)$ for $t\geq2$. 
\end{itemize}
\end{lem}
\begin{IEEEproof}
Consider (i). At $t=0$, $\epsilon_{0}\leq1/2$. Assume that $\epsilon_{t-1}\leq\frac{1}{2^{t}}$
for $t\geq1$, then 
\begin{align*}
\epsilon_{t} & =t\beta\frac{3l}{2R}(\tanh\epsilon_{t-1})^{(lR^{3})^{1/4}}\leq t\beta\frac{3l}{2R}(\epsilon_{t-1})^{(lR^{3})^{1/4}}\\
 & \leq t\beta\frac{3l}{2R}(\frac{1}{2^{t}})^{(lR^{3})^{1/4}}=\frac{t\beta\frac{3l}{R}}{2^{t((lR^{3})^{1/4}-1)}}\times\frac{1}{2^{t+1}}.
\end{align*}
The proof is complete if $t\beta\frac{3l}{R}<2^{t(\sqrt[4]{lR^{3}}-1)}$
for $t\geq1$. It is clear that this is true for $l$ large enough.
\end{IEEEproof}
Now consider (ii). Clearly for $l$ large enough such that 
\[
\delta_{2}=\exp\left(-\frac{l}{5R}\right)+\frac{l}{R}(2\sqrt{\delta_{1}})^{l-1}\leq2\exp\left(-\frac{l}{5R}\right).
\]
To complete the proof by induction, we remark that $\delta_{t}<2\exp\left(-\frac{l}{5R}t)\right)<1$
implies 
\begin{align*}
\frac{l}{R}(2\sqrt{\delta_{t}})^{l-1}<\exp\left(-\frac{l}{5R}(t+1)\right)
\end{align*}
for $l$ large enough independent of $t$. 

We now turn to the proof of Theorem \ref{thm:h_converge}. It is organized
in three steps: 
\begin{itemize}
\item 1) We first show that for any small $\delta_{1}$ and $\epsilon_{1}$,
one can find an integer $l_{1}$ such that for $l\geq l_{1}$ 
\begin{align*}
p_{1}\equiv\mathbb{P}\left\{ \vert h^{(1)}\vert\leq\frac{\epsilon_{1}}{\beta}\right\} \geq1-\delta_{1}.
\end{align*}

\item 2) We then show by induction on $t\geq1$ that 
\[
p_{t}\equiv\mathbb{P}\left\{ \vert h^{(t)}\vert<\frac{\epsilon_{t}}{\beta}\right\} \geq1-\delta_{t}.
\]

\item 3) Finally using Lemma \ref{lem:lem2} we deduce that $h^{(t)}\to0$
almost surely as $t\to+\infty$. \end{itemize}
\begin{IEEEproof}
{[}Proof of theorem \ref{thm:h_converge}{]}
\end{IEEEproof}
We begin by noting that regardless of the initial distribution, $\widehat{q}^{(t)}(\widehat{h})$
has a symmetric density due to the symmetric distribution of $J$.
Moreover, $\left|\widehat{h}^{(t)}\right|\leq1$ from \eqref{eq:DE_hhat_updates}.
Thus, $\mathbb{E}_{\widehat{q}^{(t)}}(\widehat{h}^{(t)})=0$ and ${\rm Var}(\widehat{h}^{(t)})=\mathbb{E}_{\widehat{q}^{(t)}}(\widehat{h}^{2})\leq1$.

\noindent \textbf{Step 1:} We set $P(r)=e^{-\lambda}\frac{\lambda^{r}}{r!}$
and $\lambda=l/R$. Let $h^{(r,t)}=\sum_{a=1}^{r}\widehat{h}_{a}^{(t)}$
where $\widehat{h}_{a}^{(t)}$ are i.i.d random variables with probability
density $\widehat{q}^{(t)}(\widehat{h})$. Let $\sigma_{0}^{2}=\mathbb{E}((\widehat{h}_{a}^{(0)})^{2})\leq1$.
According to \cite[Theorem 3.5.3]{durrett10} we have 
\[
\lim_{r\to\infty}\sqrt{r}\mathbb{P}\left\{ \vert h^{(r,0)}\vert<\frac{\epsilon_{0}}{\beta}\right\} =\frac{2\epsilon_{0}}{\beta\sqrt{2\pi\sigma_{0}^{2}}},
\]
for any $\epsilon_{0}>0$. Thus, there exists $r'(\epsilon_{0},\beta)\in\mathbb{N}$
such that for $r>r'$, 
\[
\mathbb{P}\left\{ \vert h^{(r,0)}\vert<\frac{\epsilon_{0}}{\beta}\right\} \geq\frac{\epsilon_{0}}{\beta\sqrt{2\pi r}}.
\]
Take $l$ such that $\lambda=l/R\geq l'/R=2r'$, then 
\begin{align*}
p_{0} & =\mathbb{P}\left\{ \vert h^{(0)}\vert<\frac{\epsilon_{0}}{\beta}\right\} \\
 & =\sum_{r=0}^{\infty}P(r)\mathbb{P}\left\{ \vert h^{(r,0)}\vert<\frac{\epsilon_{0}}{\beta}\right\} \\
 & \geq\sum_{r=\lambda/2}^{3\lambda/2}P(r)\mathbb{P}\left\{ \vert h^{(r,0)}\vert<\frac{\epsilon_{0}}{\beta}\right\} \\
 & \geq\frac{\epsilon_{0}}{\beta\sqrt{3\pi\lambda}}\sum_{r=\lambda/2}^{3\lambda/2}P(r)\\
 & >\frac{\epsilon_{0}}{\beta\sqrt{3\pi\lambda}}(1-2e^{-\frac{\lambda}{10}}).
\end{align*}
The last inequality follows from lemma \ref{lem:poisson}. Thus for
$l$ large enough 
\begin{equation}
p_{0}=\mathbb{P}\left\{ \vert h^{(0)}\vert<\frac{\epsilon_{0}}{\beta}\right\} >\frac{\epsilon_{0}}{2\beta\sqrt{3\pi\lambda}}\equiv1-\delta_{0}.\label{eq:p_0bound}
\end{equation}

Recall $\widehat{h}^{(t+1)}=\frac{1}{\beta}\tanh^{-1}\bigl(J\tanh\beta\prod_{i=1}^{l-1}\tanh\beta h_{i}^{(t)}\bigr)$.
From $\tanh^{-1}\bigl(a\tanh\beta\bigr)\leq a\beta$ for $0<a<1$,
we have 
\[
\left|\widehat{h}^{(t+1)}\right|\leq\prod_{i=1}^{l-1}\tanh\left|\beta h_{i}^{(t)}\right|.
\]
Define 
\[
Z_{l}^{(t)}\equiv\ln\left(\prod_{i=1}^{l-1}\tanh\left|\beta h_{i}^{(t)}\right|\right)=\sum_{i=1}^{l-1}\ln\left(\tanh\left|\beta h_{i}^{(t)}\right|\right).
\]
Note that $Z_{l}^{(t)}$ is always negative and if one of $h_{i}^{(t)}$
tends to zero, it diverges to $-\infty$. Consider $t=0$. We will
show that $Z_{l}^{(0)}$ has a large negative value with high probability.
Define 
\[
u_{i}\equiv\begin{cases}
u_{i-1}, & \text{if }\left|h_{i-1}^{(0)}\right|>\frac{\epsilon_{0}}{\beta},\\
u_{i-1}+\ln\tanh\epsilon_{0}, & \text{otherwise},
\end{cases}
\]
with $u_{0}=0$. One can check for later use that $Z_{l}^{(0)}\leq u_{l}$.
Moreover, because of \eqref{eq:p_0bound} one can consider $u_{l}$
as a random walk (with negative jumps), 
\[
u_{i}=\begin{cases}
u_{i-1}, & \text{with prob. }1-p_{0}\\
u_{i-1}+\ln\tanh\epsilon_{0}, & \text{with prob. }p_{0}.
\end{cases}
\]
Let $s=\ln\left(\tanh(\epsilon_{0})\right)$. Using the Chernoff's
theorem~\cite[Page 151]{billingsley95}, 
\begin{align*}
\mathbb{P}\left\{ \frac{1}{l-1}\frac{u_{l}}{s}<\lambda^{-3/4}\right\} <\exp\left(-(l-1)D(\lambda^{-3/4}||p_{0})\right),
\end{align*}
where $D(x||y)=x\ln(\frac{x}{y})+(1-x)\ln(\frac{1-x}{1-y})$. Now,
since 
\begin{align*}
 & x\ln(\frac{x}{p_{0}})>x\ln(x),\\
 & (1-x)\ln\left(\frac{1-x}{1-p_{0}}\right)>(1-x)\ln\left(\frac{1-x}{\delta_{0}}\right),
\end{align*}
we have 
\begin{align}
D(\lambda^{-3/4}||p_{0})>-H_{2}(\lambda^{-3/4})\ln(2)-(1-\lambda^{-3/4})\ln\left(\delta_{0}\right),\label{eq:D_lowerbound}
\end{align}
for $\delta_{0}$ defined in \eqref{eq:p_0bound}. By a large $\lambda$
expansion of the right hand side of \eqref{eq:D_lowerbound}: 
\begin{align*}
-H_{2}(\lambda^{-3/4})\ln2- & (1-\lambda^{-3/4})\ln\delta_{0}\\
 & =\frac{\epsilon_{0}}{2\beta\sqrt{3\pi\lambda}}+o(\frac{1}{\sqrt{\lambda}}).
\end{align*}
Thus, there exists $l''\in\mathbb{N}$ depending on $R,\beta$ and
$\epsilon_{0}$ such that for $l>l''$, 
\begin{align}
\mathbb{P}\left\{ \frac{1}{l-1}\frac{u_{l}}{s}<\lambda^{-3/4}\right\} <\exp\left(-\frac{\epsilon_{0}(l-1)}{4\beta\sqrt{3\pi\lambda}}\right).
\end{align}
By replacing $s=\ln\tanh\epsilon_{0}$ and $\lambda=\frac{l}{R}\approx\frac{l-1}{R}$
for large degrees, 
\begin{align*}
\mathbb{P}\left\{ u_{l}>(lR^{3})^{1/4}\ln\tanh\epsilon_{0})\right\} <\exp\left(-\frac{\epsilon_{0}\sqrt{Rl}}{4\beta\sqrt{3\pi}}\right),
\end{align*}
Note that the inequality in $\mathbb{P}(\dots)$ is reversed since
$s<0$. Now recall $Z_{l}^{(0)}\leq u_{l}$. Therefore, 
\begin{align*}
 & \mathbb{P}\left\{ Z_{l}^{(0)}\leq(lR^{3})^{1/4}\ln\tanh\epsilon_{0}\right\} \\
 & \geq\mathbb{P}\left\{ u_{l}\leq(lR^{3})^{1/4}\ln\tanh\epsilon_{0}\right\} \geq1-\exp\left(-\frac{\epsilon_{0}\sqrt{Rl}}{4\beta\sqrt{3\pi}}\right).
\end{align*}
Consequently, 
\begin{align*}
 & \mathbb{P}\left\{ \left|\widehat{h}^{(1)}\right|\leq\left(\tanh\epsilon_{0}\right)^{(lR^{3})^{1/4}}\right\} \\
 & \geq\mathbb{P}\left\{ Z_{l}^{(0)}\leq(lR^{3})^{1/4}\ln\tanh\epsilon_{0}\right\} \geq1-\exp\left(-\frac{\epsilon_{0}\sqrt{Rl}}{4\beta\sqrt{3\pi}}\right).
\end{align*}

From $r$, $\left|h^{(r,1)}\right|=\left|\sum_{a=1}^{r}\widehat{h}_{a}^{(1)}\right|\leq\sum_{a=1}^{r}\left|\widehat{h}_{a}^{(1)}\right|$.
we deduce 
\begin{align*}
 & \mathbb{P}\left\{ \left|h^{(r,1)}\right|\leq r(\tanh\epsilon_{0})^{(lR^{3})^{1/4}}\right\} \\
 & \geq\mathbb{P}\left\{ \left|\widehat{h}^{(1)}\right|\leq(\tanh\epsilon_{0})^{(lR^{3})^{1/4}}\right\} ^{r}\\
 & \geq\left\{ 1-\exp\left(-\frac{\epsilon_{0}\sqrt{Rl}}{4\beta\sqrt{3\pi}}\right)\right\} ^{r}\\
 & \geq1-r\exp\left(-\frac{\epsilon_{0}\sqrt{Rl}}{4\beta\sqrt{3\pi}}\right).
\end{align*}
for $l$ large enough. Therefore we have, 
\begin{align*}
 & \mathbb{P}\left\{ \left|h^{(1)}\right|\leq\frac{3}{2}\lambda(\tanh\epsilon_{0})^{(lR^{3})^{1/4}}\right\} \\
 & =\sum_{r=0}^{\infty}P(r)\mathbb{P}\left\{ \left|h^{(r,1)}\right|\leq\frac{3}{2}\lambda(\tanh\epsilon_{0})^{(lR^{3})^{1/4}}\right\} \\
 & \geq\sum_{r=0}^{3\lambda/2}P(r)\mathbb{P}\left\{ \left|h^{(r,1)}\right|\leq\frac{3}{2}\lambda(\tanh\epsilon_{0})^{(lR^{3})^{1/4}}\right\} \\
 & \geq\sum_{r=0}^{3\lambda/2}P(r)\mathbb{P}\left\{ \left|h^{(r,1)}\right|\leq r(\tanh\epsilon_{0})^{(lR^{3})^{1/4}}\right\} \\
 & \geq\sum_{r=0}^{3\lambda/2}P(r)\left(1-r\exp\left(-\frac{\epsilon_{0}\sqrt{Rl}}{4\beta\sqrt{3\pi}}\right)\right)\\
 & \geq1-\exp(-0.1\lambda)-\lambda\exp\left(-\frac{\epsilon_{0}\sqrt{Rl}}{4\beta\sqrt{3\pi}}\right).
\end{align*}
To summarize, we have obtained 
\begin{align}
p_{1}=\mathbb{P}\left\{ \left|h^{(1)}\right|\leq\frac{\epsilon_{1}}{\beta}\right\} \geq1-\delta_{1}.\label{eq:final}
\end{align}
This completes step 1.

\noindent \textbf{Step 2:} The proof is by induction. Assume that
\[
p_{t}=\mathbb{P}\left\{ \vert h^{(t)}\vert\leq\frac{\epsilon_{t}}{\beta}\right\} \geq1-\delta_{t}.
\]
We prove that this holds also for $t+1$. This mainly consists in
repeating the derivations \eqref{eq:p_0bound} to \eqref{eq:final}
for $p_{t}$, $\epsilon_{t}$ and $\delta_{t}$. We briefly repeat
them here: 
\begin{align*}
 & \mathbb{P}\left\{ \left|\widehat{h}^{(t+1)}\right|\leq\left(\tanh\epsilon_{t}\right)^{(lR^{3})^{1/4}}\right\} \\
 & \geq\mathbb{P}\left\{ Z_{l}^{(t)}\leq(lR^{3})^{1/4}\ln\left(\tanh\epsilon_{t}\right)\right\} \\
 & \geq1-\exp\left(-(l-1)D(\lambda^{-3/4}||p_{t})\right).
\end{align*}
Assume that $\delta_{t}\ll1$. From \eqref{eq:D_lowerbound}, 
\begin{align*}
D(\lambda^{-3/4}||p_{t})>-H_{2}(\lambda^{-3/4})\ln(2)-(1-\lambda^{-3/4})\ln\left(\delta_{t}\right).
\end{align*}
If $\lambda^{-3/4}<\frac{1}{2}$ (equivalently, $l>2^{4/3}R$), 
\begin{align*}
D(\lambda^{-3/4}||p_{t})>-\ln2-\frac{1}{2}\ln\delta_{t}.
\end{align*}
Thus, 
\begin{align*}
\mathbb{P}\left\{ \left|\widehat{h}^{(t+1)}\right|\leq\left(\tanh\epsilon_{t}\right)^{(lR^{3})^{1/4}}\right\} \geq1-(2\sqrt{\delta_{t}})^{l-1},
\end{align*}
and finally, 
\begin{align*}
 & \mathbb{P}\left\{ \left|\widehat{h}^{(t+1)}\right|\leq(t+1)\frac{3}{2}\lambda(\tanh\epsilon_{t})^{(lR^{3})^{1/4}}\right\} \\
 & \geq\sum_{r=0}^{3(t+1)\lambda/2}P(r)\mathbb{P}\left\{ \left|h^{(r,t+1)}\right|\leq(t+1)\frac{3}{2}\lambda(\tanh\epsilon_{t})^{(lR^{3})^{1/4}}\right\} \\
 & \geq\sum_{r=0}^{3(t+1)\lambda/2}P(r)\mathbb{P}\left\{ \left|h^{(r,t+1)}\right|\leq r(\tanh\epsilon_{t})^{(lR^{3})^{1/4}}\right\} \\
 & \geq\sum_{r=0}^{3(t+1)\lambda/2}P(r)\left(1-r(2\sqrt{\delta_{t}})^{l-1}\right)\\
 & \geq1-\exp(-(t+1)\frac{\lambda}{10})-\lambda(2\sqrt{\delta_{t}})^{l-1}.
\end{align*}
Or equivalently, 
\[
p_{t+1}=\mathbb{P}\left\{ \vert h^{(t+1)}\vert<\frac{\epsilon_{t+1}}{\beta}\right\} \geq1-\delta_{t+1}.
\]
This completes step 2.

\noindent \textbf{Step 3:} Using lemma \ref{lem:lem2}, for $l$ large
enough (depending on $\beta$ and $R$, but independent of $t$) 
\[
\mathbb{P}\left\{ \left|h^{(t)}\right|>\frac{1}{\beta2^{(t+1)}}\right\} \leq\delta_{t}\leq2\exp\left(-\frac{l}{5R}t\right).
\]
The Borel-Cantelli lemma \cite[Theorem 2.3.1]{durrett10} states that,
$h^{(t)}\to0$ almost surely if for all $\alpha>0$, 
\[
\sum_{t=1}^{\infty}\mathbb{P}\left\{ \left|h^{(t)}\right|>\alpha\right\} <+\infty.
\]
Let us verify that $h^{(t)}$ has this property. For any $\alpha$,
there is $\tau$ such that $1/2^{\tau+1}<\beta\alpha$. Therefore,
for $t\geq\tau$, 
\begin{align*}
\mathbb{P}\left\{ \left|h^{(t)}\right|>\epsilon\right\} \leq\mathbb{P}\left\{ \left|h^{(t)}\right|>\frac{1}{2^{(t+1)}\beta}\right\} <\delta_{t}
\end{align*}
and hence, 
\begin{align*}
\sum_{t=1}^{\infty}\mathbb{P}\left\{ \left|h^{(t)}\right|>\epsilon\right\}  & \leq\tau+\sum_{t=\tau}^{\infty}\mathbb{P}\left\{ \left|h^{(t)}\right|>\epsilon\right\} \\
 & <\tau+\sum_{t=\tau}^{\infty}\delta_{t}\\
 & <\tau+\sum_{t=\tau}^{\infty}2\exp\left(-\frac{l}{10R}t\right)<+\infty.
\end{align*}
This completes step 3.

\section{Proof of Theorem \ref{thm:Poisson}}
\begin{IEEEproof}
\label{proof:Poisson} We first show the property (i). Note that it
is satisfied by $\widehat{q}^{+(0)}$ and $q^{+(0)}$. The equations
\eqref{eq:DE_zero_eta_updates} and \eqref{eq:DE_zero_eta_updates1}
are density evolution equations an LDGM ensemble on the BSC. In \cite{richardson2008modern},
It is known that (i) is preserved under density evolution recursions
(see e.g. \cite{richardson2008modern} for similar properties in the
case of LDPC codes).
\end{IEEEproof}
Let us turn to the proof of (ii). First note that \eqref{eq:etahat_concen}
implies \eqref{eq:etahat_concen2}. Indeed 
\begin{align*}
\mathbb{P}\{\widehat{\eta}^{(t)}<-1+\epsilon\} & =\int_{-1}^{-1+\epsilon}\widehat{q}^{+(t)}(\widehat{\eta})\text{d}\widehat{\eta}\\
 & =\int_{1-\epsilon}^{1}e^{-2\beta\widehat{\eta}}\widehat{q}^{+(t)}(\widehat{\eta})\text{d}\widehat{\eta}\\
 & \geq e^{-2\beta}\mathbb{P}\{\widehat{\eta}^{(t)}>1-\epsilon\}\\
 & \geq\frac{1}{1+e^{2\beta}}(1-\delta).
\end{align*}
So we only have to prove \eqref{eq:etahat_concen}. We will use induction.
The induction hypothesis is \eqref{eq:etahat_concen} for some $\delta>0$
and $\epsilon>0$ at iteration $t$. It is obviously true at $t=0$.

Let us first show that 
\begin{equation}
\mathbb{E}(\eta^{(t)})=\lambda\mathbb{E}(\widehat{\eta}^{(t)})\geq2\lambda s.\label{eq:fff}
\end{equation}
for $s=\frac{1}{2}(1-\delta)(1-\epsilon)(1-e^{-2\beta(1-\epsilon)})/(1+e^{-2\beta})$.
We have 
\begin{align*}
\mathbb{E}(\widehat{\eta}^{(t)}) & =\int_{-1}^{1}\widehat{\eta}\widehat{q}^{+(t)}(\widehat{\eta})\text{d}\widehat{\eta}\\
 & =\int_{-1}^{0}\widehat{\eta}\widehat{q}^{+(t)}(\widehat{\eta})\text{d}\widehat{\eta}+\int_{0}^{1}\widehat{\eta}\widehat{q}^{+(t)}(\widehat{\eta})\text{d}\widehat{\eta}\\
 & =-\int_{0}^{1}\widehat{\eta}e^{-2\beta\widehat{\eta}}\widehat{q}^{+(t)}(\widehat{\eta})\text{d}\widehat{\eta}+\int_{0}^{1}\widehat{\eta}\widehat{q}^{+(t)}(\widehat{\eta})\text{d}\widehat{\eta}\\
 & =\int_{0}^{1}\widehat{\eta}(1-e^{-2\beta\widehat{\eta}})\widehat{q}^{+(t)}(\widehat{\eta})\text{d}\widehat{\eta}\\
 & \geq\int_{1-\epsilon}^{1}\widehat{\eta}(1-e^{-2\beta\widehat{\eta}})\widehat{q}^{+(t)}(\widehat{\eta})\text{d}\widehat{\eta}\\
 & \geq(1-e^{-2\beta(1-\epsilon)})(1-\epsilon)\int_{1-\epsilon}^{1}\widehat{q}^{+(t)}(\widehat{\eta})\text{d}\widehat{\eta}\\
 & >(1-\delta)(1-\epsilon)\frac{1-e^{-2\beta(1-\epsilon)}}{1+e^{-2\beta}}.
\end{align*}
This proves \eqref{eq:fff}.

By applying Hoeffding's inequality \cite{billingsley95} for $\lambda/2<r<3\lambda/2$,
\begin{align*}
 & \mathbb{P}\left\{ \sum_{a=1}^{r}\widehat{\eta}_{a}^{(t)}<\lambda\frac{s}{2}\right\} \\
 & =\mathbb{P}\left\{ \sum_{a=1}^{r}(\widehat{\eta}_{a}^{(t)}-\mathbb{E}(\widehat{\eta}^{(t)}))<\lambda\frac{s}{2}-r\mathbb{E}(\widehat{\eta}^{(t)})\right\} \\
 & \leq\mathbb{P}\left\{ \sum_{a=1}^{r}(\widehat{\eta}_{a}^{(t)}-\mathbb{E}(\widehat{\eta}^{(t)}))<\lambda\frac{s}{2}-2rs\right\} \\
 & \leq\mathbb{P}\left\{ \sum_{a=1}^{r}(\widehat{\eta}_{a}^{(t)}-\mathbb{E}(\widehat{\eta}^{(t)}))<-\lambda\frac{s}{2}\right\} \\
 & <\exp(-\frac{\lambda^{2}s^{2}}{8r})\\
 & <\exp(-\lambda\frac{s^{2}}{12}).
\end{align*}

From 
\begin{align*}
 & \mathbb{P}\left\{ \eta^{(t)}<\lambda\frac{s}{2}\right\} =\sum_{r=0}^{\infty}P(r)\mathbb{P}\left\{ \sum_{a=1}^{r}\widehat{\eta}_{a}^{(t)}<\lambda\frac{s}{2}\right\} \\
 & \leq\sum_{r=0}^{\lambda/2}P(r)+\sum_{r=\lambda/2}^{3\lambda/2}P(r)\mathbb{P}\left\{ \sum_{a=1}^{r}\widehat{\eta}_{a}^{(t)}<\lambda\frac{s}{2}\right\} \\
 & \,\,\,\,+\sum_{r=3\lambda/2}^{\infty}P(r).
\end{align*}
and Lemma \ref{lem:poisson}, we get 
\begin{equation}
\mathbb{P}\left\{ \eta^{(t)}>\lambda\frac{s}{2}\right\} >1-2\exp\left(-\frac{\lambda}{10}\right)-\exp(-\lambda\frac{s^{2}}{12}).\label{eq:eta_lowerbound}
\end{equation}

Now consider the density evolution equation \eqref{eq:DE_zero_eta_updates1}.
We have 
\begin{align*}
 & \mathbb{P}\left\{ \widehat{\eta}^{(t+1)}>\frac{1}{\beta}\text{atanh}\left(\tanh(\beta)\left[\tanh(\beta\lambda\frac{s}{2})\right]^{l-1}\right)\right\} \\
 & \geq\mathbb{P}\left\{ J=1,\eta_{1}^{(t)}>\frac{\lambda s}{2},\dots,\eta_{l-1}^{(t)}>\frac{\lambda s}{2}\right\} \\
 & =\frac{1+\tanh(\beta)}{2}\left(\mathbb{P}\left\{ \eta^{(t)}>\frac{\lambda s}{2}\right\} \right)^{l-1}\\
 & \geq\frac{1+\tanh(\beta)}{2}\left(1-2\exp\left(-\frac{\lambda}{10}\right)-\exp(-\lambda\frac{s^{2}}{12})\right)^{l-1}\\
 & \geq\frac{e^{2\beta}}{1+e^{2\beta}}\left(1-(l-1)\left(2\exp(-\frac{l}{10R})+\exp(-\frac{ls^{2}}{12R})\right)\right)
\end{align*}

Let 
\begin{align*}
1-\varepsilon(l,R,\beta) & =\frac{1}{\beta}\text{atanh}\left(\tanh(\beta)\left[\tanh(\beta s\frac{l}{2R})\right]^{l-1}\right),\\
\Delta(l,R) & =(l-1)\left(2\exp(-0.1\frac{l}{R})+\exp(-\frac{ls^{2}}{12R})\right).
\end{align*}
Inequality \eqref{eq:etahat_concen} holds at $t+1$ , if $\varepsilon(l,R,\beta)\leq\epsilon$
and $\Delta(l,R)\leq\delta$. This is true for $l>l_{1}$ large enough
since $\varepsilon(l,R,\beta)$ and $\Delta(l,R)$ are decreasing
functions of $l$ (for large values of $l$).

\section*{Acknowlegment}

We thank R. Urbanke for insightful discussions and encouragement during
initial stages of this work. Vahid Aref was supported by grant No.
200021-125347, and Marc Vuffray by grant No. 200020-140388 of the
Swiss National Science Foundation.\bibliographystyle{IEEEtran}
\bibliography{references}

\newcommand{\SortNoop}[1]{}
\begin{thebibliography}{10}
\providecommand{\url}[1]{#1}
\csname url@samestyle\endcsname
\providecommand{\newblock}{\relax}
\providecommand{\bibinfo}[2]{#2}
\providecommand{\BIBentrySTDinterwordspacing}{\spaceskip=0pt\relax}
\providecommand{\BIBentryALTinterwordstretchfactor}{4}
\providecommand{\BIBentryALTinterwordspacing}{\spaceskip=\fontdimen2\font plus
\BIBentryALTinterwordstretchfactor\fontdimen3\font minus
  \fontdimen4\font\relax}
\providecommand{\BIBforeignlanguage}[2]{{%
\expandafter\ifx\csname l@#1\endcsname\relax
\typeout{** WARNING: IEEEtran.bst: No hyphenation pattern has been}%
\typeout{** loaded for the language `#1'. Using the pattern for}%
\typeout{** the default language instead.}%
\else
\language=\csname l@#1\endcsname
\fi
#2}}
\providecommand{\BIBdecl}{\relax}
\BIBdecl

\bibitem{Goblick63}
T.~J. Goblick, ``Coding for discrete information source with a distortion
  measure,'' Ph.D. dissertation, MIT, 1963.

\bibitem{ViterbiTrellis74}
A.~Viterbi and J.~Omura, ``Trellis encoding of memoryless discrete-time sources
  with a fidelity criterion,'' \emph{Information Theory, IEEE Transactions on},
  vol.~20, no.~3, pp. 325--332, May 1974.

\bibitem{kostina2012fixed}
V.~Kostina and S.~Verd{\'u}, ``Fixed-length lossy compression in the finite
  blocklength regime,'' \emph{Information Theory, IEEE Transactions on},
  vol.~58, no.~6, pp. 3309--3338, 2012.

\bibitem{ArikanPolar09}
E.~Arikan, ``Channel polarization: A method for constructing capacity-achieving
  codes for symmetric binary-input memoryless channels,'' \emph{Information
  Theory, IEEE Transactions on}, vol.~55, no.~7, pp. 3051--3073, July 2009.

\bibitem{KoradaPolar10}
S.~Korada and R.~Urbanke, ``Polar codes are optimal for lossy source coding,''
  \emph{Information Theory, IEEE Transactions on}, vol.~56, no.~4, pp.
  1751--1768, April 2010.

\bibitem{Tal2013}
I.~Tal and A.~Vardy, ``How to construct polar codes,'' \emph{Information
  Theory, IEEE Transactions on}, vol.~59, no.~10, pp. 6562--6582, Oct 2013.

\bibitem{Martinian2003}
E.~Martinian and J.~Yedidia, ``Iterative quantization using codes on graph,''
  in \emph{Proc. of 41th Annual Allerton Conference on Communication, Control,
  and Computing, (Monticello, IL)}, October 2003.

\bibitem{murayama04}
\BIBentryALTinterwordspacing
T.~Murayama, ``{T}houless-{A}nderson-{P}almer approach for lossy compression,''
  \emph{Phys. Rev. E}, vol.~69, p. 035105, Mar 2004. [Online]. Available:
  \url{http://link.aps.org/doi/10.1103/PhysRevE.69.035105}
\BIBentrySTDinterwordspacing

\bibitem{ciliberti05source}
\BIBentryALTinterwordspacing
S.~Ciliberti and M.~M{\'e}zard, ``The theoretical capacity of the parity source
  coder,'' \emph{Journal of Statistical Mechanics: Theory and Experiment}, vol.
  2005, no.~10, p. P10003, 2005. [Online]. Available:
  \url{http://stacks.iop.org/1742-5468/2005/i=10/a=P10003}
\BIBentrySTDinterwordspacing

\bibitem{Wainwright10LDGManalysis}
M.~Wainwright, E.~Maneva, and E.~Martinian, ``Lossy source compression using
  low-density generator matrix codes: Analysis and algorithms,''
  \emph{Information Theory, IEEE Transactions on}, vol.~56, no.~3, pp.
  1351--1368, March 2010.

\bibitem{Filler07BPLDGM}
T.~Filler and J.~Fridrich, ``Binary quantization using belief propagation with
  decimation over factor graphs of {LDGM} codes,'' in \emph{Proc. 45th Allerton
  Conference on Coding, Communication, and Control, (Monticello, IL)},
  September 2007.

\bibitem{CG10LSC}
D.~Castanheira and A.~Gameiro, ``Lossy source coding using belief propagation
  and soft-decimation over {L}{D}{G}{M} codes,'' in \emph{Personal Indoor and
  Mobile Radio Communications (PIMRC), 2010 IEEE 21st International Symposium
  on}, Sept 2010, pp. 431--436.

\bibitem{ZigFel}
A.~Jimenez~Felstrom and K.~Zigangirov, ``Time-varying periodic convolutional
  codes with low-density parity-check matrix,'' \emph{Information Theory, IEEE
  Transactions on}, vol.~45, no.~6, pp. 2181--2191, Sep 1999.

\bibitem{IDLDPCC}
M.~Lentmaier, A.~Sridharan, D.~Costello, and K.~Zigangirov, ``Iterative
  decoding threshold analysis for {L}{D}{P}{C} convolutional codes,''
  \emph{Information Theory, IEEE Transactions on}, vol.~56, no.~10, pp.
  5274--5289, Oct 2010.

\bibitem{TerminLDPCCCthreshold}
M.~Lentmaier, A.~Sridharan, K.~Zigangirov, and D.~Costello, ``Terminated
  {L}{D}{P}{C} convolutional codes with thresholds close to capacity,'' in
  \emph{Information Theory, 2005. ISIT 2005. Proceedings. International
  Symposium on}, Sept 2005, pp. 1372--1376.

\bibitem{ProtoLDPCC}
M.~Lentmaier, D.~G.~M. Mitchell, G.~P. Fettweis, and D.~J. Costello,
  ``{A}symptotically regular {L}{D}{P}{C} codes with linear distance growth and
  thresholds close to capacity,'' in \emph{Information Theory and Applications
  Workshop (ITA)}, January 2010, pp. 1--8.

\bibitem{Aref11UniRateless}
V.~Aref and R.~Urbanke, ``Universal rateless codes from coupled {L}{T} codes,''
  in \emph{Information Theory Workshop (ITW), 2011 IEEE}, Oct 2011, pp.
  277--281.

\bibitem{kudekar2011threshold}
S.~Kudekar, T.~J. Richardson, and R.~L. Urbanke, ``Threshold saturation via
  spatial coupling: Why convolutional {LDPC} ensembles perform so well over the
  {BEC},'' \emph{Information Theory, IEEE Transactions on}, vol.~57, no.~2, pp.
  803--834, 2011.

\bibitem{kudekar2012spatially}
S.~Kudekar, T.~Richardson, and R.~Urbanke, ``Spatially coupled ensembles
  universally achieve capacity under belief propagation,'' \emph{Information
  Theory, IEEE Transactions on}, vol.~59, no.~12, pp. 7761--7813, Dec 2013.

\bibitem{Yelda12threshsat}
A.~Yedla, Y.-Y. Jian, P.~Nguyen, and H.~Pfister, ``A simple proof of threshold
  saturation for coupled scalar recursions,'' in \emph{Turbo Codes and
  Iterative Information Processing (ISTC), 2012 7th International Symposium
  on}, Aug 2012, pp. 51--55.

\bibitem{kumar2012proof}
S.~Kumar, A.~Young, N.~Macris, and H.~Pfister, ``A proof of threshold
  saturation for irregular {LDPC} codes on {BMS} channels,'' in \emph{Proc. of
  50th Annual Allerton Conference on Communication, Control, and Computing,
  (Monticello, IL)}, October 2012.

\bibitem{Hassani10Couplgraphical}
S.~Hassani, N.~Macris, and R.~Urbanke, ``Coupled graphical models and their
  thresholds,'' in \emph{Information Theory Workshop (ITW), 2010 IEEE}, Aug
  2010, pp. 1--5.

\bibitem{HMU10}
\BIBentryALTinterwordspacing
S.~H. Hassani, N.~Macris, and R.~Urbanke, ``Chains of mean-field models,''
  \emph{Journal of Statistical Mechanics: Theory and Experiment}, vol. 2012,
  no.~02, p. P02011, 2012. [Online]. Available:
  \url{http://stacks.iop.org/1742-5468/2012/i=02/a=P02011}
\BIBentrySTDinterwordspacing

\bibitem{Hassani11SAT}
\BIBentryALTinterwordspacing
S.~Hamed~Hassani, N.~Macris, and R.~Urbanke,
  ``\BIBforeignlanguage{English}{Threshold saturation in spatially coupled
  constraint satisfaction problems},''
  \emph{\BIBforeignlanguage{English}{Journal of Statistical Physics}}, vol.
  150, no.~5, pp. 807--850, 2013. [Online]. Available:
  \url{http://dx.doi.org/10.1007/s10955-012-0664-x}
\BIBentrySTDinterwordspacing

\bibitem{Kudekar10compress}
S.~Kudekar and H.~Pfister, ``The effect of spatial coupling on compressive
  sensing,'' in \emph{Proc. of 48th Annual Allerton Conference on
  Communication, Control, and Computing, (Monticello, IL)}, Aug. 2010, pp. 347
  --353.

\bibitem{Krzakala2012}
\BIBentryALTinterwordspacing
F.~Krzakala, M.~M\'ezard, F.~Sausset, Y.~F. Sun, and L.~Zdeborov\'a,
  ``Statistical-physics-based reconstruction in compressed sensing,''
  \emph{Phys. Rev. X}, vol.~2, p. 021005, May 2012. [Online]. Available:
  \url{http://link.aps.org/doi/10.1103/PhysRevX.2.021005}
\BIBentrySTDinterwordspacing

\bibitem{donoho2012information}
D.~Donoho, A.~Javanmard, and A.~Montanari, ``Information-theoretically optimal
  compressed sensing via spatial coupling and approximate message passing,''
  \emph{Information Theory, IEEE Transactions on}, vol.~59, no.~11, pp.
  7434--7464, Nov 2013.

\bibitem{amuv2012}
V.~Aref, N.~Macris, R.~Urbanke, and M.~Vuffray, ``Lossy source coding via
  spatially coupled {L}{D}{G}{M} ensembles,'' in \emph{Information Theory
  Proceedings (ISIT), 2012 IEEE International Symposium on}, July 2012, pp.
  373--377.

\bibitem{amv2013}
V.~Aref, N.~Macris, and M.~Vuffray, ``Approaching the rate-distortion limit by
  spatial coupling with belief propagation and decimation,'' in
  \emph{Information Theory Proceedings (ISIT), 2013 IEEE International
  Symposium on}.\hskip 1em plus 0.5em minus 0.4em\relax IEEE, 2013, pp.
  1177--1181.

\bibitem{mezard09information}
M.~M{\'e}zard and A.~Montanari, \emph{Information, physics, and
  computation}.\hskip 1em plus 0.5em minus 0.4em\relax Oxford University Press,
  2009.

\bibitem{cover12elements}
T.~M. Cover and J.~A. Thomas, \emph{Elements of information theory}.\hskip 1em
  plus 0.5em minus 0.4em\relax Wiley-interscience, 2012.

\bibitem{CilMezZec06}
S.~Ciliberti, M.~M{\'e}zard, and R.~Zecchina, ``Message-passing algorithms for
  non-linear nodes and data compression,'' \emph{Complexus}, vol.~3, no. 1-3,
  pp. 58--65, 2006.

\bibitem{gmu13}
A.~Giurgiu, N.~Macris, and R.~Urbanke, ``And now to something completely
  different: Spatial coupling as a proof technique,'' in \emph{Information
  Theory Proceedings (ISIT), 2013 IEEE International Symposium on}, July 2013,
  pp. 2443--2447.

\bibitem{mezard2001cavity}
M.~M{\'e}zard and G.~Parisi, ``The {B}ethe lattice spin glass revisited,''
  \emph{The European Physical Journal B-Condensed Matter and Complex Systems},
  vol.~20, no.~2, pp. 217--233, 2001.

\bibitem{franz2003replica}
S.~Franz and M.~Leone, ``Replica bounds for optimization problems and diluted
  spin systems,'' \emph{Journal of Statistical Physics}, vol. 111, no. 3-4, pp.
  535--564, 2003.

\bibitem{guerra2002interpolation}
F.~Guerra and F.~L. Toninelli, ``The thermodynamic limit in mean field spin
  glass models,'' \emph{Communications in Mathematical Physics}, vol. 230,
  no.~1, pp. 71--79, 2002.

\bibitem{iyengar11windowed}
A.~Iyengar, P.~Siegel, R.~Urbanke, and J.~Wolf, ``Windowed decoding of
  spatially coupled codes,'' \emph{Information Theory, IEEE Transactions on},
  vol.~59, no.~4, pp. 2277--2292, April 2013.

\bibitem{hassan12window}
N.~ul~Hassan, A.~Pusane, M.~Lentmaier, G.~Fettweis, and D.~Costello, ``Reduced
  complexity window decoding schedules for coupled {L}{D}{P}{C} codes,'' in
  \emph{Information Theory Workshop (ITW), 2012 IEEE}, Sept 2012, pp. 20--24.

\bibitem{georgii2011gibbs}
H.-O. Georgii, \emph{Gibbs measures and phase transitions}.\hskip 1em plus
  0.5em minus 0.4em\relax Walter de Gruyter, 2011, vol.~9.

\bibitem{semerjian}
\BIBentryALTinterwordspacing
A.~Montanari, F.~Ricci-Tersenghi, and G.~Semerjian, ``Clusters of solutions and
  replica symmetry breaking in random {K}-satisfiability,'' \emph{Journal of
  Statistical Mechanics: Theory and Experiment}, vol. 2008, no.~04, p. P04004,
  2008. [Online]. Available:
  \url{http://stacks.iop.org/1742-5468/2008/i=04/a=P04004}
\BIBentrySTDinterwordspacing

\bibitem{Mezard2006}
\BIBentryALTinterwordspacing
M.~Mézard and A.~Montanari, ``\BIBforeignlanguage{English}{Reconstruction on
  trees and spin glass transition},''
  \emph{\BIBforeignlanguage{English}{Journal of Statistical Physics}}, vol.
  124, no.~6, pp. 1317--1350, 2006. [Online]. Available:
  \url{http://dx.doi.org/10.1007/s10955-006-9162-3}
\BIBentrySTDinterwordspacing

\bibitem{guerra2004high}
\BIBentryALTinterwordspacing
F.~Guerra and F.~Toninelli, ``\BIBforeignlanguage{English}{The high temperature
  region of the {V}iana--{B}ray diluted spin glass model},''
  \emph{\BIBforeignlanguage{English}{Journal of Statistical Physics}}, vol.
  115, no. 1-2, pp. 531--555, 2004. [Online]. Available:
  \url{http://dx.doi.org/10.1023/B%3AJOSS.0000019815.11115.54}
\BIBentrySTDinterwordspacing

\bibitem{Urruty2001}
J.-B.~H. Urruty and C.~Lemar{\'e}chal, \emph{Fundamentals of convex
  analysis}.\hskip 1em plus 0.5em minus 0.4em\relax Springer, 2001.

\bibitem{KYMP2014}
S.~Kumar, A.~Young, N.~Macris, and H.~Pfister, ``Threshold saturation for
  spatially-coupled {L}{D}{P}{C} and {L}{D}{G}{M} codes on {B}{M}{S}
  channels,'' \emph{Information Theory, IEEE Transactions on}, vol.~60, no.~12,
  pp. 7389--7415, 2014.

\bibitem{Tersenghi}
F.~Ricci-Tersenghi and G.~Semerjian, ``On the cavity method for decimated
  random constraint satisfaction problems and the analysis of belief
  propagation guided decimation algorithms,'' \emph{Journal of Statistical
  Mechanics: Theory and Experiment}, vol. P09001, 2009.

\bibitem{coja2011belief}
A.~Coja-Oghlan, ``On belief propagation guided decimation for random {K-SAT},''
  in \emph{Proceedings of the Twenty-Second Annual ACM-SIAM Symposium on
  Discrete Algorithms}.\hskip 1em plus 0.5em minus 0.4em\relax SIAM, 2011, pp.
  957--966.

\bibitem{durrett10}
R.~Durrett, \emph{Probability: theory and examples}.\hskip 1em plus 0.5em minus
  0.4em\relax Cambridge university press, 2010.

\bibitem{billingsley95}
P.~Billingsley, \emph{Probability and measure}.\hskip 1em plus 0.5em minus
  0.4em\relax John Wiley \& Sons, 1995.

\bibitem{richardson2008modern}
T.~Richardson and R.~L. Urbanke, \emph{Modern coding theory}.\hskip 1em plus
  0.5em minus 0.4em\relax Cambridge University Press, 2008.

\end{thebibliography}

\end{document}